\documentclass[rmp,aps,epsfig,twocolumn,amssymb,amsfonts,nofootinbib,yfonts,eufrak,eqsecnum]{revtex4}
\newcommand{\be}{\begin{equation}}
\newcommand{\ee}{\end{equation}}
\newcommand{\bea}{\begin{eqnarray}}
\newcommand{\eea}{\end{eqnarray}}
\newcommand{\br}{{\bf r}}

\usepackage{graphicx}
\usepackage{mathrsfs}
\usepackage{bm}
\usepackage{yfonts}
\usepackage{chapterbib}
\usepackage{amsmath}
\usepackage{amssymb}
\usepackage{amsfonts}
\usepackage{upgreek}
\usepackage{wasysym}
\usepackage{mathbbol}

\DeclareMathAlphabet{\mathpzc}{OT1}{pzc}{m}{it}

\begin{document} 


\title{Anderson Transitions 
}
  
\author{Ferdinand Evers$^{1,2}$ and Alexander~D.~Mirlin$^{1,2,3}$\thanks{Also
  at}}
\date{\today}

\affiliation{ 
$^{1}$ Institut f\"ur Nanotechnologie, Forschungszentrum
Karlsruhe, 76021 Karlsruhe, Germany \\
$^{2}$ Institut f\"ur Theorie der Kondensierten Materie, 
Universit\"at Karlsruhe, 76128 Karlsruhe, Germany
\\$^3$ Petersburg Nuclear Physics Institute, 188300 St.~Petersburg, Russia} 

\begin{abstract}
The physics of Anderson transitions between
localized and metallic phases in disordered systems is reviewed.
The term ``Anderson transition'' is understood in a broad sense,
including both metal-insulator transitions and quantum-Hall-type
transitions between phases with localized states.   
The emphasis is put on recent developments, which include: multifractality 
of critical wave functions, criticality in the power-law random banded
matrix model, symmetry classification of disordered electronic
systems, mechanisms of criticality in quasi-one-dimensional and
two-dimensional systems and survey of corresponding critical
theories, network models, and random Dirac Hamiltonians. Analytical approaches
are complemented by advanced numerical simulations.

\end{abstract}

\maketitle

\tableofcontents

\section{Introduction}
\label{s1}

It is known since the seminal work of \textcite{anderson58}
that disorder can
localize a quantum particle 
inspite of quantum tunneling processes and even if the 
particle is not localized classically.
For a given energy and disorder strength the quantum 
states are either all localized or all delocalized. 
This implies the existence of transitions between
localized and metallic phases in disordered electronic systems, known as
{\em Anderson transitions}. A great progress in understanding of the
corresponding 
physics was achieved in the seventies and the eighties, due to the
developments of 
scaling theory and field-theoretical approaches to localization,
which demonstrated connections between the Anderson transition and
conventional second-order phase transitions. These results were summarized in
several review articles: \cite{lee85,kramer93}, and, in the context of the
quantum Hall transitions, \cite{huckestein95}, as well as in the book
\cite{efetov97}.  

During the last ten years  considerable progress in the field has
been 
made in several research directions. This has strongly advanced the
understanding of the Anderson localization phenomenon and the associated 
quantum phase transition physics in
disordered electronic systems  and allows us to view it
nowadays in a considerably broader and more general context. These
important advances have motivated us in writing the present review.
While the article will also include a brief overview of basic earlier  
results, the main emphasis will be put on recent
developments and, in particular, on novel types of critical systems. 
In this review we understand 
the term ``Anderson transition'' in a broad sense, 
including not only metal-insulator transitions 
but also critical points separating phases with localized states (most
prominently, quantum-Hall-type transitions).

We now list the key developments in the field that took place during
the last decade and constitute the main subject of the review. 

\subsection{Symmetry classification and universality classes}
\label{s1.1}

Within the early classification scheme, three universality classes for the
Anderson transition were identified -- orthogonal, unitary, and symplectic --  
in correspondence with the Wigner-Dyson
classification of random matrix theory ensembles.
Two basic symmetries of this scheme are the invariance of the Hamiltonian
under time reversal and spin rotations.  
More recent research has shown, however,
that this picture is in fact by far incomplete, for two reasons: (i) there
exist more symmetry classes of disordered systems, and (ii) in many cases, 
the symmetry class does not uniquely determine the universality class of the
transition.  

\subsubsection{Additional symmetries.}
\label{s1.1.1}

It has been understood that a complete set of random matrix theories includes, 
in addition to the three Wigner-Dyson classes, three chiral
ensembles and four Bogoliubov-de Gennes ensembles. 
The additional ensembles are characterized by one of the additional
symmetries -- the {\it chiral} or the {\it particle-hole} one.
The field theories ($\sigma$-models) 
associated with these new symmetry classes have in fact been 
considered already in the eighties \cite{hikami82,wegner88}. 
However, it was only after their physical
significance had been better understood that the new symmetry classes were
studied systematically. For the chiral ensembles, important
contributions in this direction were made in
\textcite{gade93,gade91,slevin93,verbaarschot93}.
The particle-hole symmetric ensembles
were called into life several years later \cite{altland97}.
Zirnbauer has also established a relation between
random matrix theories, $\sigma$-models
and Cartan's classification of symmetric spaces (Sec.~\ref{s:SymDisSys}),
which provides the mathematical basis for the statement of completeness
of the new random matrix
classification \cite{zirnbauer96b,heinzner05}. 

Amongst other things,
these developments have led the theorists to predict  
two novel quantum Hall effects, the spin quantum Hall effect (SQHE)
\cite{senthil98,kagalovsky99}, Sec.~\ref{s6.4},  
and the thermal quantum Hall effect (TQHE)
\cite{senthil99,chalker00}, Sec.~\ref{ss:tqhe}. 
Both should occur
in materials with paired fermions where the particle-hole symmetry is
realized.  

\subsubsection{From symmetry classes to universality classes}
\label{s1.1.2}

The classification of fixed points governing the localization transitions 
in disordered metals
has turned out to be much richer than that of
symmetries of random matrix ensembles (or field theories).
The first prominent example of this was in fact given more than 20 years ago by
\textcite{pruisken84} who showed that the quantum Hall transition is
described by a $\sigma$-model with an additional, topological, term. However,
it is only recently that the variety of types of criticality -- particularly
rich in 2D systems -- was fully appreciated:

(i) In several symmetry classes, the field theory ($\sigma$-model) allows
 for inclusion of the topological $\theta$-term (responsible for the quantum
 Hall criticality) or of the Wess-Zumino (WZ) term. 

(ii) The phase diagram may depend on the type of disorder. 
The class D represents a
   prominent example, with three different 
   network-model realizations yielding vastly different phase
   diagrams, Sec.~\ref{ss:tqhe}. 

(iii) In some cases, the field theory may possess a line of fixed points, since
  the coupling constant corresponding to the conductivity is truly marginal. 
This situation is in particular realized in the chiral symmetry classes
  BDI, AIII, and CII, Sec.~\ref{s6.6}. 

(iv)  In some cases, the symmetry of the $\sigma$-model may get
enhanced under renormalization, so that the ultimate fixed-point
theory may have a different form. A paradigm for this behavior is
provided by the $\text{S}^2$ sphere $\sigma$-model with $\theta=\pi$
topological term  (describing a spin-${1\over 2}$ antiferromagnet)
which flows into a $\text{SU}(2)$ Wess-Zumino-Witten (WZW) 
model. It was conjectured that a
similar mechanism may be relevant to some $\sigma$-models of
localization, including the critical theory of the integer quantum
Hall effect (IQHE).

(v) It is possible that the same critical theory is shared by
  systems belonging to different symmetry classes.
  This type of "super-universality" has been proposed to occur in
  disordered wires with critical states, Sec.~\ref{s5.4}, \ref{s5.5}. 
   
(vi) It was recently discovered that Griffiths effects can render 
the conventional RG analysis of a $\sigma$-model insufficient.
In the framework of the RG calculations
the result can be recovered if 
infinitely many relevant couplings are kept, Sec.~\ref{s6.6.2}.


\subsubsection{Many-channel disordered wires}
\label{s1.1.3}

Common wisdom has it that all states in one-dimensional disordered systems are
localized. However, for several symmetry classes wires with critical
states and even such with perfectly transmitting
eigenchannels were identified recently. The emergence of criticality
depends crucially on whether the number of channels is even or odd. 
These developments make a survey of disordered wires (Sec.~\ref{s5}) a
natural part of this review. 

\subsection{Multifractality of wave functions}
\label{s1.2}

It was appreciated by the beginning of nineties that  
wavefunctions at the Anderson transition exhibit strong amplitude fluctuations
that can be characterized as wave function multifractality. The corresponding
results are summarized in the review papers \cite{janssen94a,huckestein95}
published about a decade ago. In more recent years a considerable
progress in the understanding of wave function statistics in metallic
samples \cite{mirlin00} and at criticality
\cite{mudry96,evers00,evers01} has been achieved. 
The multifractality implies the presence of
infinitely many relevant operators, which is a peculiarity of the Anderson
transition critical point, and
the spectrum of multifractal exponents constitutes a crucially important
characteristics of the fixed point governing the transition. 
  The understanding of general
properties of the statistics of critical wave functions and their
multifractality (see Sec.~\ref{s2.3}) 
was complemented by a detailed study -- analytical as well as
numerical -- for a number of localization 
critical points, such as conventional Anderson transition in various
dimensionalities, Dirac fermions in a random vector potential, 
IQHE, SQHE, and symplectic-class Anderson transition in 2D,
as well as the power-law random banded matrix model.

In several situations, the characterization of a critical point by its
multifractality spectrum has turned out to be particularly important.
Specifically, a great deal of recent research activity has been devoted to
conformal theories governing Anderson critical points in 2D systems.
Further, in systems of the new symmetry
classes, peculiar critical points have been found 
that correspond to strong disorder \cite{motrunich02,carpentier00},
such that critical wavefunctions show at the same time some kind of
localization. 
Entrance of a system into such a strong-coupling regime  
manifests itself as a phase transition in the multifractality spectrum
(the ``freezing transition'').

Finally, the notion of multifractality was very recently extended
onto a boundary of a critical system, yielding a novel independent set of
surface critical exponents. The importance of this notion has been confirmed
by analytical and numerical studies of the surface multifractality for several
models at criticality (Sec.~\ref{s2.3.7}).

\subsection{Quantitative understanding of critical behavior}
\label{s1.3}

For several types of Anderson transitions, 
very detailed studies using both analytical and numerical tools have been
performed during the last few years. As a result, 
a fairly comprehensive quantitative understanding 
of the localization critical phenomena has been achieved. 
The following developments played a particularly important role in this
context: 

\subsubsection{Power-law random banded matrix model}
\label{s1.3.1}

An ensemble of power-law random banded matrices (PRBM), 
which can be viewed as a 1D
system with long-range hopping, has been analytically solved on its critical
line \cite{mirlin00a}. This allowed, in particular, a detailed study of
the statistics of wave functions (in particular, multifractality) and
energy levels at criticality.  The PRBM model
serves as a ``toy model'' for the Anderson criticality.
This model possesses a truly marginal coupling, thus yielding a line of
critical points and allowing to study the evolution of critical properties in
the whole range from weak- to strong-coupling fixed points.

\subsubsection{Network models}
\label{s1.3.2}

Formulations of quantum dynamics in terms of network models, pioneered in 
\textcite{chalker88} in the IQHE context, 
have been developed and systematically exploited for both analytical studies
and computer simulations. Such network models have played a key role in
advancement of understanding of Quantum Hall critical points, including the
conventional IQHE and  the systems of
unconventional symmetries -- SQHE and
TQHE \cite{cho97,chalker00,gruzberg99,read01}. 
In particular, the investigation of the network model of SQHE
has led to an analytical understanding of the critical behavior of
a number of most important  physical observables, Sec.~\ref{s6.4}.

\subsubsection{Progress in numerical simulations}
\label{s1.3.3}

During the last ten years
numerical mathematicians have developed highly efficient routines
for diagonalizing sparse matrices.
Combined with the increase in computer power and an improved
understanding of finite size effects, this development has recently paved
the way for highly accurate numerical studies of critical behavior
for a variety of Anderson critical points. 

\subsubsection{Field theories: $\sigma$-models and Dirac fermions}
\label{s1.3.4}

The development of the symmetry classification of disordered systems
has allowed to classify also the corresponding field
theories having a form of nonlinear $\sigma$-models defined on different
symmetric spaces. The renormalization-group
(RG) method was used to analyze them at and near two dimensions. 
A complementary  approach is based on the
analysis of 2D disordered Dirac fermions 
subjected to different types of disorder \cite{nersesyan95,ludwig94}. 
Analytical methods have allowed to
identify fixed points and determine the critical behavior for some types of
disorder corresponding to unconventional symmetry classes. The
interest to the random 
Dirac fermion models has been largely motivated by their applications to
disordered $d$-wave superconductors, see \textcite{altland02} for review. 
Recent breakthrough in the fabrication of
monoatomic graphene sheets and corresponding transport measurements
\cite{novoselov04,novoselov05,zhang05} has
greatly boosted the theoretical activity in this field.


\section{Anderson transitions in conventional symmetry classes} 
\label{s2}

\subsection{Scaling theory, observables, and critical behavior}
\label{s2.1}

Quantum interference can
completely suppress the diffusion of a particle in random potential, a
phenomenon known as {\em Anderson localization}
\cite{anderson58}. When the energy or the disorder strength is 
varied, the system can thus undergo a transition from the metallic
phase with delocalized eigenstates to the insulating phase, where
eigenfunctions are exponentially localized,
\be
\label{e2.1}
|\psi^2(\bf r)| \sim \exp(-|\bf r - {\bf r_0}|/\xi),
\ee
and $\xi$ is the localization length.
The character of this transition remained, however, unclear for
roughly 20 years, until Wegner conjectured, developing earlier ideas of
\textcite{thouless74}, a close connection between the
Anderson transition and  the scaling theory of critical phenomena
\cite{wegner76}.  Three years later,
Abrahams, Anderson, Licciardello, and Ramakrishnan 
formulated a 
{\em scaling theory} of localization \cite{abrahams79}, which
describes the flow of the dimensionless conductance $g$ with the system size
$L$,
\be
\label{e2.2}
d\ln g /d \ln L = \beta(g).
\ee 
This phenomenological theory was put on a solid basis after Wegner's 
discovery of the field-theoretical description of the 
localization problem in terms of a nonlinear $\sigma$-model \cite{wegner79},
Sec.~\ref{s2.2}.  
This paved the way for the resummation of singularities in perturbation 
theory at or near two dimensions \cite{gorkov79,vollhardt80} 
and allowed to cast the scaling in the systematic form of a field-theoretical
RG. A microscopic derivation of the $\sigma$-model worked out 
in a number of papers \cite{schaefer80,juengling80,efetov80}
has completed a case for it as the field theory of the
Anderson localization. 

To analyze the transition, one starts from
the Hamiltonian $\hat{H}$  consisting of the free part
$\hat{H}_0$ and the disorder potential $U({\bf r})$:
\begin{equation}
\hat{H}=\hat{H}_0+U({\bf r})\ ;\qquad \hat{H}_0 =
\hat{{\bf p}}^2 / 2m. 
\label{e2.2a}
\end{equation}
The disorder is defined by the correlation function $\langle U({\bf
  r})U({\bf r'})\rangle$; we can assume it to be of the white-noise
type for definiteness,
\begin{equation}
\langle U({\bf r})U({\bf r'})\rangle=
(2\pi\rho\tau)^{-1}\delta({\bf r}-{\bf r'}). 
\label{e2.2b}
\end{equation}
Here, $\rho$ is the density of states, $\tau$ the mean free
time and $\langle\ldots\rangle$ denote the disorder average.
It may be shown that models with finite-range and/or 
anisotropic disorder correlations
are equivalent with respect to the long-time and long-distance
behavior (hydrodynamics) to the white noise model 
with renormalized parameters 
(tensor of diffusion coefficients) 
\cite{woelfle84}.

More convenient for numerical simulations is the lattice version of
(\ref{e2.2a}), (\ref{e2.2b}) known as the Anderson tight-binding model,
\be
\label{e2.2c}
\hat{H} = t \sum_{\langle ij \rangle} c_i^\dagger c_j + 
\sum_i u_i c_i^\dagger c_i\ ,
\ee
where the sum $\langle ij \rangle$ goes over nearest neighbor sites and
the random site 
energies $u_i$ are chosen from some distribution ${\cal P}(u)$; the
standard choice is the uniform distribution over an interval
$[-W/2;\:W/2]$ (``box distribution'').

The physical observables whose scaling at the transition point is of primary
importance is the localization length $\xi$ 
on the insulating side (say, $E<E_c$)
and the DC conductivity $\sigma$ on the metallic side ($E>E_c$),
\bea
\label{e2.3}
&& \xi \propto (E_c-E)^{-\nu}, \\
&& \sigma \propto (E-E_c)^s.
\label{e2.3a}
\eea
The corresponding critical indices $\nu$ and $s$ satisfy the scaling relation
$s=\nu (d-2)$, first derived in \textcite{wegner76}.

On a more technical level, the localization transition manifests
itself in a change of the behavior of the diffusion propagator,
\be
\label{e2.4}
\Pi({\bf r_1},{\bf r_2}; \omega) = \langle G^R_{E+\omega/2}({\bf r_1}, {\bf
  r_2})  G^A_{E-\omega/2}({\bf r_2}, {\bf r_1})\rangle,
\ee
where $\langle\ldots\rangle$ denotes the disorder averaging and $G^R$,
$G^A$ are retarded and advanced Green functions, 
\be
\label{e2.5}
G^{R,A}_{E}({\bf r}, {\bf r'}) = \langle {\bf r} |
(E-\hat{H}\pm i\eta)^{-1}| {\bf r'}\rangle, \qquad \eta\to +0.
\ee
In the delocalized regime $\Pi$ has the familiar diffusion form (in
the momentum space),
\be
\label{e2.6}
\Pi({\bf q},\omega) = 2\pi \rho(E) / (Dq^2-i\omega),
\ee
where $\rho$ is the density of states and $D$ is the diffusion
constant, related to the conductivity via the Einstein relation 
$\sigma = e^2 \rho D$. In
the insulating phase, the propagator ceases to have the Goldstone form
(\ref{e2.6}) and becomes massive,
\be
\label{e2.7}
\Pi ({\bf r_1},{\bf r_2}; \omega) \simeq {2\pi\rho \over -i\omega} {\cal
  F} (|{\bf r_1}-{\bf r_2}|/\xi),
\ee
with the function ${\cal F}({\bf r})$ decaying exponentially on the
scale of the localization length, ${\cal F}(r/\xi) \sim \exp(-r/\xi)$.
It is worth emphasizing that the localization length $\xi$ obtained
from the averaged correlation function $\Pi = \langle G^RG^A\rangle$,
Eq.~(\ref{e2.4}), is in general different from the one governing the
exponential decay of the typical value $\Pi_{\rm typ} = \exp \langle
\ln G^RG^A\rangle$. For example, in quasi-1D systems the 
two lengths differ by a factor of 4. However, this is 
usually not important for the definition of the 
critical index $\nu$.\footnote{A remarkable exception is the behavior
  of $\xi$ in disordered wires of the chiral symmetry, see
  Eqs.~(\ref{e5.19}), (\ref{e5.20}).}   
We will return to observables that are related to critical
fluctuations of wave functions and discuss the corresponding family of
critical exponents in Sec.~\ref{s2.3}.

\subsection{Field-theoretical description}
\label{s2.2}

\subsubsection{Effective field theory: Non-linear $\sigma$-model}
\label{s2.2.1}

In the original derivation of the $\sigma$-model
\cite{wegner79,schaefer80,juengling80,efetov80},  the
replica trick was used to perform the disorder
averaging. Within this approach, $n$ copies of the system are considered, with
fields $\phi_\alpha$, $\alpha =1,\ldots,n$ describing the particles,
and the replica limit $n\to 0$ is taken in the end. The resulting
$\sigma$-model is defined on the $n\to 0$ limit of either
non-compact or compact symmetric space, depending on whether the
fields $\phi_\alpha$ are considered as bosonic or fermionic. As an
example, for the unitary symmetry class (A), which
corresponds to a system with broken time-reversal invariance, the
$\sigma$-model target manifold is $\text{U}(n,n)/\text{U}(n)\times
\text{U}(n)$ in the first case 
and $\text{U}(2n)/\text{U}(n)\times \text{U}(n)$ in the second case,
with $n \to 0$. A   
supersymmetric formulation given by \textcite{efetov83} combines
fermionic and bosonic degrees of freedom, with the field
$\Phi$ becoming a supervector. The resulting $\sigma$-model is defined
on a supersymmetric coset space,
e.g. $\text{U}(1,1|2)/\text{U}(1|1)\times \text{U}(1|1)$ 
for the unitary class. This manifold combines compact and non-compact
features and represents a product of the hyperboloid $\text{H}^2 =
\text{U}(1,1)/\text{U}(1)\times \text{U}(1)$ and the sphere
$\text{S}^2=\text{U}(2)/\text{U}(1)\times \text{U}(1)$ 
``dressed'' by anticommuting (Grassmannian) variables. 
For a detailed presentation of the
supersymmetry formalism and its applications to mesoscopic systems, 
see \textcite{efetov83,verbaarschot85,efetov97,fyodorov95a,fyodorov97,
guhr98,mirlin00,mirlin00b,zirnbauer04}. While being equivalent
to the replica version on the level of the perturbation theory
(including its RG resummation), the supersymmetry formalism allows also for a
non-perturbative treatment of the theory, which is particularly
important for the analysis of the energy level and eigenfunction
statistics, properties of quasi-1D systems, topological effects, etc.

We briefly sketch the key steps in the conventional 
derivation of the $\sigma$-model; to be specific, we consider the unitary
symmetry class.  
One begins by expressing the  product of the retarded and advanced 
Green functions in terms
of the integral over a supervector field
$\Phi=(S_1,\chi_1,S_2,\chi_2)$: 
\begin{eqnarray}
&& \!\!\!\!\! G^R_{E+\omega/2}({\bf r_1},{\bf r_2}) G^A_{E-\omega/2}({\bf
r_2},{\bf r_1} ) \nonumber \\ 
&& \!\!\!\!\! = \int D\Phi\, D\Phi^\dagger S_1({\bf r_1})
S_1^*({\bf r_2})  S_2({\bf r_2})S_2^*({\bf r_1}) \nonumber\\
&&\!\!\!\!\!
\times  \exp\left\{i\int d{\bf r} \Phi^\dagger({\bf r})
[(E-\hat{H})\Lambda+ {\omega\over 2} +i\eta ]
\Phi({\bf r})\right\},\ \ \ \
\label{e2.8}
\end{eqnarray}
where $\Lambda=\mbox{diag}\{1,1,-1,-1\}$.
After disorder averaging, the resulting quartic term is decoupled via the
Hubbard-Stratonovich transformation, by
introducing a $4\times 4$ supermatrix variable
${\cal R}_{\mu\nu}({\bf r})$ conjugate to the tensor product
$\Phi_\mu({\bf r}) \Phi^\dagger_\nu({\bf r})$. 
Integrating out the $\Phi$ fields, one gets the action in terms of the 
${\cal R}$  fields,
\be
S[{\cal R}] = \pi\rho\tau\! \int\! d^d{\bf r}\, {\rm Str}{\cal R}^2  
+ {\rm Str} \ln [E+({\omega\over 2}+i\eta)\Lambda - \hat{H}_0-{\cal R}],
\label{e2.8a}
\ee
where Str denotes the supertrace. The next step is 
to use the saddle-point approximation, which leads to the
following equation for ${\cal R}$:
\begin{eqnarray}
&& {\cal R}({\bf r})= (2\pi\rho\tau)^{-1} 
\langle{\bf r}|(E-\hat{H}_0-{\cal R})^{-1}|{\bf r}\rangle 
\label{e2.10}
\end{eqnarray}
The relevant set of the solutions (the saddle-point manifold)  has the
form:
\begin{equation}
{\cal R}=\Sigma\cdot I - (i / 2\tau) Q
\label{e2.10a}
\end{equation}
where $I$ is the unity matrix, $\Sigma$ is certain constant, 
and the $4\times 4$ supermatrix 
$Q=T^{-1}\Lambda T$ satisfies the condition $Q^2=1$ and
belongs to the $\sigma$-model target space 
described above. Finally, one performs the gradient
expansion of the second term in (\ref{e2.8a}), for ${\cal R}$ having the
form (\ref{e2.10a}) with a slowly varying $Q({\bf r})$.    
The expression for the propagator $\Pi$, Eq.~(\ref{e2.4}), then reads,
\be
\label{e2.11}
\Pi({\bf r_1},{\bf r_2}; \omega) = \int DQ\:  Q_{12}^{bb}({\bf r_1})
Q_{21}^{bb}({\bf r_2}) e^{-S[Q]},
\ee
where $S[Q]$ is the $\sigma$-model action
\be
\label{e2.12}
S[Q] = {\pi\rho \over 4}\int d^d{\bf r} \:
\mbox{Str}\: [-D(\nabla Q)^2 - 2i\omega\Lambda Q],
\ee 
The size $4$ of the matrix is due to
(i) two types of the Green functions (advanced and retarded),  
and (ii) necessity to introduce bosonic and fermionic
degrees of freedom to represent these Green's function in terms of a
functional integral. The matrix $Q$ consists thus of four $2\times 2$
blocks according to its advanced-retarded structure, 
each of them being a supermatrix in the boson-fermion space. In
particular, $Q_{12}^{bb}$ is the boson-boson element of the RA block,
and so on. 
One can also consider an average of the product of $n$
retarded and $n$ advanced Green functions, which will generate a
$\sigma$-model defined on a larger manifold, with the base being a
product of $\text{U}(n,n)/\text{U}(n)\times \text{U}(n)$ and
$\text{U}(2n)/\text{U}(n)\times \text{U}(n)$ 
(these are the same structures as in the replica formalism, but now
{\it without} the $n\to 0$ limit).  

For other symmetry classes, the symmetry of the $\sigma$-model is
different but the general picture is the same. For example, for the
orthogonal class (AI) the $8\times 8$ 
$Q$-matrices span the manifold whose base is
the product of the non-compact space 
$\text{O}(2,2)/\text{O}(2)\times \text{O}(2)$ and the 
compact space $\text{Sp}(4)/\text{Sp}(2)\times \text{Sp}(2)$.  
The $\sigma$-model symmetric
spaces for all the classes (Wigner-Dyson as well as unconventional)
are listed in Sec.~\ref{s:SymDisSys}.

\subsubsection{RG in $2+\epsilon$ dimensions; $\epsilon$-expansion} 
\label{s2.2.2}

The $\sigma$-model is the effective low-momentum, low-frequency theory
of the problem, describing the dynamics of interacting soft modes --
diffusons and cooperons. Its RG treatment yields a flow equation of
the form (\ref{e2.2}), thus justifying the scaling theory of
localization. The $\beta$-function $\beta(t) \equiv - dt/d\ln L$
can be calculated perturbatively in
the coupling constant $t$ inversely proportional to the dimensional
conductance,  $t=1/2\pi g$.\footnote{For spinful systems, $g$ here
does not include summation over spin projections.}
 This allows one to get the
$\epsilon$-expansion for the critical exponents in $2+\epsilon$
dimensions, where the transition takes place at $t_*\ll 1$. In
particular, for the orthogonal symmetry class (AI) one finds
\cite{wegner88} 
\be
\label{e2.13}
\beta(t) = \epsilon t - 2t^2 - 12 \zeta(3)t^5 +O(t^6).
\ee
The transition point $t_*$ is given by the zero of the $\beta$-function, 
\be 
\label{e2.14}
t_* = {\epsilon\over 2}  - {3\over 8}\zeta(3)\epsilon^4 + O(\epsilon^5).
\ee
The localization length exponent $\nu$ is determined by the derivative
\be 
\label{e2.15}
\nu = -1/\beta'(t_*) = \epsilon^{-1} - {9\over 4} \zeta(3)\epsilon^2 +
O(\epsilon^3),
\ee
 and the conductivity exponent $s$ is
\be 
\label{e2.16}
s = \nu \epsilon = 1 - {9\over 4} \zeta(3)\epsilon^3 +
O(\epsilon^4).
\ee
Numerical simulations of localization on fractals with dimensionality
slightly above 2 give the behavior of $\nu$ that is in good agreement with
Eq.~(\ref{e2.15}) \cite{schreiber96}. 
For the unitary symmetry class (A), the corresponding results read
\bea
\label{e2.17}
&& \hspace*{-1cm} \beta(t) = \epsilon t - 2t^3 - 6t^5 +O(t^7); \\
&&  \hspace*{-1cm}  t_* = \left( {\epsilon \over 2}\right)^{1/2} - {3\over 2} 
 \left( {\epsilon \over 2}\right)^{3/2} + O(\epsilon^{5/2});
\label{e2.18} \\
&& \hspace*{-1cm}
  \nu = {1\over 2\epsilon} - {3\over 4} +O(\epsilon)\ ; \qquad 
s = {1\over 2} - {3\over 4}\epsilon +O(\epsilon^2).
\label{e2.19}
\eea
In 2D ($\epsilon=0$) the fixed point $t_*$ in both cases 
becomes zero: the $\beta$-function
is negative for any $t>0$, implying that all states are localized. The
situation is qualitatively different for the third Wigner-Dyson class -- the
symplectic one. The corresponding $\beta$-function is related to that for the
orthogonal class via $\beta_{\rm Sp}(t) = - 2 \beta_{\rm O} (-t/2)$,
yielding\footnote{Here $t=1/\pi g$, where $g$ is the total
  conductance of the spinful system.} 
\be
\label{e2.20}
\beta(t) = \epsilon t + t^2 - {3\over 4} \zeta(3) t^5 +O(t^6).
\ee
In 2D  the $\beta$-function (\ref{e2.20}) is positive at 
sufficiently small $t$,
implying the existence of a truly metallic phase at $t< t_*$, with an Anderson
transition at certain $t_* \sim 1$, Sec.~\ref{s6.2}. This peculiarity
of the symplectic class 
represents one of possible mechanisms of the emergence of criticality in 2D,
see Sec.~\ref{s6.1}.  The results for the $\beta$-functions in all the
symmetry classes will  be given in Sec.~\ref{s4.6}.

\subsubsection{Additional comments}
\label{s2.2.3}

A few interrelated comments are  in order here.

(i) {\it $\epsilon$-expansion vs 3D exponents.}
The $\epsilon$-expansion is of asymptotic character, yielding
numerically accurate values of the critical exponents only in the
limit of small $\epsilon$. It is thus not surprising that, if
Eqs.~(\ref{e2.15}), (\ref{e2.16}) are used to estimate the indices in
3D ($\epsilon=1$), the best thing to do is to keep just the
leading (one-loop) term, yielding a quite substantial error ($\nu=1$
instead of $\nu\simeq 1.57 \pm 0.02$ \cite{slevin99}  known from numerical
simulations). 
Rather unexpectedly, the agreement turns out to
be remarkably good for the multifractal exponents, see
Sec.~\ref{s2.3.5a}. Independently of the accuracy of these numbers, the
$\epsilon$-expansion plays a major role in understanding of qualitative
properties of the transition. 

(ii) {\it Composite operators.}
The RG can also be used to calculate scaling dimensions of
composite operators. In particular, the operators of the type
$(Q\Lambda)^n$  determine multifractal fluctuations of wave functions
at criticality that will be discussed in Sec.~\ref{s2.3}. Another class of
operators -- those with high derivatives, $({\bf \nabla} Q)^{2n}$ --
were studied in \textcite{kravtsov88,lerner90} and found to
have the scaling dimensions
\be
\label{e2.21}
y_n = d -2n +2 t_*n(n-1) + O(t_*^2),
\ee
where $t_*$ is given by Eq.~(\ref{e2.14}) for the orthogonal class and 
Eq.~(\ref{e2.18}) for the unitary class. The fact that the one-loop
result (\ref{e2.21}) becomes positive for a sufficiently large number of
gradients, $n > t_*^{-1}$ (suggesting that the corresponding operators
are relevant and might drive the system into an unknown fixed point)
has launched a debate about the stability of the $\sigma$-model and
the one-parameter scaling. This question is not specific to the
localization problem but is equally applicable to a broader class of
$\sigma$-models, including the $\text{O}(n)$ model of a Heisenberg
ferromagnet \cite{wegner90,castilla93}. It was, however, pointed out
in  \textcite{brezin97,derkachov97} that since the expansion (\ref{e2.21})
is asymptotic and its true parameter is $t_*n$, the behavior of the
one-loop result at  $t_*n\gtrsim 1$ does not allow to make any
reliable conclusions.

(iii) {\it Order parameter.}
In view of the analogy with continuous thermodynamic 
phase transitions, it is
natural to ask what is the order parameter for the Anderson
transition. While naively Eq.~(\ref{e2.12}) suggests that it is the
expectation value of $Q$, the latter is in fact uncritical. To
describe the transition in terms of  symmetry breaking, one has to
introduce an order parameter {\it function} (OPF) $F(Q)$ resulting from
integrating out $Q$-fields at all points except one (${\bf r_0}$), with
given $Q({\bf r_0})\equiv Q$
\cite{zirnbauer86a,zirnbauer86b,efetov87}. 
One can introduce an OPF $F(\Phi)$ with similar properties also within the
supervector formalism \cite{mirlin91}. It was shown
\cite{mirlin94a,mirlin94b} that the OPF is closely related to the
distribution of one-site Green functions, in particular local density
of states (LDOS), and wave function amplitudes.
In the framework of scattering theory, this suggests an inerpretation
of the Anderson transition as a phenomenon of spontaneous breakdown of 
$S-$matrix unitarity \cite{fyodorov03}.

(iv) {\it Upper critical dimension.}
For conventional critical phenomena, there exists an upper
critical dimension $d_c$ above which the transition is governed by a
Gaussian fixed point, with  exponents being $d$-independent and
given by their mean-field values. As a consequence, an
$\epsilon$-expansion near $d_c$ (in the most standard case, in
$4-\epsilon$ dimensions) exists, alternative to $2+\epsilon$
expansion. One can ask whether this is also the case for the Anderson
localization transition. The answer is negative: there is no conventional mean
field theory for the Anderson transition, and it was argued that the  
the upper critical dimension is $d_c=\infty$
\cite{mirlin94a,mirlin94b}. The closest existing analog of the
mean-field theory is the model on the Bethe lattice corresponding to
$d=\infty$; we will discuss it in more detail in Sec.~\ref{s2.4}.

\subsection{Critical wave functions: Multifractality}
\label{s2.3} 

\subsubsection{Scaling of inverse participation ratios}
\label{s2.3.1}

Multifractality of wave functions, describing their strong
fluctuations at criticality, is a striking feature of the Anderson
transitions \cite{wegner80,castellani86}.   
Multifractality as a concept has been introduced by 
\textcite{mandelbrot74}. Multifractal structures 
are characterized by an infinite set of
critical exponents describing the scaling of the moments of some
distribution. This feature has been observed in various complex
objects, such as the energy dissipating set in turbulence, strange
attractors in chaotic dynamical systems, and the growth probability
distribution in diffusion-limited aggregation. For the present
problem, the underlying normalized measure is just $|\psi^2({\bf r})|$ and
the corresponding moments are the inverse participation ratios (IPR)
\footnote{Strictly speaking, $P_q$ as defined by Eq.~(\ref{e2.22}),
diverges for sufficiently negative $q$ ($q\le -1/2$ for real $\psi$
and $q\le -3/2$ for complex $\psi$), because of zeros of wave
functions related to their oscillations on the scale of the wave
length. To find $\tau_q$ for such negative $q$, one should first
smooth $|\psi^2|$ by averaging over some microscopic volume (block  of
several neighboring sites in the discrete version).}
\be
\label{e2.22}
P_q = \int d^d{\bf r} |\psi({\bf r})|^{2q}.
\ee
At criticality, $P_q$ show an anomalous scaling with the system size
$L$,
\be
\label{e2.23}
\langle P_q\rangle = L^d  \langle|\psi({\bf r})|^{2q}\rangle \sim
L^{-\tau_q},
\ee
governed by a continuous set of exponents $\tau_q$. One often
introduces fractal dimensions $D_q$ via $\tau_q=D_q(q-1)$. In a metal
$D_q=d$, in an insulator $D_q=0$, while at a critical point $D_q$ is a
non-trivial function of $q$, implying wave function
multifractality. Splitting off the normal part, one defines the
anomalous dimensions $\Delta_q$, 
\be
\label{e2.24}
\tau_q\equiv d(q-1)+\Delta_q,
\ee
which distinguish the critical point from the metallic phase and
determine the scale dependence of the wave function correlations.
Among them, $\Delta_2\equiv -\eta$ plays
the most prominent role, governing the spatial correlations of
the ``intensity'' $|\psi|^2$,
\be
L^{2d} \langle |\psi^2({\bf r})\psi^2({\bf r}')|\rangle 
\sim (|{\bf r} - {\bf r}'|/L)^{-\eta}.
\label{e2.25}
\ee
Equation (\ref{e2.25}) can be obtained from (\ref{e2.23}) by using the fact
that the wave function amplitudes become essentially uncorrelated at
$|\br-\br'|\sim L$. Scaling behavior of higher order spatial
correlations, 
$\langle|\psi^{2q_1}({\bf r_1})\psi^{2q_2}({\bf r_2})\ldots
\psi^{2q_n}({\bf r_n})|\rangle$ can be found in a similar way, e.g.
\begin{eqnarray}
&& L^{d(q_1+q_2)} 
\langle |\psi^{2q_1}({\bf r_1})\psi^{2q_2}({\bf r_2})|\rangle
\nonumber \\
&& \sim L^{-\Delta_{q_1}-\Delta_{q_2}}
(|{\bf r_1} - {\bf r_2}|/L)^{\Delta_{q_1+q_2}-\Delta_{q_1}-\Delta_{q_2}}.
\label{e2.25a}
\end{eqnarray}
Correlations of two different (but close in energy) eigenfunctions  
possess the same scaling properties,
\be
\left.\begin{array}{l}
L^{2d} \langle |\psi_i^2({\bf r})\psi_j^2({\bf r}')|\rangle \\
L^{2d} \langle \psi_i({\bf r})\psi_j^*({\bf r})
\psi_i^*({\bf r}')\psi_j({\bf r}')\rangle 
\end{array}  \right\}
\sim \left( {|{\bf r} - {\bf r}'| \over L_\omega}\right)^{-\eta},
\label{e2.26}
\ee
where $\omega=\epsilon_i-\epsilon_j$, 
$L_\omega\sim (\rho\omega)^{-1/d}$, $\rho$ is the density of states, 
and $|{\bf r} - {\bf r}'| < L_\omega$. 
For conventional classes, where the DOS is uncritical, the diffusion
propagator (\ref{e2.4}) scales in the same way. 

In the field-theoretical language (Sec.~\ref{s2.2}), $\Delta_q$ are the
leading anomalous dimensions of the operators ${\rm Tr} (Q\Lambda)^q$
(or, more generally,  ${\rm Tr} (Q\Lambda)^{q_1} \ldots {\rm Tr}
(Q\Lambda)^{q_m} $ with $q_1+\ldots+ q_m =q$) \cite{wegner80}. The
strong multifractal fluctuations of wave functions at criticality are
related to the fact that $\Delta_q<0$ for $q>1$, so that the
corresponding operators increase under RG. In this formalism, the
scaling of correlation functions [Eq.~(\ref{e2.25}) and its
higher-order generalizations] results from an operator product
expansion \cite{wegner85,duplantier91,mudry96}. 

\subsubsection{Singularity spectrum $f(\alpha)$}
\label{s2.3.2}

The average IPR $\langle P_q\rangle$ are (up to the normalization
factor $L^d$) the moments of the distribution function ${\cal
  P}(|\psi|^2)$ of the eigenfunction intensities. The behavior
(\ref{e2.23}) of the moments corresponds to the intensity distribution
function of the form
\begin{equation}
\label{e2.27}
{\cal P}(|\psi^2|) \sim {1\over
|\psi^2|}L^{-d + f (-{\ln |\psi^2|\over \ln L} )}
\end{equation}
Indeed, calculating the moments $\langle |\psi^{2q}|\rangle$ with the
distribution function (\ref{e2.27}), one finds
\begin{equation}
\label{e2.28}
\langle P_q\rangle = L^d \langle |\psi^{2q}|\rangle \sim \int d\alpha\,
L^{-q\alpha+f(\alpha)}\ ,
\end{equation}
where we have introduced $\alpha=-\ln |\psi^2|/ \ln L$. Evaluation of
the integral by the saddle-point method (justified in the limit of
large $L$) reproduces the result
(\ref{e2.23}), with the exponent $\tau_q$ related to the
singularity spectrum $f(\alpha)$ via the Legendre transformation,
\begin{equation}
\label{e2.29}
\tau(q)=q\alpha-f(\alpha)\: , \qquad q=f'(\alpha)\: , \qquad \alpha =\tau'_q. 
\end{equation}
The meaning of the function $f(\alpha)$ is as follows: it 
is the fractal dimension of the set of those points ${\bf r}$ where the
eigenfunction intensity is $|\psi^2({\bf r})| \sim L^{-\alpha}$. 
In other words, in a lattice version of the model the number of such points 
scales as $L^{f(\alpha)}$ \cite{halsey86}.

General properties of the functions $\tau_q$ and $f(\alpha)$ follow
from their definitions and the wave function normalization:

(i) $\tau_q$ is a non-decreasing, convex function ($\tau_q'\ge 0$,
$\tau_q''\le 0$ ), with $\tau_0=-d$, $\tau_1=0$;

(ii) $f(\alpha)$ is a convex function ($f''(\alpha)\le 0$) defined on
the semiaxis $\alpha\ge 0$ with a maximum at some point $\alpha_0$
(corresponding to $q=0$ under the Legendre transformation) and
$f(\alpha_0)=d$. Further, for the point $\alpha_1$ (corresponding to
$q=1$) we have $f(\alpha_1) = \alpha_1$ and $f'(\alpha_1)=1$. 

\begin{figure}[btp]
    \begin{center}\leavevmode
     \includegraphics[width=0.7\linewidth]{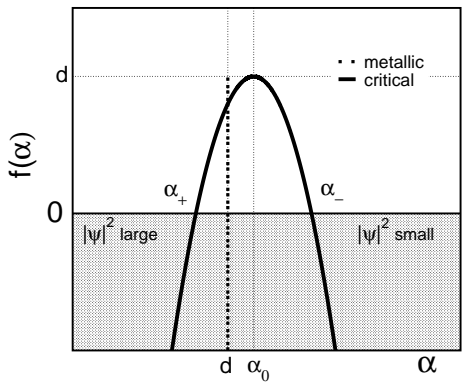}
    \end{center}
\vspace*{-5mm}
      \caption{Schematic plot of the multifractal spectrum,
        $f(\alpha)$. A metal is represented by a ``needle'',
        i.e. $f(\alpha)$ having zero width, at $\alpha=d$.
        At criticality $f(\alpha)$ acquires a finite width
        and the apex, $\alpha_0$, shifts to a value larger than $d$.
        The negative parts of $f(\alpha)$ (grey area) correspond to 
        rare events -- values of the wavefunction
        amplitude that typically do not occur in a single sample.}
\vspace*{-2mm}
      \label{f2.1}
\end{figure}

If one formally defines $f(\alpha)$ for a metal, it will be
concentrated in a single point $\alpha=d$, with $f(d)=d$ and
$f(\alpha)= - \infty$ otherwise. On the other hand, at criticality
this ``needle'' broadens and the maximum shifts to a position
$\alpha_0>d$, see Fig.~\ref{f2.1}. 

\subsubsection{Weak multifractality: approximately parabolic spectrum}
\label{s2.3.3}

One of the situations in which the $\tau_q$ spectrum can be evaluated
analytically is the regime of weak multifractality, when the critical
point is, in a sense, close to a metal. This happens, in particular,
for the Anderson transition in $2+\epsilon$ dimensions with
$\epsilon\ll 1$, see Sec.~\ref{s2.3.5a}, 
and in the PRBM model with $b\gg 1$,
Sec.~\ref{s3.2}. 
In this situation, one finds generically a spectrum of the form
\be
\label{e2.30}
\tau_q \simeq d(q-1) - \gamma q(q-1), \qquad \gamma \ll 1,
\ee
i.e. the anomalous dimension $\Delta_q \simeq \gamma q(1-q)$. (We
remind that $\Delta_0=\Delta_1=0$ by definition.)
The approximation (\ref{e2.30}) is valid in general as long as the
second term ($\Delta_q$) is small compared to the first one, i.e. for
$q\ll d/\gamma$. After the Legendre transformation Eq.~(\ref{e2.30})
yields
\be
\label{e2.31}
f(\alpha) \simeq d - {(\alpha-\alpha_0)^2\over 4 (\alpha_0-d)}; \qquad
\alpha_0 = d+\gamma.
\ee

In some specific cases, the parabolic form of the spectrum
(\ref{e2.30}), (\ref{e2.31}) is not just an approximation but rather
an exact result. This happens, in particular, for the random vector
potential model, see Sec.~\ref{s6.7.3}.  There is a conjecture corroborated by
numerical simulations that this is also the case for the IQHE
transition, see Sec.~\ref{s6.3.7}.  
Note that exact parabolicity cannot extend
to all $q$: at $q_c = (d+\gamma)/2\gamma$ the derivative $\tau_q'$
becomes zero (i.e. the corresponding $\alpha=0$), so that $\tau_q$
should stay constant for larger $q$. We will discuss this issue, known
as ``termination'' of the multifractal spectra, in Sec.~\ref{s2.3.6}.

\subsubsection{Symmetry of the multifractal spectra}
\label{s2.3.4}

\begin{figure}
\begin{center}
\includegraphics[width=0.7\columnwidth]{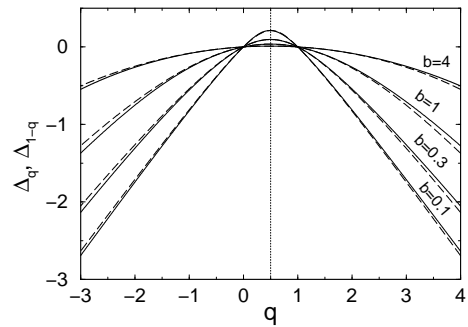}
\vspace*{-0.2cm}
\caption{Multifractal exponents $\Delta_q$ for the PRBM model with
$b=4$, 1, 0.3, 0.1. The symmetry (\ref{e2.32}) with respect to the point
$q=1/2$ is evident. A small difference between $\Delta_q$ (full
line) and $\Delta_{1-q}$ (dashed) is due to numerical errors.
\protect\cite{mildenberger07a}.}
\label{f2.1a}
\end{center}
\vspace*{-0.5cm}
\end{figure}

\begin{figure}
\begin{center}
\includegraphics[width=0.7\columnwidth]{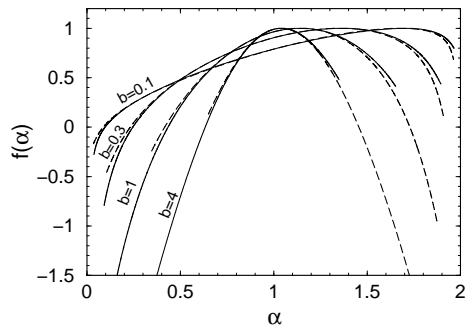}
\vspace*{-0.2cm}
\caption{The data of Fig.~\ref{f2.1a} in terms of the singularity
  spectrum $f(\alpha)$. Dashed lines represent $f(2-\alpha)+\alpha-1$,
  demonstrating the validity of
  Eq.~(\ref{e2.33}). \protect\cite{mildenberger07a}.} 
\label{f2.1b}
\end{center}
\vspace*{-0.5cm}
\end{figure}

As was recently shown \cite{mirlin06}, the multifractal exponents for the
Wigner-Dyson classes satisfy an exact symmetry relation\footnote{If
  the multifractal spectrum possesses a termination (non-analyticity) 
 point $q_c$, Sec.~\ref{s2.3.6},
  the status of the relation (\ref{e2.32}) beyond this point is not clear.} 
\be
\label{e2.32}
\Delta_q = \Delta_{1-q}\ ,
\ee
connecting  exponents with $q<1/2$ (in particular, with negative
$q$) to those with $q>1/2$. In terms of the singularity spectrum, this
implies
\be
\label{e2.33}
f(2d-\alpha) = f(\alpha) + d -\alpha.
\ee
The analytical derivation of Eqs.~(\ref{e2.32}), (\ref{e2.33})
is based on the supersymmetric $\sigma$-model; 
it has been confirmed by numerical simulations on the
PRBM model at criticality \cite{mildenberger07a}, see Figs.~\ref{f2.1a},
\ref{f2.1b} and Sec.~\ref{s3}, and the 2D
Anderson transition of the symplectic class \cite{mildenberger07,obuse07},
Sec.~\ref{s3}.

\subsubsection{Role of ensemble averaging}
\label{s2.3.5}

\paragraph{Average vs. typical spectra.}
It should be stressed that the definition (\ref{e2.23}) of $\tau_q$ is
based on the ensemble-averaged IPRs, $\langle P_q\rangle$. On the
other hand, until
recently most numerical studies of multifractality were
dealing with properties of a single (representative) wave function. 
Formally, this corresponds to an analysis of the typical IPR,
\be
\label{e2.34}
P_q^{\rm typ} = \exp \langle  \ln P_q \rangle.
\ee
Similarly to Eq.~(\ref{e2.23}), one can define the exponents
$\tau_q^{\rm typ}$,
\be
\label{e2.35}
P_q^{\rm typ} \sim  L^{- \tau_q^{\rm typ}},
\ee
and introduce the spectrum $f^{\rm typ}(\alpha)$ as the Legendre
transform of $\tau_q^{\rm typ}$. The relation between $\tau_q$,
$f(\alpha)$, on one side, and $\tau_q^{\rm typ}$, $f^{\rm
typ}(\alpha)$, on the other side, was analyzed in detail in
\textcite{evers00,mirlin00a}.\footnote{In \textcite{evers00,mirlin00a}
 different notations were used: $\tau_q$, $f(\alpha)$ for the typical
 spectra, and  $\tilde{\tau}_q$, $\tilde{f}(\alpha)$ for the averaged
 spectra.}    The function  $\tau_q^{\rm typ}$ has the form
\be
\label{e2.36}
\tau_q^{\rm typ} = \left\{ \begin{array}{ll}
q\alpha_-\ , \qquad & q<q_- \\
\tau_q\ ,    \qquad & q_- < q < q_+ \\
q\alpha_+\ , \qquad & q> q_+\ ,
\end{array}
\right.
\ee
where $\alpha_\pm$ are determined by the condition $f(\alpha)=0$, and
$q_\pm$ are the corresponding values of $q$, with
$q_-<q_+$.\footnote{It is tacidly assumed here, that
  $q_\pm$ and 
  $\alpha_{-}$ actually exist, i.e. are not infinite. To the best of
  our knowledge, an example to the opposite has never been encountered
in the context of the Anderson transitions.}
The singularity spectrum  $f^{\rm typ}(\alpha)$ is defined on the
interval $[\alpha_+,\alpha_-]$, where it is equal to $f(\alpha)$. The
information about the negative part of $f(\alpha)$ (on
$\alpha<\alpha_+$ and $\alpha>\alpha_-$), or, equivalently, about the
part of $\tau_q$ with $q$ outside the range $[q_-,q_+]$, gets lost when
one considers a single wave function. This is because the average
number of points with such a singularity $\alpha$ for a single
eigenfunction is $L^{f(\alpha)}\ll 1$, so that the ensemble averaging
is of crucial importance for determination of this part of the
multifractal spectrum, see Fig.~\ref{f2.1}.

\paragraph{IPR distribution and tail exponents.}
A closely related issue is that of the distribution function of the
IPR $P_q$. It was conjectured in \textcite{fyodorov95} and shown in
\textcite{evers00,mirlin00a} 
that the distribution function of the IPR normalized to its
typical value $P_q^{\rm typ}$ has a scale invariant form 
${\cal P}(P_q / P_q^{\rm typ})$ at criticality. In other words, the
distribution function of the IPR logarithm, ${\cal P}(\ln P_q)$,
preserves its form and 
only shifts along the $x$ axis with increasing $L$. On the large-$P_q$
side, this distribution develops a power-law tail,
\be
\label{e2.37}
{\cal P}(P_q / P_q^{\rm typ}) \propto (P_q / P_q^{\rm typ})^{-1-x_q},
\qquad
P_q \gg P_q^{\rm typ}.
\ee
The upper cutoff of this tail, $(P_q / P_q^{\rm typ})_{\rm max}$
depends on the system size $L$, moving to infinity with $L\to\infty$. 
It is clear that the relation between  $\tau_q^{\rm typ}$ and $\tau_q$
depends crucially on the power-law exponent $x_q$. If $x_q>1$,
the two definitions of the fractal
exponents are identical, $\tau(q) = \tau^{\rm typ}(q)$.  
On the other hand, if $x_q<1$, the average
$\langle P_q\rangle$ is determined by the upper cut-off of
the power-law tail, which depends on $L$. As a result,
$\langle P_q\rangle$ shows scaling with an exponent $\tau^{\rm typ}_q$
different from ${\tau}_q$. In this situation the average value  
$\langle P_q\rangle$ is not representative and is determined by rare
realizations of disorder. Thus, $x_q=1$  for $q=q_\pm$, $x_q>1$ for
$q_-<q<q_+$, and $x_q<1$ otherwise. Furthermore, it was found 
\cite{mirlin00a}  that the power-law-tail index $x_q$ is related to the
fractal exponents as follows:
\be
\label{e2.38}
x_q\tau^{\rm typ}_q = \tau_{qx_q}\ .
\ee
More precisely, Eq.~(\ref{e2.38}) was proven for the case when
$x_q$ is an integer. Also, it was  shown to hold for the small-$b$
limit of the PRBM model, at $q>1/2$. The generic validity of this
formula remains a conjecture. In the range of non-self-averaging IPR
it yields
\be
\label{e2.39}
x_q = q_+/q\ , \qquad q>q_+ ,
\ee
and similarly for $q<q_-$. As to the range $q_-<q<q_+$, the behavior
of $x_q$ depends on the specific form of $\tau_q$. In the particular
case of the weak multifractality, Eq.~(\ref{e2.30}),  the solution of
Eq.~(\ref{e2.38}) reads
\be
\label{e2.40}
x_q \simeq (q_+/q)^2\ , \qquad q_- < q < q_+\ ,
\ee
with 
\be
\label{e2.41}
q_\pm = \pm (d/\gamma)^{1/2}.
\ee

\subsubsection{Dimensionality dependence of the wave function
  statistics at the Anderson transition}
\label{s2.3.5a}

In this subsection, which is largely based on \textcite{mildenberger02}, we
summarize the results for the Anderson transition in $d$ dimensions, obtained
by analytical and numerical means. This allows us to analyze the evolution of
the critical statistics from the weak-multifractality regime in $d=2+\epsilon$
dimensions to the strong multifractality at $d\gg 1$. 

In $2+\epsilon$ dimensions with $\epsilon\ll 1$ the multifractality exponents
can be obtained within the $\epsilon$-expansion, Sec.~\ref{s2.2.2}. The 4-loop
results for the orthogonal and unitary symmetry classes read \cite{wegner87}
\bea
 \Delta_q^{(O)} = q(1-q)\epsilon + {\zeta(3)\over
  4}q(q-1)(q^2-q+1)\epsilon^4 && \nonumber \\ + O(\epsilon^5); && 
\label{e2.42}\\
\Delta_q^{(U)} = q(1-q)(\epsilon/2)^{1/2} -
{3\over 8}q^2(q-1)^2\zeta(3)\epsilon^2 && \nonumber \\
+ O(\epsilon^{5/2}). &&
\label{e2.43}
\eea 
Keeping only the leading (one-loop) term on the r.h.s. of Eqs.~(\ref{e2.42})
and (\ref{e2.43}), we get the parabolic approximation for $\tau_q$,
Eq.~(\ref{e2.30}), and $f(\alpha)$, Eq.~(\ref{e2.31}), with $\gamma =
\epsilon$ for the orthogonal class and $\gamma=(\epsilon/2)^{1/2}$ for the
unitary class. The IPR fluctuations (Sec.~\ref{s2.3.5})
can be studied analytically as well. In particular, in the orthogonal symmetry
class, the variance $\sigma_q$ of the distribution ${\cal P}(\ln P_q)$ is
given, to the leading order in $\epsilon\ll 1$, by 
\be
\label{e2.44}
\sigma_q = 8\pi^2 a\epsilon^2 q^2(q-1)^2\ , \qquad |q|\ll q_+ \ ,
\ee
where $a\simeq 0.00387$ for the periodic boundary conditions. The values
$q_\pm$ of $q$ beyond which the typical and the average IPR scale differently
are given by Eq.~(\ref{e2.41}) with $d\simeq 2$ and the above values of
$\gamma$. The power-law exponent $x_q$ of the IPR distribution is given by 
Eqs.~(\ref{e2.39}), ({\ref{e2.40}). At $q\gg q_+$ the variance $\sigma_q$ is
  governed by the slowly decaying power-law tail, yielding
\be
\label{e2.45}
\sigma_q \simeq x_q^{-1} \simeq q/q_+ \ .
\ee

\begin{figure}
  \begin{center}
\includegraphics[width=0.7\columnwidth,clip]{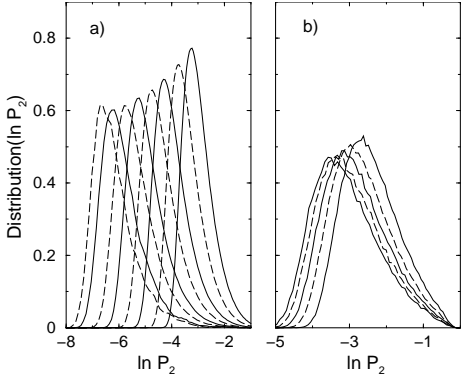}
\vspace*{-0.2cm}
\caption{IPR distribution at the Anderson transition (GOE) (a) 3D (system
  sizes 
         $L=8,11,16,22,32,44,64,80$) and (b) 4D
         ($L=8,10,12,14,16$). \cite{mildenberger02}}
\label{f2.2}
  \end{center}
\vspace*{-0.5cm}
\end{figure}

\begin{figure}
  \begin{center}
\includegraphics[width=0.7\columnwidth]{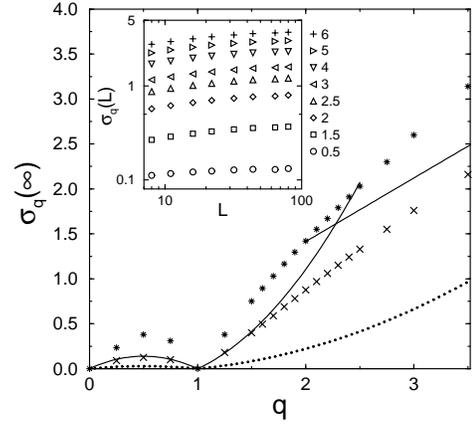}
\vspace*{-0.2cm}
\caption{The rms deviation $\sigma_q$ of $\ln P_q$ extrapolated to
$L\to\infty$ in 3D ($\times$) and 4D ($\ast$). The dotted line is the
analytical result (\ref{e2.44}) for $\epsilon=0.2$; the full lines
represent Eqs.~(\ref{e2.44}), (\ref{e2.45}) with $\epsilon=1$.
Inset: evolution of $\sigma_q$ with $L$ in 3D for  values
of $q=0.5,1.5,2,2.5,3,4,5,6$. The leading finite
size correction of all data has the form
$L^{-y}$ with $y=0.25\div0.5$ for 3D and $y=0.1\div0.4$ in 4D.
\cite{mildenberger02}}
\label{f2.3}
  \end{center}
\vspace*{-0.5cm}
\end{figure}

Results of numerical simulations of the wave function statistics in 3D and 4D
for the orthogonal symmetry class are shown in Figs.~\ref{f2.2}--\ref{f2.5},
in comparison with the one-loop analytical results of the $2+\epsilon$
expansion. Figure \ref{f2.2} demonstrates that the critical IPR distribution
${\cal P}(\ln P_q)$ acquires the scale-invariant form (as also found
in \cite{cuevas01}). The
corresponding variance is shown in Fig.~\ref{f2.3}; in 3D it is
described very well by the analytical formulas with $\epsilon=1$.
The evolution from the weak- to strong-multifractality with increasing $d$ is
nicely seen in Fig.~\ref{f2.4} for $f(\alpha)$. As is demonstrated in the
inset, the one-loop result of the $2+\epsilon$ expansion with $\epsilon=1$
describes the 3D singularity spectrum with a remarkable accuracy (though with
detectable deviations). In particular, the position of the maximum,
$\alpha_0=4.03\pm 0.05$, is very close to its value $\alpha_0=d+\epsilon$
implied by Eq.~(\ref{e2.31}). As expected, in 4D the deviations from
parabolic shape are much more pronounced and $\alpha_0=6.5\pm 0.2$ differs
noticeably from 6.

\begin{figure}
  \begin{center}
\includegraphics[width=0.7\columnwidth,clip]{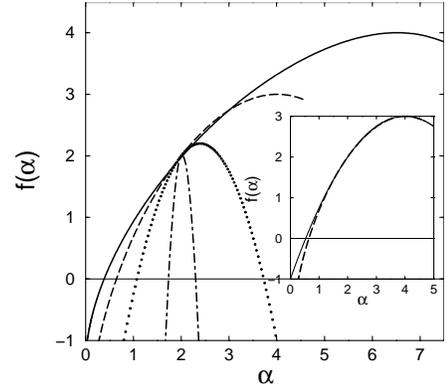}
\vspace*{-0.2cm}
\caption{Singularity spectrum $f(\alpha)$ in 3D (dashed) and
4D (full line). To illustrate the evolution of the spectrum from $d=2$ to
$d=4$, 
analytical results for $d=2+\epsilon$ are shown for $\epsilon=0.2$
(dotted) and $\epsilon=0.01$ (dot-dashed). Inset: comparison between
$f(\alpha)$ for 3D and the one-loop result of the $2+\epsilon$
expansion with $\epsilon=1$ (solid). \cite{mildenberger02}}
\label{f2.4}
  \end{center}
\vspace*{-0.5cm}
\end{figure}

\begin{figure}
  \begin{center}
\includegraphics[width=0.7\columnwidth,clip]{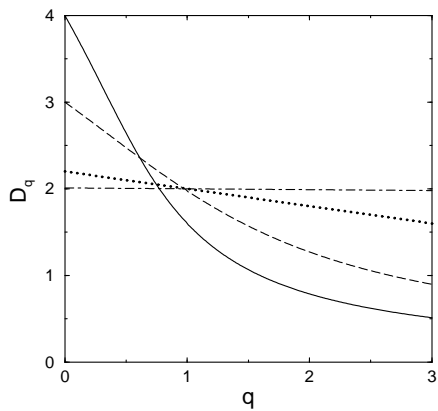}
\vspace*{-0.2cm}
\caption{Fractal dimensions $D_q/q(q-1)$ in 3D (dashed) and 4D (full
line). Analytical results for $d=2+\epsilon$ with $\epsilon=0.2$ (dotted) and
$\epsilon=0.01$ (dot-dashed) are also shown. \cite{mildenberger02}}
\label{f2.5}
  \end{center}
\vspace*{-0.5cm}
\end{figure}

Evolution of the fractal dimension $D_q\equiv \tau_q/(q-1)$ with $d$ is shown
in Fig.~\ref{f2.5}. It is seen that the fractal dimensions $D_q$ with
$q\gtrsim 1$ decrease with increasing $d$. As an example, for $q=2$ we have
$D_2\simeq 2-2\epsilon$ in $2+\epsilon$ dimensions, $D_2=1.3\pm 0.05$ in 3D,
and $D_2=0.9\pm 0.15$ in 4D. This confirms the expectation based on the
Bethe-lattice results  (Sec.~\ref{s2.4}) that $\tau_q\to 0$ at $d\to\infty$ for
$q>1/2$. Such a behavior of the multifractal exponents is a manifestation of a
very sparse character of critical eigenstates at $d\gg 1$, formed by rare
resonance spikes. In combination with the symmetry  relation (\ref{e2.32})
this implies the limiting form of the multifractal spectrum at $d\to\infty$, 
\be
\label{e2.46}
\tau_q=\left\{ \begin{array}{ll}
0\ , & \qquad q\ge 1/2 \\
2d(q-1/2)\ , & \qquad q\le 1/2\ .
\end{array}
\right.
\ee
This corresponds to $f(\alpha)$ of the form
\be
\label{e2.47}
f(\alpha) = \alpha/2\ , \qquad 0<\alpha<2d\ ,
\ee
dropping to $-\infty$ at the boundaries of the interval $[0,2d]$.
In \textcite{mildenberger02}  arguments were given that the way the
multifractality spectrum approaches this limiting form with increasing $d$ is
analogous to the behavior found in the PRBM model with $b\ll 1$,
Sec.~\ref{s3.3}.

\subsubsection{Possible singularities in multifractal spectra:
  termination and freezing}
\label{s2.3.6}

In this subsection, we discuss what kinds of singularities may be
typically encountered in the multifractality spectra $f(\alpha)$ and
$\tau_q$. First of all, we recall that the spectrum $\tau_q^{\rm typ}$
of a typical eigenfunction has non-analyticity points at $q_\pm$,
corresponding to the termination of $f^{\rm typ}(\alpha)$ at its zero
$\alpha_\pm$, see Sec.~\ref{s2.3.5}. However, the ensemble-averaged
spectra $\tau_q$ and  $f(\alpha)$ (that we are considering throughout)
do not have any singularity there.

Singularities in  $\tau_q$ and  $f(\alpha)$ may arise, depending on the
behavior of $f(\alpha)$ at $\alpha=0$ in the particular critical
system under investigation. 
One possibility is that $f(\alpha)$ approaches the $\alpha=0$
axis continuously, with $f(\alpha)\to -\infty$ as $\alpha\to 0$
(Fig.~\ref{f2.6}a). Then $\tau_q$ increases monotonically with $q$, without
any non-analyticities. Such a situation is realized e.g. in the PRBM
model, see Sec.~\ref{s3.3}. An alternative option is that $f(0)$ is finite, see
Fig.~\ref{f2.6}b. This generically implies that $\tau_q$ has a discontinuity
in the second derivative at certain $q_c\equiv f'(\alpha)|_{\alpha\to
  0}$ and is strictly constant, $\tau_q=-f(0)$ at $q\ge q_c$. Such a
behavior of the multifractality spectrum at $q=q_c$ is called
``termination''. In particular, it takes place unavoidably if the
spectrum is exactly parabolic, as is the case, e.g., for the random
vector potential problem, Sec.~\ref{s6.7.3}. From the point of view of the
underlying field theory, termination implies that there is a
qualitative change in properties of the operators ${\cal O}_q$
describing the moments of the density of states. An explicit example
of how this may happen is provided by the 2D Liouville field theory
\cite{seiberg90,zamolodchikov96,kogan96} 
(closely related to the random vector potential
problem), where the operators  ${\cal O}_q$ cease to be local for
$q>q_c$. 

A spectrum with termination may show another peculiarity. While
normally $f(0)$ is negative, one can also imagine a situation with
$f(0)=0$, Fig.~\ref{f2.6}c (corresponding to $q_c\le 1$). In fact, this is
exactly what happens in the random vector potential problem,
Sec.~\ref{s6.7.3}, 
when the disorder strength exceeds a certain critical value. The
transition into this phase is termed ``freezing transition''. In the
``frozen'' phase the wave functions combine properties of
localized and critical states: while the wave function normalization
is governed by a vicinity of one or few (of order unity) points, the
tails away from these points show multifractal fluctuations and
correlations. 

\begin{figure}
  \begin{center}
\includegraphics[width=1.0\columnwidth,clip]{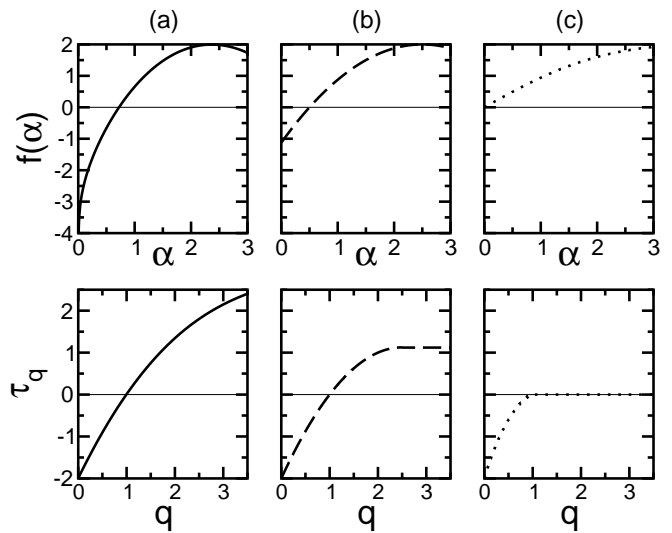}
\caption{Possible behavior of the singularity spectrum $f(\alpha)$ 
at $\alpha\to 0$:
 (a) no singularity, (b) termination, (c) freezing. The corresponding
 behavior of $\tau_q$ is shown as well.}
\label{f2.6}
  \end{center}
\vspace*{-0.5cm}
\end{figure}

\subsubsection{Surface vs. bulk multifractality}
\label{s2.3.7}

Recently, the concept of wave function multifractality was
extended \cite{subramaniam06} to
the surface of a system at the critical  point of an Anderson
transition. It was shown that the fluctuations of
critical wave functions at the surface are characterized by a new set of
exponent $\tau_q^{\rm s}$ (or, equivalently, anomalous exponents
$\Delta_q^{\rm s})$, which are in general independent from
their bulk counterparts, 
\begin{eqnarray}\label{tau-q-s}
&& L^{d-1} \langle |\psi({\bf r})|^{2q} \rangle \sim L^{-\tau_q^{\rm
s}}, \\
&& \tau_q^{\rm s} = d(q-1) + q\mu + 1 + \Delta_q^{\rm s}.
\label{delta-q-s}
\end{eqnarray}
Here $\mu$ is introduced for generality, in order to account for
a possibility of  non-trivial scaling of the average
value, $\langle|\psi({\bf r})|^2\rangle
\propto L^{-d-\mu}$,  at the boundary in unconventional symmetry
classes. For the Wigner-Dyson classes, $\mu=0$.  
The normalization factor
$L^{d-1}$ is chosen such that Eq.~(\ref{tau-q-s}) yields the
contribution of the surface to the inverse participation ratios
$\langle P_q\rangle = \langle \int d^d {\bf r} |\psi({\bf
r})|^{2q} \rangle $. The exponents $\Delta_q^{\rm s}$ as defined in
Eq.~(\ref{delta-q-s})  vanish in a metal and 
govern statistical fluctuations of wave
functions at the boundary, $\langle |\psi({\bf r})|^{2q} \rangle /
\langle |\psi({\bf r})|^2 \rangle^q \sim L^{-\Delta_q^{\rm s}}$,
as well as their spatial correlations, e.g. $L^{2(d+\mu)} \langle
|\psi^2({\bf r})\psi^2({\bf r}')|\rangle \sim (|{\bf r} - {\bf
r}'|/L)^{\Delta_2^{\rm s}}$.

Wave function fluctuations are much stronger at the edge than in the bulk.
As a result, surface exponents are important even if
one performs a multifractal analysis for the whole sample, without
separating it into ``bulk'' and ``surface'', despite the fact that the
weight of surface points is down by a factor $1/L$. 
This was analytically demonstrated in
\textcite{subramaniam06}, using a model of a 2D weakly
localized metallic system (large dimensionless conductance $g\gg
1$), which shows weak multifractality  on length scales below the
localization length $\xi\sim e^{(\pi g)^\beta}$, where $\beta = 1$
(2) for systems with preserved (resp. broken) time-reversal
symmetry. With minor modifications, the formulas below describe
also the Anderson transition in $2 + \epsilon$ dimensions.

For the bulk multifractal spectrum one gets the result 
(\ref{e2.30}) with $\gamma = (\beta\pi g)^{-1} \ll 1$
\cite{wegner80,altshuler86,falko95a,falko95b}; 
generalization of this result to the surface case reads:
\begin{equation}\label{2d-surface-tauq}
\tau_q^{\rm s} = 2(q-1) + 1 + 2\gamma q (1-q).
\end{equation}
The corresponding $f(\alpha)$-spectra have the form:
\begin{eqnarray} \label{2d-bulk-fa}
f^{\rm b}(\alpha) &=& 2 - (\alpha-2-\gamma)^2/4\gamma , \\
f^{\rm s}(\alpha) &=& 1 - (\alpha-2-2\gamma)^2/8\gamma .
\label{2d-surface-fa}
\end{eqnarray}
These results are illustrated in Fig.~\ref{f2.7}. When the
multifractality in 
the whole sample is analyzed, the lowest of the $\tau_q$ exponents
``wins''. The surface effects become
dominant outside the range $q^{\rm bs}_- < q < q^{\rm bs}_+$, 
where $q^{\rm bs}_\pm \simeq \pm \gamma^{-1/2}$ 
are the roots of the equation $\tau_q^{\rm
b}=\tau_q^{\rm s}$. The lower panel of Fig.~\ref{f2.7} shows how
this is translated into the $f(\alpha)$ representation. The total
singularity spectrum is given by the bulk function $f^{\rm
b}(\alpha)$ only for $\alpha_+^{\rm b} < \alpha < \alpha_-^{\rm
b}$, where $\alpha_\pm^{\rm b}-2\simeq \mp 2 \gamma^{1/2}$.
Outside this range the surface effects are important.
Specifically, $f(\alpha)$ is equal to the surface spectrum $f^{\rm
s}(\alpha)$ for $\alpha < \alpha_+^{\rm s}$ and $\alpha >
\alpha_-^{\rm s}$, where $\alpha_\pm^{\rm s}-2\simeq \mp 4
\gamma^{1/2}$, while in the intermediate intervals $\alpha_+^{\rm
s} < \alpha < \alpha_+^{\rm b}$ and $\alpha_-^{\rm b} < \alpha <
\alpha_-^{\rm s}$ its dependence on $\alpha$ becomes linear (shown
by dashed lines). The latter behavior is governed by intermediate
(between  ``bulk'' and ``surface'') points with a distance from
the surface $r\sim L^\beta$, $0 < \beta < 1$; their $f(\alpha)$
spectrum is found to be $f_\beta(\alpha)=\beta f^{\rm
b}(\alpha) + (1-\beta) f^{\rm s}(\alpha)$. Note that in this case
the surface effects modify $f(\alpha)$ in the whole range below
$f(\alpha)\simeq 1$. Therefore, the surface exponents affect the
multifractal spectrum of the sample not only for rare realizations
of disorder (governing the negative part of $f(\alpha)$) but
already in a typical sample.

\begin{figure}
\begin{center}
\includegraphics[width=0.7\columnwidth]{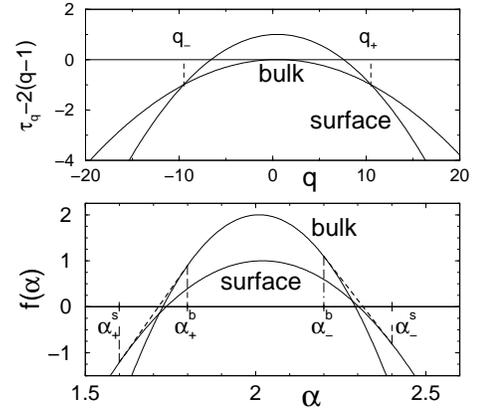}
\vspace*{-0.3cm} \caption{Surface and bulk multifractal spectra
$\tau_q$ and $f(\alpha)$ for a 2D metal with $\gamma=0.01$. For
details see text. \cite{subramaniam06}. } 
\label{f2.7}
\end{center}
\vspace*{-0.7cm}
\end{figure}

The boundary multifractality was also
explicitly studied, analytically as well as numerically, for 
several other systems at criticality:
the 2D spin quantum Hall transition \cite{subramaniam06}, 
the Anderson transition in a 2D system with spin-orbit coupling  
\cite{obuse07}, and the PRBM model \cite{mildenberger07a}, see 
Sec.~\ref{s6.4.5}, \ref{s6.2.4}, and \ref{s3.5}, respectively.  

In  \textcite{obuse07} the notion of surface multifractality was
further generalized to a corner of a critical system. It was shown that
for a 2D system at criticality conformal invariance leads to the
following dependence of the corresponding anomalous exponent
$\Delta_q^\theta$ on the opening angle $\theta$,
\be
\label{e2.48}
\Delta_q^\theta = (\pi /\theta) \Delta_q^{\rm s}.
\ee
More carefully, for $\theta < \pi$ the spectrum will terminate 
at some $q_\theta$ (see Sec.~\ref{s2.3.6}, Fig.~\ref{f2.6}b), even if
the surface spectrum showed no singularity, as in Fig.~\ref{f2.6}a.
Equation (\ref{e2.48}) holds then for $q\le q_\theta$; for larger $q$
the exponent $\tau_q$ is constant.

\subsubsection{Manifestations of multifractality in other observables.}
\label{s2.3.8}

The multifractal structure of wave functions at criticality manifests
itself also in other physical characteristics of the system. In
particular, one can open the system by attaching a local lead at some
point ${\bf r}$. 
The system can then be characterized by the Wigner
delay time $t_W$ (energy derivative of the scattering phase shift),
whose statistical properties in chaotic and disordered systems have
become a subject of research activity in recent years
\cite{fyodorov97,kottos05}.  
At criticality, the moments of the inverse delay time show a scaling
behavior  \cite{ossipov05,mendez-bermudez05,mirlin06,fyodorov03},
\be
\label{e2.49}
\langle t_W^{-q}\rangle \propto L^{-\gamma_q}.
\ee
It was shown that for all  Wigner-Dyson classes the exponents
$\gamma_q$ are linked to the wave function exponents $\tau_q$ by an
exact relation \cite{ossipov05,mirlin06},
\begin{equation}
\label{e2.50}
    \gamma_q = \tau_{1+q}.
\end{equation}

If a second  local lead is attached to the system, the
statistics of the two-point conductance $g({\bf r}, {\bf r'})$ can be
studied. Specifically, one can analyze the scaling of the moments 
$\langle g^q(\br',\br)\rangle$ with the distance 
$|{\bf r}-{\bf r'}|$ between the
contacts \cite{zirnbauer94,janssen99,zirnbauer99}, 
\be
\label{e2.51}
\langle g^q(\br,\br')\rangle \sim |\br-\br'|^{-X_q}.
\ee 
For the case of the unitary symmetry class (A), a relation linking the
exponents $X_q$ to the wave function anomalous dimensions $\Delta_q$
[and based on a result of \textcite{klesse00}]
was obtained  \cite{evers01},    
\be
\label{e2.52}
X_q=\left\{ \begin{array}{ll}
\Delta_q+\Delta_{1-q}\ , & \qquad q<1/2 \\
2\Delta_{1/2}        \ , & \qquad q>1/2. 
\end{array} \right. 
\ee       
In view of (\ref{e2.32}), the first line of Eq.~(\ref{e2.52}) can be
equivalently written as $X_q=2\Delta_q$, which has also been
proposed in \cite{janssen99}. 

A relation between the exponents $X_q$ and $\Delta_q$ was also derived
for the case of the spin quantum Hall transition (Sec.~\ref{s6.4.5}),
belonging 
to the unconventional symmetry class C. In contrast to critical points
of the Wigner-Dyson classes, the density of states $\rho$ in this case
is critical, i.e. it has a non-trivial scaling dimensions $\rho\propto
L^{-x_\rho}$ with $x_\rho > 0$. It was shown in \textcite{mirlin02}
(see also \textcite{bernard02}) that in this case
\be
\label{e2.53}
X_q=\left\{ \begin{array}{ll}
2qx_\rho + 2\Delta_q\ , & \qquad q \le q_0\ , \\
X_{q_0}        \ , & \qquad q> q_0 \ ,
\end{array} \right. 
\ee       
where $q_0$ is the point at which $2qx_\rho + 2\Delta_q$ reaches its
maximum. It is plausible that the relation (\ref{e2.53}) (which
reduces to (\ref{e2.52}) for the Wigner-Dyson classes) holds in fact
for critical points of all symmetry classes.

\subsection{Anderson transition in $d=\infty$: Bethe lattice}
\label{s2.4} 

The Bethe lattice (BL) is a tree-like lattice with a fixed coordination
number. Since the number of sites at a distance $r$ increases
exponentially with $r$ on the BL, it effectively corresponds to
the limit of high dimensionality $d$. As has been already mentioned in
Sec.~\ref{s2.2.3}, the BL models are the closest existing analogs of
the mean-field theory for the case of the Anderson transition. 

The Anderson tight-binding model (lattice version of
Eqs.~(\ref{e2.2a}), (\ref{e2.2b})) on the BL was studied for the first
time in \textcite{abou-chacra73}, where the existence of the
metal-insulator transition was proven and the position of the mobility
edge was determined. The analytical results were confirmed by
numerical simulations \cite{abou-chacra74,girvin80}. 
In later works the BL versions of the
$\sigma$-model (\ref{e2.12})
\cite{efetov85,zirnbauer86a,zirnbauer86b,efetov87} 
and of the tight-binding model \cite{mirlin91} were studied within the
supersymmetry formalism, which allowed to determine the critical
behavior. It was found that the localization length diverges in the
way usual for BL models, $\xi\propto |E-E_c|^{-1}$, where $E$ is a
microscopic parameter driving the transition. When reinterpreted
within the effective-medium approximation  \cite{efetov90,fyodorov92}, 
this yields the conventional mean-field value of the
localization length exponent, $\nu=1/2$. On the other hand, the
critical behavior of other observables is very peculiar. The
inverse participation ratios $P_q$ with $q>1/2$ have a finite limit at
$E\to E_c$ when the critical point is approached from the localized
phase and then jump to zero. By comparison with the scaling formula,
$P_q\propto \xi^{-\tau_q}$, this can be interpreted as $\tau_q=0$ for
all $q\ge 1/2$. Further, in the delocalized phase the diffusion
coefficient vanishes exponentially when the critical point is
approached, 
\begin{eqnarray}
\label{e2.54}
&& D \propto \Omega^{-1} \ln^3 \Omega\ ; \\
&& \Omega \sim \exp\{{\rm const}\:|E-E_c|^{-1/2}\},
\label{e2.55}
\end{eqnarray}
which can be thought as corresponding to the infinite value,
$s=\infty$, of the critical index $s$. The distribution function of
the LDOS $v\equiv\rho(r)/\langle\rho\rangle$ (normalized to its
average value for convenience) was found to be of the form
\be
\label{e2.56}
{\cal P}(v) \propto \Omega^{-1/2} v^{-3/2}\ , \qquad \Omega^{-1} \ll v
\ll \Omega\ , 
\ee
and exponentially small outside this range. 
Equation (\ref{e2.56}) implies the following behavior of the LDOS
moments:
\be
\label{e2.57}
\langle v^q\rangle \propto \Omega^{|q-1/2|-1/2}.
\ee
The physical reason for the unconventional critical behavior was
unravelled in \textcite{mirlin94a,mirlin94b}. 
It was shown that the exponential largeness of the factor
$\Omega$ reflects the spatial structure of the BL: the ``correlation
volume'' $V_\xi$ (number of sites within a distance $\xi$ from the
given one) on such a lattice is exponentially large. On the other
hand, for any finite dimensionality $d$ the correlation volume has a
power-law behavior, $V_d(\xi)\propto \xi^d \propto |E-E_c|^{\nu d}$,
where $\nu\simeq 1/2$ at large $d$. Thus, the scale $\Omega$ cannot
appear for finite $d$ and, assuming some matching between the BL and
large-$d$ results, will be replaced by $V_d(\xi)$. Then
Eq.~(\ref{e2.57}) yields the following high-$d$ behavior of the
anomalous exponents $\Delta_q$ governing the scaling of the LDOS
moments (Sec.~\ref{s2.3}), 
\be
\label{e2.58}
\Delta_q \simeq d(1/2 - |q-1/2|)\ ,
\ee
or, equivalently, the results (\ref{e2.46}), (\ref{e2.47}) for the
multifractal spectra $\tau_q$, $f(\alpha)$. These formulas describe
the strongest possible multifractality. 

The critical behavior of the conductivity, Eq.~(\ref{e2.54}), is
governed by the same exponentially large factor $\Omega$. When it is replaced
by the correlation volume $V_d(\xi)$, the power-law
behavior at finite $d\gg 1$ is recovered, 
$\sigma\propto|E-E_c|^s$ with $s\simeq d/2$. 
The result for the exponent $s$ agrees (within its accuracy,
i.e. to the leading order in $d$) with the scaling relation
$s=\nu(d-2)$.

\subsection{Level statistics at criticality}
\label{s2.5}

We restrict ourselves here to a brief account of key results on the
critical level statistics; more detailed exposition and list of
references can be found in the review \cite{mirlin00}. The primary
quantity is the two-level correlation function (the superscript
``c'' standing for the connected part) 
\be
\label{e2.60}
R_2^{(c)}(s) = \langle \rho\rangle^{-2} 
 \langle \rho(E-\omega/2) \rho(E+\omega/2) \rangle -1 \ ,
\ee
where $\rho(E) = V^{-1}{\rm Tr}\:\delta(E-\hat{H})$ is the fluctuating
DOS, $V$ the system volume, $s=\omega/\Delta$, and $\Delta =
1/\langle\rho\rangle V $ the mean level spacing. In a metallic system 
$R_2^{(c)}(s)$ is given, as a first approximation, by the RMT
(Wigner-Dyson statistics; see
\textcite{mirlin00} for analysis of deviation and limits of
applicability), while in the insulating limit the levels are
uncorrelated,  $R_2^{(c)}(s)= \delta(s)$ (Poisson statistics).
In a critical point the level statistics takes an intermediate
scale-invariant form. Specifically, $R_2^{(c)}(s)$ (and
higher-order correlation functions) is independent under rescaling of
the sample, although it does depend on the sample shape and the
boundary conditions, 
see Fig. \ref{f2.7b}.
\begin{figure}
\begin{center}
\includegraphics[width=0.7\columnwidth]{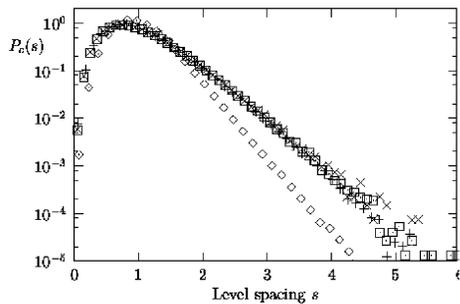}
\vspace*{-0.3cm} 
\caption{Critical level spacing distribution of 2D 
symplectic systems for periodic (PBC) and Dirichlet (DBC) 
boundary conditions and various system sizes $L$.
PBC: $L{=}20$ ($\diamond$); DBC: $L{=}40,80,120$ $(+,\Box,\times)$. 
Adapted from \protect\textcite{schweitzer99}; see also
\textcite{schweitzer97b}.
}
\label{f2.7b}
\end{center}
\vspace*{-0.7cm}
\end{figure}
A closely related quantity is the variance 
$\langle \delta N (E)^2\rangle$ of the number of levels $N(E)$ in a
spectral window of the width $E$,
\be
\label{e2.61}
\langle \delta N (E)^2\rangle = \int _{-\langle N(E)\rangle}^{\langle
  N(E)\rangle} ds (\langle N(E)\rangle - |s|) R_2^{(c)}(s)\ .
\ee
In the RMT limit the level number variance increases logarithmically,
$\langle \delta N^2\rangle =(2/\pi^2\beta) \ln \langle N\rangle $ at
$\langle N\rangle\gg 1$, while in the Poisson limit   
$\langle \delta N^2\rangle = \langle N\rangle$. At criticality,
$\langle \delta N^2\rangle$ shows an intermediate linear behavior
\be
\label{e2.62}
\langle \delta N^2\rangle = \chi \langle N\rangle\ ,
\ee
with a coefficient $0<\chi<1$ called ``spectral compressibility''. The
parameter $\chi$ is a universal characteristics of the critical
theory, i.e. it has a status analogous to critical indices. Evolution
of $\chi$  from the ``quasi-metallic'' ($\chi\ll 1$) to the
``quasi-insulating''  ($\chi$ close to 1) criticality can be
explicitly analyzed for the family of critical PRBM theories,
Sec.~\ref{s3}. It is expected that similar evolution  takes place for the
Anderson transition in $d$ dimensions when $d$ is changed from
$d=2+\epsilon$ to $d\gg 1$, see a related discussion of the wave
function statistics in Sec.~\ref{s2.3.5a}. The ``quasi-metallic''
$d=2+\epsilon$ limit can be studied analytically with the result
$\chi=t^*/\beta$, where the critical coupling $t^*$ is given 
by Eqs.~(\ref{e2.14}) or (\ref{e2.18}), depending on the symmetry
class. The approach of the critical statistics to the Poisson
limit at large $d$ was clearly demonstrated in the recent
numerical work \cite{garcia-garcia06}, 
where systems of dimensionality up to $d=6$ were
studied (with $\chi$ reaching the value $\simeq 0.8$ in 6D). 

In systems of unconventional symmetry classes already the one-point
correlation function (average DOS) is non-trivial and acquires, in
analogy with 
the two-level correlation function discussed above, a scale-invariant
form at criticality. In particular, this will be shown in
Sec.~\ref{s6.4.6} for the case of the SQH transition (class C).


\section{Criticality in the power-law random banded matrix (PRBM)
  model}
\label{s3}

\subsection{Definition and generalities}
\label{s3.1}

The PRBM model is defined  \cite{mirlin96}
as the ensemble of random $L\times L$ Hermitean matrices $\hat H$ 
(real for $\beta=1$ or complex for $\beta=2$). 
The matrix elements $H_{ij}$ are independently distributed
Gaussian variables with zero
mean $\langle H_{ij}\rangle=0$ and with variance 
\begin{equation}
\label{e3.1}
\langle |H_{ij}|^2\rangle \equiv J_{ij} =a^2(|i-j|)\ ,
\end{equation}
where $a(r)$ is given by
\begin{equation}
a^2(r)= [1+(r/b)^{2\alpha}]^{-1}\ .
\label{e3.2}
\end{equation}
At $\alpha=1$ the model undergoes an Anderson transition from the
localized ($\alpha>1$) to the delocalized ($\alpha<1$) phase. Below, 
we concentrate on the critical value $\alpha=1$, when 
$a(r)$ falls down as $a(r)\propto 1/r$ at $r\gg b$. 

In a straightforward interpretation, the PRBM model describes a 1D
sample with random long-range hopping, the hopping amplitude decaying
as $1/r^\alpha$ with the distance. Also, such an ensemble
arises as an effective description in a number of physical contexts,
see \textcite{evers00}  for relevant references.
At $\alpha=1$ the PRBM model is critical for arbitrary
value of $b$ and
shows all the key features of the Anderson critical point, including
multifractality of eigenfunctions and non-trivial spectral
compressibility. The existence of the parameter
$b$ which labels the critical point is a distinct feature of the PRBM
model: Eq.~(\ref{e3.1}) defines a whole family of critical
theories parametrized by $b$.  The limit $b\gg 1$ represents a
regime of weak multifractality, analogous to  
the conventional Anderson transition in $d=2+\epsilon$ with
$\epsilon\ll 1$. This limit allows for a
systematic analytical treatment via a mapping onto a supermatrix
$\sigma$-model and a weak-coupling expansion 
\cite{mirlin96,mirlin00,evers00,mirlin00a}. 
The opposite limit $b\ll 1$ is characterized by very strongly
fluctuating eigenfunctions, similarly to the Anderson transition in $d\gg 1$,
where the transition takes place in the strong disorder (``strong
coupling'' in the field-theoretical language) regime.
It is also accessible to an
analytical treatment using a real-space renormalization-group (RG) 
method \cite{mirlin00a} introduced earlier for related models
by \textcite{levitov90}. 

In addition to the feasibility of the systematic analytical treatment
of both the 
weak-coupling and strong-coupling regimes, the PRBM model is very well
suited for numerical simulations in a broad range of
couplings. For these reasons, it has attracted a considerable interest
in the last few years as a model for the
investigation of various properties of the Anderson critical point,
see Sec.~\ref{s3.5}, \ref{s3.6}.

\subsection{Weak multifractality, $b\gg 1$}
\label{s3.2}


The quasi-metallic regime $b\gg 1$ can be studied 
\cite{mirlin96,mirlin00,evers00,mirlin00a}
via mapping onto the
supermatrix $\sigma$-model, cf. Sec.~\ref{s2.2.1}, 
\begin{equation}
\label{e3.8}
S[Q] = {\pi\rho\beta\over 4}{\rm Str} \left[ \pi \rho  \sum_{rr'}
J_{rr'} Q(r) Q(r') -
i \omega  \sum_r Q(r) \Lambda \right], 
\end{equation}
In momentum ($k$) space and in the low-$k$ limit, 
the action takes the form
\begin{equation}
\label{e3.19}
S[Q]=\beta\, {\rm Str}\left [-{1\over t}\int {dk\over 2\pi}|k| Q_k
Q_{-k} - {i\pi\rho\omega\over 4} Q_0\Lambda\right]\ ,
\end{equation}
where $Q_k=\sum_re^{ikr}Q(r)$ and $Q(r)$ is a $4\times 4$ ($\beta=2$)
or $8\times 8$ ($\beta=1$) supermatrix field constrained by
$Q^2(r)=1$, see Sec.~\ref{s2.2.1}, $\rho$ is 
the density of states given by the Wigner semicircle law
\begin{equation}
\label{e3.20}
\rho(E)=(1 /2\pi^2 b) (4\pi b - E^2)^{1/2}\ , \qquad 
|E|<2\sqrt{\pi b}\ ,
\end{equation}
and $t\ll 1$ is the coupling constant,
\begin{equation}
\label{e3.21}
1/t = (\pi/4)(\pi\rho)^2b^2=(b/ 4)(1- E^2/4\pi b)\ .
\end{equation}
 The main difference between the
action (\ref{e3.19}) and that of the diffusive $\sigma$-model,
Eq.~(\ref{e2.12}) is in the
replacement of the diffusion operator ${\pi\rho\over 8}Dk^2$ by
${1\over t}|k|$. Consequently, all calculations within the weak
coupling expansion of the $\sigma$-model are generalized to the PRBM
case by substituting $\Pi(k)=t/8|k|$ for the diffusion propagator
$\Pi(k)=1/\pi\rho Dk^2$. The $1/|k|$ behavior of $\Pi(k)$ 
implies that the kinetics is superdiffusive, also known as L\'evy
flights, and  leads to criticality in 1D \cite{bouchaud90}. 
 In particular, calculating the average IPR
$\langle P_q\rangle$, one finds the following result for the fractal
dimensions 
\begin{equation}
\label{e3.22}
\tau_q \simeq (q-1)(1 - q t / 8\pi\beta)\ ,\qquad 
q \ll {8\pi\beta / t}\ ,
\end{equation}
i.e. the weak-multifractality-results (\ref{e2.30}), (\ref{e2.31}) for
$\tau_q$ and $f(\alpha)$ with $d=1$ and $\gamma = t/8\pi\beta$. 
For definiteness, we focus below on the band center, $E=0$, where
$\gamma=1/2\pi\beta b$. 

These results are in good agreement with numerical simulations, 
Figs. \ref{fig3.00}, \ref{fig3.1}. 
The deviations from the asymptotic
(parabolic) form in Fig.~\ref{fig3.1}, which are particularly
pronounced at $b=1$,
are a precursor of the crossover to the small-$b$ regime
(Sec.~\ref{s3.3}), where the parabolic approximation breaks down
completely. 

\begin{figure}
\includegraphics[width=0.7\columnwidth,clip]{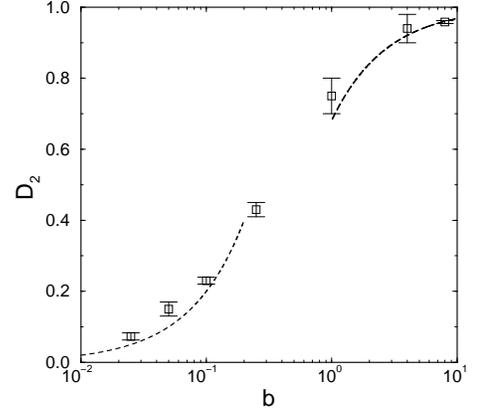}
\vspace*{-0.3cm}
\caption{Fractal dimension $D_2$ as a function of the parameter $b$ of
the PRBM ensemble. The data points are the results of the numerical
simulations, while the lines represent the $b\gg 1$ and $b\ll 1$
analytical asymptotics, $D_2=1-1/\pi b$ [Eq.~(\ref{e3.22})] and
$D_2=2b$ [Eq.~(\ref{e3.47})]. \cite{mirlin00a}.}
\label{fig3.00} 
\end{figure}

\begin{figure}
\includegraphics[width=0.7\columnwidth,clip]{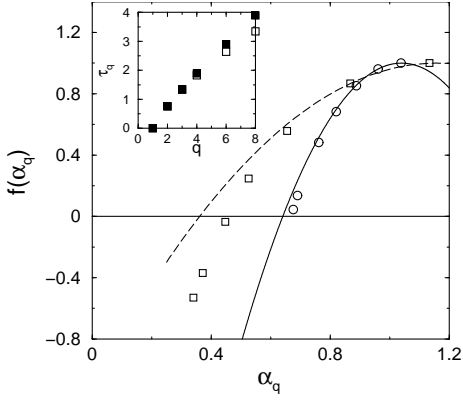}
\vspace*{-0.3cm}
\caption{Multifractal spectrum $f(\alpha)$ for $b=1$ ($\square$) and
$b=4$ ($\circ$). Solid line indicates the parabolic approximation
Eq.~(\ref{e2.31}). Inset: exponent $\tau_q$ ($\square$) and
$\tau^{\rm typ}_q$ ($\blacksquare$)
for $b=1$.
 \cite{mirlin00a}.} 
\label{fig3.1} 
\end{figure}

The IPR fluctuations are also found \cite{evers00,mirlin00a} by 
generalizing the results obtained for metallic samples
\cite{fyodorov95,prigodin98,mirlin00}. 
In particular, the IPR variance is
given for $q\ll q_+(b)\equiv (2\beta\pi b)^{1/2}$ by  
\be
{\rm var}(P_q)/ \langle P_q\rangle^2  
=  q^2(q-1)^2 /  24\beta^2 b^2\ ,
\label{e3.25a}
\ee
cf. Eq.~(\ref{e2.44}).
Calculating higher cumulants, one 
can restore the corresponding scale-invariant distribution function,
\be
{\cal P}(\tilde{P})
= e^{-\tilde{P}-{\bf C}}\exp(-e^{-\tilde{P}-{\bf C}})\ ,
\label{e3.28}
\ee
where 
\begin{equation}
\label{e3.26}
\tilde{P}=\left[{P_q\over \langle P_q\rangle}-1\right]{2\pi\beta
b\over q(q-1)}\ ,
\end{equation}
and ${\bf C}\simeq 0.5772$ is the Euler constant. 
Equation (\ref{e3.28}) is valid for $P_q/\langle P_q\rangle-1\ll 1$.
At $P_q/\langle P_q\rangle -1\sim 1$ the exponential falloff
(\ref{e3.28}) crosses over to a power-law tail 
\begin{equation}
\label{e3.29}
{\cal P}(P_q) \sim (P_q/\langle P_q\rangle)^{-1-x_q}\ .
\end{equation}
with
\begin{equation}
\label{e3.32}
x_q={2\pi\beta b / q^2}\ ,\qquad q^2<2\pi\beta b\ ,
\end{equation}
see the discussion of general properties of the IPR distribution in
Sec.~\ref{s2.3.5}.

\begin{figure}
\begin{center}
\includegraphics[width=0.7\columnwidth,clip]{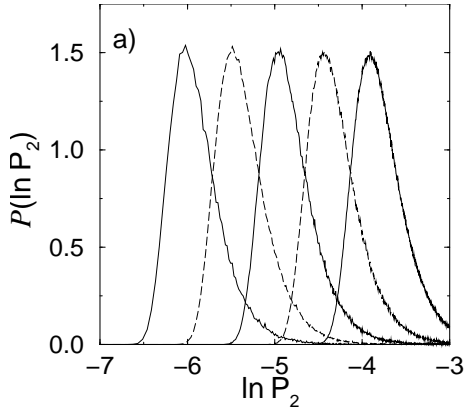} 
\end{center}
\vspace{-5mm}
\caption{Evolution of the distribution ${\cal P}(\ln P_2)$ for $b=1$ 
with the system size $L$ (from
left to right: $L=4096,2048,1024,512,(256)$). The scale invariance of
the IPR distribution is clearly seen. \cite{mirlin00a}. }
\label{fig3.06} 
\end{figure}

\begin{figure}
\includegraphics[width=0.7\columnwidth,clip]{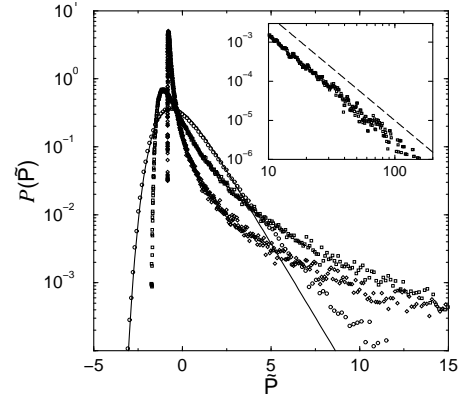}
\vspace{-3mm}
\caption{Distribution function ${\cal P}(\tilde P_q)$ at $q=2$
($\circ$), 4 ($\square$), and 6 ($\diamond$) at
$b=4$ for systems of size $L=4096$. The solid line
represents the analytical result Eq. (\ref{e3.28}). 
Inset: Asymptotic of ${\cal P}(\tilde P_4)$.
Dashed line indicates power law with exponent $x_4=1.7$.
 \cite{mirlin00a}.}
\label{fig3.2} 
\end{figure}

The analytical results on the IPR distribution are confirmed by
numerical simulations. Figure \ref{fig3.06} demonstrates the scale
invariance of the distribution. 
Figure \ref{fig3.2} shows results for the 
distribution of the IPRs $P_q$ with $q=2$, 4, and 6 at $b=4$ (the
corresponding value of $q_+$ being $q_+=(8\pi)^{1/2}\simeq 5$). It is
seen that at $q=2$ the analytical formula (\ref{e3.28}) nicely describes
the ``main body'' of the distribution, with the upward deviations at
large $\tilde{P}$ indicating the crossover to the power-law tail
(\ref{e3.29}). The asymptotic behavior (\ref{e3.29}) is outside the reach
of the numerical simulations for $q=2$, however, since the condition
of its validity $\tilde{P}\gg 2\pi\beta/q(q-1)\simeq 12.5$ corresponds
to very small values of the distribution function ${\cal
P}(\tilde{P})\ll 10^{-5}$, and its clear resolution would require a
much larger statistical ensemble. The situation changes, however, with
increasing $q$ (see the data for $q=4$, 6 in
Fig.~\ref{fig3.2}). Equation (\ref{e3.28}) becomes inapplicable (since the
condition of its validity $q\ll q_+$ is no longer met), and the
power-law asymptotic behavior (\ref{e3.29}) becomes clearly seen. In
particular, the inset of Fig.~\ref{fig3.2} shows the tail for $q=4$; the
extracted value of the index $x_4\simeq 1.7$ is in good agreement with
the prediction of the $b\gg 1$ theory, $x_4=\pi/2$.

\subsection{Strong multifractality, $b\ll 1$}
\label{s3.3}

In the quasi-insulating case $b\ll 1$ the problem can be studied
\cite{mirlin00a} via the
RG method earlier developed for related problems by 
\textcite{levitov90,levitov99a}.  
The idea of the method is as follows. One starts from the diagonal
part of the matrix $\hat{H}$, each eigenstate with an energy
$E_i=H_{ii}$ being localized on a
single site $i$. Then one includes into consideration non-diagonal matrix
elements  $H_{ij}$ with $|i-j|=1$. 
Most of these matrix elements are essentially irrelevant, since their
typical value is $\sim b$, while the energy difference $|E_i-E_j|$ is
typically of order unity. Only with a small probability ($\sim b$) is
$|E_i-E_j|$ also of the order of $b$, so that the matrix element mixes
strongly the two states, which are then said to be in resonance. 
In this case one is led to consider a 
two-level problem 
\begin{equation}
\label{e3.34}
\hat{H}_{\rm two-level}=\left(\begin{array}{cc} 
E_i & V \\ V & E_j
\end{array}\right)\ ;\qquad V=H_{ij}\ .
\end{equation}
The corresponding eigenfunctions and eigenenergies are
\begin{eqnarray}
\label{e3.35}
&& \psi^{(+)}=\left(\begin{array}{c} \cos\theta\\ 
                                   \sin\theta
\end{array}\right)\ ; \qquad
\psi^{(-)}=\left(\begin{array}{c} -\sin\theta\\ 
                                   \cos\theta
\end{array}\right)\: \\
\label{e3.36}
&& E_{\pm}=(E_i+E_j)/2 \pm|V|\sqrt{1+\tau^2}\ ,
\end{eqnarray}
where
$\tan\theta =-\tau +\sqrt{1+\tau^2}$ and $\tau= (E_i-E_j)/2V$.

In the next RG step the matrix elements $H_{ij}$ with $|i-j|=2$ are
taken into account, then those with $|i-j|=3$, and so forth. 
Each time a resonance is encountered, the Hamiltonian is
re-expressed in terms of the new states. Since the probability of a
resonance at a distance $r$ is $\sim b/r$, the typical scale
$r_2$ at which a resonance state formed at a scale $r_1$ will be again
in resonance satisfies
\begin{equation}
\label{e3.39}
\ln (r_2 / r_1) \sim 1/b\ ,
\end{equation}
so that $r_2$ is much larger than $r_1$. Therefore, 
when considering the resonant two-level system at the scale $r_2$, one
can treat the $r_1$-resonance state as point-like. 

This leads to the following evolution equation for the IPR
distribution (for $\beta=1$):
\begin{eqnarray}
\label{e3.44}
&&{\partial\over\partial \ln r}f(P_q,r) =  {2b\over\pi}\int_0^{\pi/2}
{d\theta\over\sin^2\theta\cos^2\theta} \nonumber \\
&& \times  [-f(P_q,r) +\int dP_q^{(1)} dP_q^{(2)} f(P_q^{(1)},r)
f(P_q^{(2)},r) \nonumber \\
&& \times \delta(P_q - P_q^{(1)}\cos^{2q}\theta -
P_q^{(2)}\sin^{2q}\theta)]\ .
\end{eqnarray}
Eq.~(\ref{e3.44}) is a kind of kinetic equation (in the fictitious time
$t= b\ln r$), with the two terms in the square brackets
describing the scattering-out and scattering-in processes,
respectively.

 Multiplying
Eq.~(\ref{e3.44}) by $P_q$ and then integrating over $P_q$, we get the
evolution equation for $\langle P_q\rangle$
\begin{equation}
\label{e3.45}
{\partial\langle P_q\rangle / \partial \ln r} = -2b T (q) 
\langle P_q\rangle
\end{equation}
with
\begin{eqnarray}
\label{e3.46}
T(q) & =& {1\over\pi}\int_0^{\pi/2}
{d\theta\over\sin^2\theta\cos^2\theta}
(1-\cos^{2q}\theta-\sin^{2q}\theta)  \nonumber \\
& = & {2\over\sqrt{\pi}}{\Gamma(q-1/2)\over \Gamma(q-1)} 
\equiv {1\over 2^{2q-3}} {\Gamma(2q-1)\over \Gamma(q)\Gamma(q-1)}\ .
\end{eqnarray}
Integrating (\ref{e3.45}) from $r=1$ to
$r\sim L$, we find the multifractal behavior $\langle P_q\rangle\sim
L^{-\tau_q}$ with the exponents
\begin{equation}
\label{e3.47}
\tau_q = 2b T(q)\ .
\end{equation}
It is assumed here that $q>1/2$, which is the condition of the existence
of the integral in Eq.~(\ref{e3.46}). For $q<1/2$ the resonance
approximation breaks down; the exponents can be found then from the
symmetry relation (\ref{e2.32}).
The function $T(q)$ is shown in Fig.~\ref{fig3.4a}.
Its asymptotics are 
\begin{eqnarray}
\label{e3.48}
& T(q)\simeq -1/[\pi(q-1/2)]\ , \qquad & q\to 1/2\ ; \\
\label{e3.49}
& T(q) \simeq (2 / \sqrt{\pi}) q^{1/2}\ , \qquad & q\gg 1\ .
\end{eqnarray}
We see that the fractal exponents are proportional to the small
parameter $b$. This is characteristic of wave functions that are very
small, typically, with rare and strong peaks (resonances).
In the limit $b\rightarrow 0$ the fractal exponents tend to their
insulator value $\tau_q=0$ for all $q>1/2$. 

\begin{center}
\begin{figure}
\includegraphics[width=0.7\columnwidth,clip]{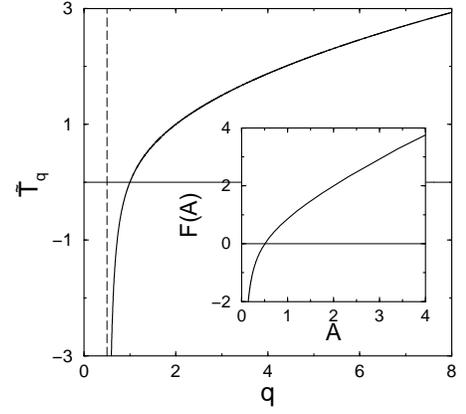}
\vspace{-3mm}
\caption{Universal function $T(q)$ characterizing the exponents
$\tau(q)$ via $\tau(q) = 2b T(q)$ at $b\ll 1$.
Dashed line indicates the pole position.
Inset: Legendre transform $F(A)$ describing the multifractal spectrum
via $f(\alpha )=2bF(\alpha/2b)$.  \cite{mirlin00a}.} 
\label{fig3.4a} 
\end{figure}
\end{center}

Legendre transformation of (\ref{e3.47}) produces the
$f(\alpha)$-spectrum of the form 
\begin{equation}
\label{e3.50}
f(\alpha)=2bF(A)\ ;\qquad A=\alpha/2b\ ,
\end{equation}
where $F(A)$ is the Legendre transform of $T(q)$. The function
$F(A)$ is shown in the inset of  Fig. \ref{fig3.4a}, its asymptotics are
\begin{eqnarray}
\label{e3.51}
& F(A)\simeq - 1/\pi A\ , \qquad  & A\to 0\ ; \\
\label{e3.52}
& F(A)\simeq A/2 \ , \qquad & A \to\infty\ .
\end{eqnarray}
Furthermore, it changes sign at $A_-\simeq 0.5104$, corresponding to
$q_c\simeq 2.4056$. 
These analytical findings are fully supported by numerical
simulations, Fig.~\ref{fig3.4}. The above results,
Eqs.~(\ref{e3.47})--(\ref{e3.52}) are valid for $q>1/2$, which
corresponds to $\alpha<1$. The other part of the spectrum can be
obtained via the symmetry relation (\ref{e2.32}), (\ref{e2.33}), which
is also confirmed by numerical results, see
Figs.~\ref{f2.1a}, \ref{f2.1b}, \ref{fig3a.3}, \ref{fig3a.4}.

\begin{center}
\begin{figure}
\includegraphics[width=0.7\columnwidth,clip]{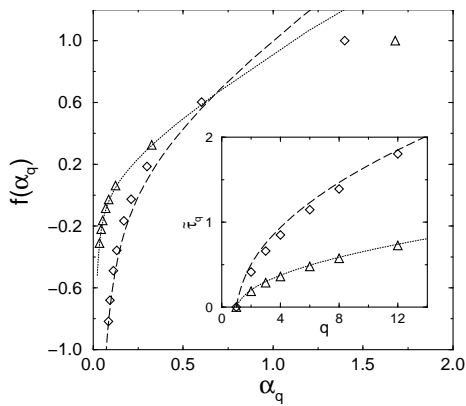}
\vspace{-3mm}
\caption{Multifractal spectrum $f(\alpha)$ for $b=0.25$ ($\Diamond$) and
$b=0.1$ ($\triangle$). Inset: exponent $\tau(q)$.
Dashed and dotted lines indicate the analytical
results Eqs. (\ref{e3.50}) and (\ref{e3.47}).  \cite{mirlin00a}. } 
\label{fig3.4} 
\end{figure}
\end{center}

We return now to the IPR distribution function.
Figure \ref{fig3.3} shows the results of the numerical integration of
Eq.~(\ref{e3.44}) for $q=2$ with the initial condition
$f(P_2)=\delta(P_2-1)$ at $t=0$. It is seen that at sufficiently large $t$
the distribution of $\ln P_2$ acquires a limiting form, shifting with
$t$ without changing its shape. This conclusion of scale-invariance of
the IPR distribution is also supported by analytical
arguments: it is found that the fixed-point distribution has the form
\begin{equation}
\label{e3.53}
f(P_q,r)=r^{\tau^{\rm typ}_q}f_0(P_q r^{\tau^{\rm typ}_q})
\end{equation}
with $\tau^{\rm typ}_q$ as defined in Sec.~\ref{s2.3.5}. 
In agreement with a general discussion in Sec.~\ref{s2.3.5}
the distribution is found to have a power-law tail, 
$f_0(\tilde{P}_q)\sim\tilde{P}^{-x_q-1}$, with the index $x_q$ given
by Eqs.~(\ref{e2.38}), (\ref{e2.39}).

\begin{figure}
\includegraphics[width=0.7\columnwidth,clip]{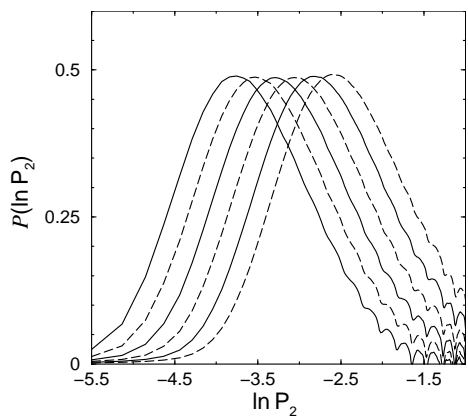}
\vspace{-3mm}
\caption{Flow of the distribution of $\ln P_2 $ calculated from the
kinetic equation (\ref{e3.44})
at $t=b\ln r =1.2\ldots 1.7$ (from right to left). The
oscillations near $\ln P_2 =-1.5$ are
numerical artifacts due to rounding errors. \cite{mirlin00a}. }
\label{fig3.3} 
\end{figure}

All the formulas of this subsection remain valid for the
case $\beta=2$, with a replacement 
$b \longrightarrow (\pi/ 2\sqrt{2})b$. 

\subsection{Levels statistics}
\label{s3.4}

In the $b\gg 1$ regime the two-level correlation function
(\ref{e2.60}) is
obtained  by an appropriate generalization of the earlier findings for
the diffusive samples. In particular, considering for
simplicity the $\beta=2$ ensemble at the band center, the level
correlation function has the form\footnote{The precise form of the
  level correlation function $R_2$ depends on the boundary conditions,
  which are chosen to be periodic in this subsection. The value of the
  spectral compressibility $\eta$ is independent on the boundary
  conditions.} \cite{kravtsov97,mirlin00,mirlin00a} 
\begin{equation}
\label{e3.62}
R_2^{(c)}(s)=\delta(s)-{\sin^2(\pi s)\over (\pi s)^2}
{(\pi s/4b)^2\over \sinh^2(\pi s/4b)}\ .
\end{equation}
The correlation function (\ref{e3.62}) follows the RMT result
$R_2^{(c)}(s)=\delta(s)-\sin^2(\pi s)/ (\pi s)^2$ up to the scale
$s\sim b$ (playing the role of the Thouless energy here), and then
begins to decay exponentially. The spectral compressibility at
$b\gg 1$ is given by 
\begin{equation}
\label{e3.63}
\chi\simeq {1 /  2\pi\beta b}\ ,\qquad b\gg 1\ .
\end{equation}

In the opposite limit, $b\ll 1$, the evolution equation for
$R_2(\omega,r)$ can be written down in analogy with Eq.~(\ref{e3.44})
\cite{mirlin00a}, with the result (for $\beta=1$)
\begin{equation}
\label{e3.66}
R_2^{(c)}(s)=\delta(s)-{\rm erfc}(|s|/ 2\sqrt{\pi}b)\ , 
\end{equation}
where ${\rm erfc}(x)=(2/\sqrt{\pi})\int_x^\infty\exp(-t^2)dt$ is the
error function. This yields the spectral compressibility 
\begin{equation}
\label{e3.67}
\chi\simeq 1-4b\ ,\qquad b\ll 1\ .
\end{equation}
The results for the $\beta=2$ case are qualitatively similar,
\begin{eqnarray}
\label{e3.69}
&& R_2^{(c)}(s)=\delta(s)-\exp(-s^2/ 2\pi b^2)\\
\label{e70}
&& \chi\simeq 1-\pi\sqrt{2}\, b\ , \qquad b\ll 1\ .
\end{eqnarray}
Thus, in the limit of small $b$ the level repulsion
is efficient in a narrow region $|s|\lesssim b$ only, and the spectral
compressibility tends to the Poisson value $\chi=1$.
The physical reason for the reduced range of the level repulsion is
quite transparent. Consider two nearby in energy states separated by a
typical distance $r\sim L$ in the coordinate space. If their energy
difference $s\lesssim b$, such two states will form a resonance pair,
so that their levels will repel. On the other hand, if $s\gg b$, these
two states will not be in resonance, their wave functions remain
weakly overlapping, and the level repulsion between them will be
inefficient.

\begin{figure}
\includegraphics[width=0.7\columnwidth,clip]{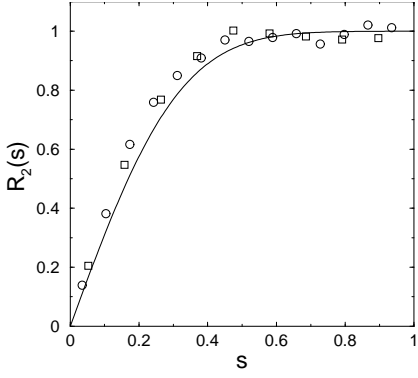}
\vspace{-3mm}
\caption{Two-level correlation function $R_2(s)$ for two system sizes
$L=256$ ($\circ$) and $L=512$ ($\square$) at $b=0.1$. The solid
line indicates the theoretical result (\ref{e3.66}).
 \cite{mirlin00a}.}
\label{fig3.9} 
\end{figure}

\begin{figure}
\includegraphics[width=0.7\columnwidth,clip]{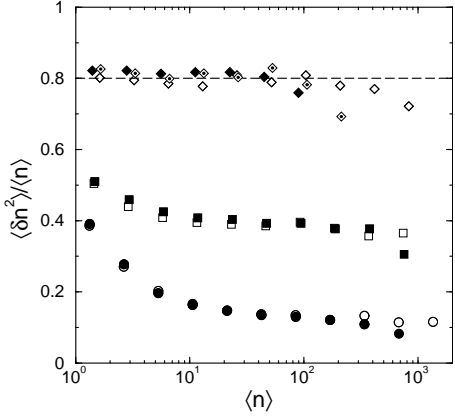}
\vspace{-3mm}
\caption{Variance of the number of levels $\langle \delta n^2 \rangle$
 as a function of 
the energy width of the interval parametrized by the mean
level number $\langle n \rangle$.
Traces correspond to $b=1$ (open $\circ$: $L=4096$, filled: $L=2048$),
$b=0.25$ (open $\square$: $L=4096$, filled: $L=2048$) and
$b=0.05$ (open $\Diamond$: $L=4096$, $\Diamond$ with dot: $L=1024$,
filled: $L=512$). The dashed line indicates the
analytical prediction Eq. (\ref{e3.67}).  \cite{mirlin00a}.}
\label{fig3.5} 
\end{figure}

These results are fully supported by the numerical data. 
Figure \ref{fig3.9} represents the level correlation function
$R_2(s)$ at $b=0.1$, in agreement with Eq.~(\ref{e3.66}). 
The level number variance ${\rm var}[n({\cal E})]$ is plotted
versus the average $\langle n({\cal E})\rangle$ in Fig.~\ref{fig3.5};
the data show an extended plateau region in 
${\rm var}[n({\cal E})]/\langle n({\cal E})\rangle$, determining
$\chi$. The upper bound for this region is set by the matrix size $L$,
while the lower bound is $\sim b$.  
The numerically obtained spectral compressibility
in a broad range of $b$ is shown in Fig.~\ref{fig3.0}; in the
large-$b$ and small-$b$ regions it agrees well with the corresponding
analytical asymptotics. 

\begin{figure}
\includegraphics[width=0.7\columnwidth,clip]{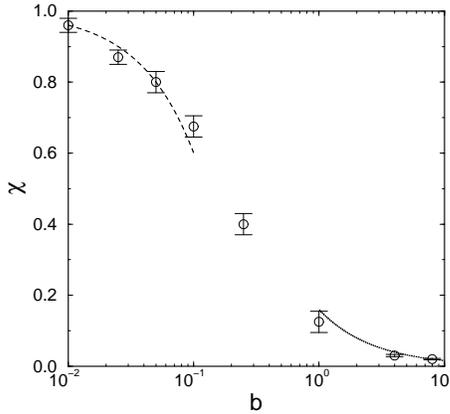}
\vspace{-3mm}
\caption{Spectral compressibility $\chi$ as a function of $b$:
crossover from the ``quasi-metallic'' ($b\gg 1$) 
to the ``quasi-insulating'' ($b\ll 1$) behavior. 
The lines indicate the analytical results for $b\gg 1$ and $b\ll 1$,
Eqs. (\ref{e3.63}) and (\ref{e3.67}).  \cite{mirlin00a}.} 
\label{fig3.0} 
\end{figure}

\subsection{Boundary criticality}
\label{s3.5}

The presentation in this subsection follows
\textcite{mildenberger07a}.  In spirit of Sec.~\ref{s2.3.7}, 
we consider the critical PRBM model model with a boundary at
$i=0$, which means that the matrix element $H_{ij}$ is zero whenever
one of the indices is negative.  The implementation of the
boundary is, however, not unique. The important point is that 
this degree of freedom affects
the boundary criticality. Specifically, one should specify what
happens with a particle which ``attempts to hop'' from a site $i\ge
0$ to a site $j<0$, which is outside of the Hilbert space. One
possibility is that such hops are simply discarded, so that the
matrix element variance $\langle|H_{ij}|^2\rangle \equiv J_{ij}$ 
is simply given by $J_{ij} = [1+(i-j)^2/b^2]^{-1}$ for $i,j\ge 0$. More
generally, the particle may be reflected by the boundary with
certain probability $p$ and ``land'' on the site $-j>0$. This leads
to the following formulation of the model,
\be
\label{e3a.7} 
J_{ij} = \dfrac{1}{1+ \left| i-j \right|^2/b^2} +
\dfrac{p}{1+ \left| i+j \right|^2/ b^2}.
\ee
While the above probability interpretation restricts $p$ to the
interval $[0,1]$, the model is defined for all $p$ in the range
$-1 < p < \infty$. The newly introduced parameter $p$ is immaterial
in the bulk, where $i,j \gg |i-j|$ and the second term in
Eq.~(\ref{e3a.7}) can be neglected. Therefore, the bulk exponents
$\tau_q^{\rm b}$ depend on $b$ only (and not on $p$), and their
analysis in Sec.~\ref{s3.2}, \ref{s3.3} remains applicable
without changes. On the other hand, as discussed below, 
the surface exponents $\tau_q^{\rm s}$ are function of two parameters,
$b$ and $p$. 

Equation (\ref{e3a.7}) describes a semi-infinite system with one
boundary at $i=0$. For a finite system of a length $L$ (implying
that the relevant coordinates are restricted to $0\le i,j \le L$)
another boundary term, $p'/[1+ (i+j-2L)^2 /b^2]$, is to be included
on the right-hand side of Eq.~(\ref{e3a.7}). In general, the parameter
$p'$ of this term may be different from $p$. This term, however,
does not affect the boundary criticality at the $i=0$ boundary, so
it is discarded below.

The regime of weak criticality, $b\gg 1$, can be studied via a
mapping onto the supermatrix $\sigma$-model as in the bulk case,
Sec.~\ref{s3.2}. This results again in an approximately parabolic
spectrum, which, however, differs by a constant factor $R_p$ from its
bulk counterpart, 
\begin{equation}
\label{e3a.20} 
\Delta_q^{\rm s} = [q(1-q) / 2\pi\beta b] R_p \equiv
\Delta_q^{\rm b} R_p.
\end{equation}
This factor is determined from the solution of the classical
Levy flight problem with boundary,
\begin{equation}
\label{e3a.13} 
\frac{\partial W_i(t)}{\partial t} + \pi \rho \sum_{j=0}^{\infty}
\left[\delta_{ij} J_0^{(i)} - J_{ij}  \right] W_j(t) = 0,
\end{equation}
where
$J_0^{(i)} = \sum_{k=0}^{\infty}J_{ik}$, with the initial condition  
$W_i(0) = \delta_{ir}$, where $r$ is near the boundary. 
Specifically, the return probability $W_r(t)$ decays with time as
$1/t$, and the corresponding prefactor yields the fractal  exponents,
\begin{equation}
\label{e3a.21} 
\Delta_q^{\rm s}/ q(1-q) = \beta^{-1} t W_{r}(t)|_{t\to\infty}.
\end{equation}
The evolution equation (\ref{e3a.13})
was solved numerically; the obtained results for
$R_p$ are shown in Fig.~\ref{fig3a.1}.
For $p=1$ the evolution equation  can be obtained from its bulk
counterpart (defined on the whole axis, $-\infty < i 
<\infty$) by ``folding the system'' on the semiaxis $i>0$ according
to $W_i(t)+W_{-i}(t) \longrightarrow W_i(t)$, leading to the
analytical result 
\be
\label{e3a.22}
R_1=2. 
\ee

\begin{figure}
\begin{center}
\includegraphics[width=0.7\columnwidth,clip]{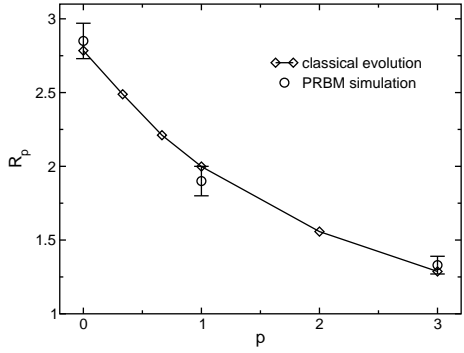}
\caption{The ratio $R_p = \Delta_q^{\rm s}(b,p) / \Delta_q^{\rm
b}(b)$ of the surface and bulk anomalous exponents for
large $b$, as
a function of the reflection parameter $p$. Diamonds represent the
results of the $\sigma$-model analysis with a numerical solution of
the corresponding classical evolution equation (\ref{e3a.13}). 
Circles represent a direct computer simulation of
the PRBM model with $b=8$. 
The ratio $R_p$ has been evaluated for the range $0 < q
< 1$, where the numerical accuracy of the anomalous exponents is the
best. Within this interval the obtained $R_p$ is $q$-independent
(within numerical errors) in agreement with
Eq.~(\ref{e3a.20}). \cite{mildenberger07a}. } 
\label{fig3a.1}
\end{center}
\end{figure}

As in the bulk case, 
the regime of small $b$ can be studied via 
the real-space RG method (Sec.~\ref{s3.3}). 
The multifractal exponents are found to be 
\begin{equation}
\label{e3a.28} 
\tau_q^{\rm s} = (1+p)^{1/2} b T(q) = [(1+p)^{1/2} /
2] \tau_q^{\rm b},
\end{equation}
with $T(q)$ given by Eq.~(\ref{e3.46}), i.e. they differ from the bulk
exponents by the factor $(1+p)^{1/2}$. In full analogy with the bulk
formula (\ref{e3.47}), the result  (\ref{e3a.28}) is valid for
$q>1/2$,  where the
multifractal exponent $\tau_q$ is small. The results
can, however be extended to the range of $q<1/2$ by using the
symmetry relation (\ref{e2.32}). For all $q$ the
obtained relation between the surface and the bulk multifractal
spectra can be formulated in the following way:
\begin{equation}
\label{e3a.30} 
\tau_q^{\rm s} (b,p) = \tau_q^{\rm b} (b \longrightarrow
b(1+p)^{1/2}/2).
\end{equation}

\begin{figure}
\begin{center}
\includegraphics[width=0.7\columnwidth,clip]{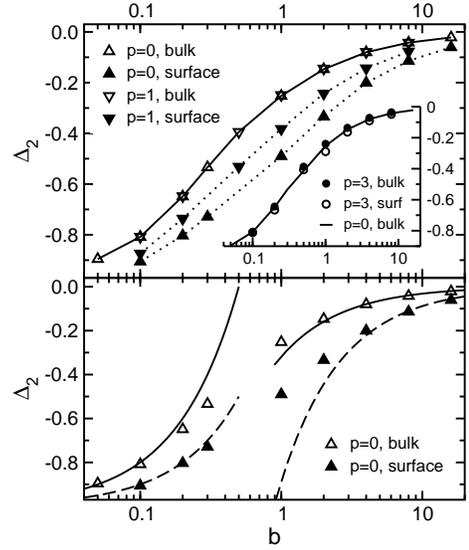}
\caption{{\it Upper panel:} Anomalous exponent $\Delta_2 \equiv D_2-1$ as
a function of $b$ from numerical simulations in the bulk and at the
boundary for the reflection parameter $p=0$ and $1$. The inset shows
data for $p=3$ compared to the $p=0$ bulk values. {\it Lower panel:}
Surface and bulk data for $p=0$ compared with analytical results for
small and large $b$ (using $R_0=2.78$), Eqs.~(\ref{e3.22}),
(\ref{e3a.20}), (\ref{e3.47}), (\ref{e3a.28}). \cite{mildenberger07a}.} 
\label{fig3a.2}
\end{center}
\end{figure}

\begin{figure}
\begin{center}
\includegraphics[width=0.7\columnwidth,clip]{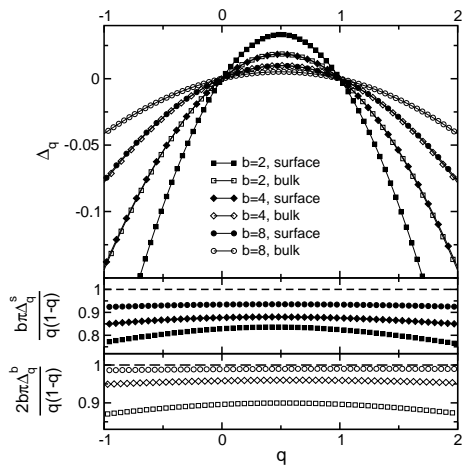}
\caption{{\it Upper panel:} Boundary and bulk
  multifractal spectra, $\Delta_q^{\rm s}$
  and $\Delta_q^{\rm b}$, at $b=2$, $4$, and $8$ for the
  reflection parameter $p=1$. In accordance with Eq.~(\ref{e3a.22}),
  the surface multifractality spectrum is enhanced by a factor
  close to two compared to the bulk. 
  {\it Middle panel:} Surface spectrum divided by the analytical
  $b\gg 1$ result, Eq.~(\ref{e3a.20}). With increasing $b$, the
  numerical data converges towards the analytical result (dashed line).
  {\it Lower panel:} Analogous plot for the bulk spectrum.
  \cite{mildenberger07a}.
} \label{fig3a.3}
\end{center}
\end{figure}

These predictions were corroborated by numerical simulations of the
PRBM model.  Figure \ref{fig3a.2} shows 
the dependence of the anomalous dimension $\Delta_2 \equiv D_2
-1$ on $b$ in the bulk and at the boundary, for three different
values of the reflection parameter $p$.
In Fig.~\ref{fig3a.3} the whole multifractal spectra $\Delta_q$ are
shown for fixed large values of
$b$. Specifically, the anomalous dimensions
$\Delta_q^{\rm s}$ and $\Delta_q^{\rm b}$ are presented for
$b=2,\,4$ and 8, with the reflection parameter chosen to be $p=1$.
For all curves the $q$ dependence is approximately parabolic, as
predicted by the large-$b$ theory, Eq.~(\ref{e3a.20}),
with the prefactor inversely proportional to $b$.  It is also
seen in Fig.~\ref{fig3a.3} that the bulk multifractality spectrum for
$b=4$ and the surface spectrum for $b=8$ are almost identical, in
agreement with Eq.~(\ref{e3a.22}).   
The same is true for the relation
between the bulk spectrum for $b=2$ and the surface spectrum for
$b=4$. The ratio of the large-$b$ surface and
bulk anomalous dimensions, $R_p = \Delta_q^{\rm s} / \Delta_q^{\rm
b}$, is in good agreement with
the $\sigma$-model predictions for $R_p$, 
as shown in Fig.~\ref{fig3a.1}.
In Fig.~\ref{fig3a.4} the surface and bulk multifractal spectra are
shown for the case of small $b$. While the spectra are strongly
non-parabolic in this limit, they clearly exhibit the symmetry $q\to
1-q$, Eq.~(\ref{e2.32}). The data are in good agreement with the
real-space RG
results.

\begin{figure}
\begin{center}
\includegraphics[width=0.7\columnwidth,clip]{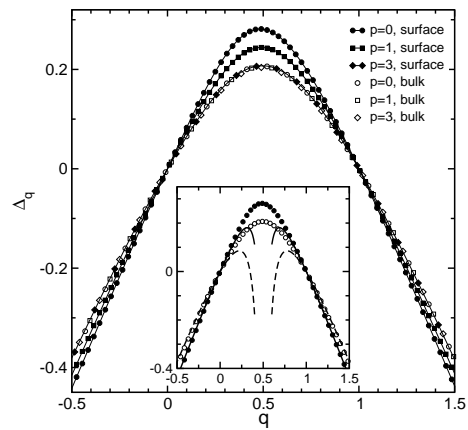}
\caption{{\it Main panel:} Numerically determined boundary and bulk
anomalous dimensions $\Delta_q$ at $b=0.1$ for $p=0$, $1$, and $3$.
As expected, the bulk anomalous dimension is independent of the
value of $p$. In accordance with Eq.~(\ref{e3a.30}), for $p=3$ surface
and bulk dimensions have
the same values.
{\it Inset:} The $p=0$ data compared to the analytical results, surface
[solid line, Eq.~(\ref{e3a.28})] and bulk [dashed line,
Eq.~(\ref{e3.47})]. Analytical data have been calculated for $q \geq
0.6$ and mirrored for $q \leq 0.4$ by using the symmetry relation
$\Delta_q = \Delta_{1-q}$. In the vicinity of $q=1/2$, at
$|q-1/2|\lesssim 1/\ln b^{-1}$,  the analytical result
(\ref{e3.47}) breaks down. \cite{mildenberger07a}.}  
\label{fig3a.4}
\end{center}
\end{figure}

\subsection{Further related activities}
\label{s3.6}

As a model of the Anderson critical point, the PRBM ensemble has 
attracted a lot of interest during recent years.
In view of space limitations, we can only briefly mention some of
these works. The corresponding research directions include:
relation to the system of interacting fermions in 1D (Luttinger liquid)
\cite{kravtsov00}; a generalization of the model to two dimensions
\cite{potempa02}; wave function statistics
\cite{cuevas01,varga02,cuevas02c,cuevas03a};
level correlations
\cite{cuevas05,garcia-garcia06a}; 
virial expansion for level statistics of almost diagonal random
matrices \cite{yevtushenko03,yevtushenko04,kravtsov06};
chiral PRBM and possible applications to quantum chromodynamics
\cite{garcia-garcia04}, and
manifestations of multifractality in scattering characteristics 
of an open system \cite{mendez-bermudez05,mendez-bermudez06}.
Rotationally-invariant 
random matrix ensembles with level statistics largely similar to
that in the critical PRBM ensembles were studied in
\textcite{muttalib93,moshe94,muttalib01}; a relation between these
ensembles and the PRBM model is discussed in \cite{kravtsov97,mirlin00}.

\section{Symmetries of disordered systems}
\label{s:SymDisSys}

In this section we briefly review the symmetry classification of
disordered systems based on the relation to the classical symmetric
spaces, which was established in \textcite{altland97,zirnbauer96b}.
For a detailed presentation of the scheme and the underlying
mathematical structures the reader is referred to a recent review by
\textcite{caselle04}. A mathematical proof of the completeness of the
classification was given in \textcite{heinzner05}.

\subsection{ Wigner-Dyson classes}
\label{s4.1}

The random matrix theory (RMT) was introduced into physics
by \textcite{wigner51}. Developing
Wigner's ideas, \textcite{dyson62} put forward a 
classification scheme of ensembles of random Hamiltonians.  
This scheme takes into account the invariance of the system under time
reversal and spin rotations, yielding three symmetry classes: unitary,
orthogonal and symplectic. 

If the time-reversal invariance ($T$) is broken, the Hamiltonians are just
arbitrary Hermitian matrices, 
\be
\label{e4.1}
H=H^\dagger\ ,
\ee 
with no further
constraints. This set of matrices is invariant with respect to
rotations by unitary matrices; hence the name ``unitary ensemble''. 
In this situation, the presence or absence of spin rotation invariance ($S$)
is not essential: if the spin is conserved, $H$ is simply a 
spinless unitary-symmetry Hamiltonian times the unit matrix in the spin space.
In the RMT one considers most frequently an ensemble of matrices with
independent, Gaussian-distributed random entries, which is called the
Gaussian unitary ensemble (GUE). While disordered systems have, of
course,  much richer physics than the Gaussian ensembles, their
symmetry classification is inherited from the RMT.   

Let us now turn to the systems with preserved time-reversal
invariance. The latter is represented by an antiunitary operator, 
$T=KC$, where $C$ is the operator of complex conjugation and $K$ is
unitary. The time-reversal invariance thus implies 
$H=KH^{\rm T}K^{-1}$ (we used the Hermiticity, $H^* = H^{\rm T}$). 
Since acting twice with $T$ should leave the physics
unchanged, one infers that $K^*K = p$, where $p=\pm 1$.
As was shown by Wigner, the two cases correspond to systems with
integer ($p=+1$) and half-integer ($p=-1$) angular momentum. 
If $p=1$, a representation can be chosen where $K=1$, so that
\be
\label{e4.2}
H=H^{\rm T}\ .
\ee 
The set of Hamiltonians thus spans the space of real
symmetric matrices in this case. This is the orthogonal symmetry
class; its representative is the Gaussian orthogonal ensemble (GOE). 
For disordered electronic systems this class is realized when spin
is conserved, as the Hamiltonian then reduces to that for spinless
particles (times unit matrix in the spin space). 

If  $T$ is preserved but $S$ is broken, we have $p=-1$. In the
standard representation, $K$ is then realized by the second Pauli
matrix, $K =i\sigma_y$, so that the Hamiltonian satisfies
\be
\label{e4.3}
H = \sigma_y H^{\rm T} \sigma_y.
\ee 
It is convenient to split the $2N \times 2N$ 
Hamiltonian in $2\times 2$ blocks (quaternions) in spin
space. Each of them then is of the form $q = q_0\sigma_0 +
iq_1\sigma_x +i q_2\sigma_y + i q_3\sigma_z$ (where $\sigma_0$ is the
unit matrix and $\sigma_{x,y,z}$ the Pauli matrices), with 
real $q_\mu$, which defines a real quaternion. This set of
Hamiltonians is invariant with respect to the group of unitary
transformations 
conserving $\sigma_y$, $U \sigma_y U^T = \sigma_y$, which is the
symplectic group $\text{Sp}(2N)$. The corresponding symmetry class is
thus called symplectic, and its RMT representative is the Gaussian
symplectic ensemble (GSE). 

\subsection{ Relation to symmetric spaces}
\label{s4.2}

Before discussing the relation to the families of symmetric
spaces, we briefly remind the reader how the latter are constructed
\cite{helgason78,caselle04}. 
Let G be one of the compact Lie groups $\text{SU}(N)$,
$\text{SO}(N)$, $\text{Sp}(2N)$, and $\mathfrak{g}$ the corresponding
Lie algebra. Further, let $\theta$ be an involutive automorphism 
$\mathfrak{g} \to \mathfrak{g}$ such that $\theta^2=1$ but $\theta$ is
not identically equal to unity. It is clear that $\theta$ splits
$\mathfrak{g}$ in two complementary subspaces, 
$\mathfrak{g} = \mathfrak{K} \oplus \mathfrak{P}$, such that 
$\theta(X)=X$ for $X\in\mathfrak{K}$ and 
$\theta(X)=-X$ for $X\in\mathfrak{P}$. It is easy to see that
the following Lie algebra multiplication relations holds:
\be
\label{e4.4}
[\mathfrak{K},\mathfrak{K}]\subset \mathfrak{K},\qquad 
[\mathfrak{K},\mathfrak{P}]\subset \mathfrak{P},\qquad 
[\mathfrak{P},\mathfrak{P}]\subset \mathfrak{K}.
\ee
This implies, in particular,  that $\mathfrak{K}$ is a subalgebra,
whereas $\mathfrak{P}$ is not. The coset space $G/K$ (where $K$ is the
Lie group corresponding to $\mathfrak{K}$) is then a compact symmetric
space. The tangent space to $G/K$ is  $\mathfrak{P}$. 
One can also construct an associated non-compact space. For this
purpose, one first defines the Lie algebra
$\mathfrak{g}^*=\mathfrak{K} \oplus i\mathfrak{P}$, which differs from   
$\mathfrak{g}$ in that the elements in $\mathfrak{P}$ are multiplied
by $i$. Going to the corresponding group and dividing $K$ out, one
gets a non-compact symmetric space $G^*/K$.

The groups $G$ themselves are also symmetric spaces and can be viewed
as coset spaces $G\times G/G$. The corresponding non-compact space is
$G^\mathbb{C}/G$, where $G^\mathbb{C}$ is the complexification of $G$
(which is obtained by taking the Lie algebra $\mathfrak{g}$, promoting
it to the algebra over the field of complex numbers, and then
exponentiating).  

The connection with symmetric spaces is now established in the following
way \cite{altland97,zirnbauer96b}. Consider first the unitary symmetry
class. Multiplying a Hamiltonian matrix by $i$, we get an
antihermitean matrix $X=iH$. Such matrices form the Lie algebra
$\mathfrak{u}(N)$. Exponentiating it, one gets the Lie group
$\text{U}(N)$, which is the compact symmetric space of class A in
Cartan's classification. 
For the orthogonal class, $X=iH$ is purely
imaginary and symmetric. The set of such matrices is a linear complement 
$\mathfrak{P}$ of the algebra $\mathfrak{K}=\mathfrak{o}(N)$ of imaginary
antisymmetric matrices  in the algebra $\mathfrak{g}=\mathfrak{u}(N)$
of antihermitean matrices. The corresponding symmetric space is 
$G/K = \text{U}(N)/\text{O}(N)$, which  is termed AI in Cartan's
classification.   For the symplectic ensemble the same consideration
leads to the symmetric space $\text{U}(N)/\text{Sp}(N)$, which is the
compact space of the class AII. If we don't multiply $H$ by $i$ but
instead proceed with $H$ in the analogous way, we end up with
associated non-compact spaces $G^*/K$. To summarize, the linear space 
$\mathfrak{P}$ of Hamiltonians can be considered as a tangent space to
the compact $G/K$ and non-compact $G^*/K$ symmetric spaces of the
appropriate symmetry class.

\begin{table*}
\begin{tabular}{|c|c|c|c|c|c|c|}
\hline
Ham. & RMT & T \ \  S & compact & non-compact & $\sigma$-model &
$\sigma$-model compact 
\\
 class &    &  & symmetric space & symmetric space & B$|$F & sector ${\cal M}_F$
\\
\hline
\multicolumn{7}{l}{Wigner-Dyson classes}\\
\hline
A & GUE  & $-$ \ \  $\pm$ & $\text{U}(N){\times}\text{U}(N)/\text{U}(N) 
\equiv \text{U}(N) $ & $\text{GL}(N,\mathbb{C})/\text{U}(N)$ & AIII$|$AIII
& $\text{U}(2n)/\text{U}(n){\times}\text{U}(n)$
\\
\hline
AI & GOE & $+$ \ \ $+$ &  $\text{U}(N)/\text{O}(N)$ & 
 $\text{GL}(N,\mathbb{R})/\text{O}(N)$
 & BDI$|$CII & $\text{Sp}(4n)/\text{Sp}(2n){\times}\text{Sp}(2n)$
\\
\hline
AII & GSE & $+$ \ \ $-$ & $\text{U}(2N)/\text{Sp}(2N)$ &
 $\text{U}^*(2N)/\text{Sp}(2N)$ &  CII$|$BDI & 
$\text{O}(2n)/\text{O}(n){\times}\text{O}(n)$
\\
\hline 
\multicolumn{7}{l}{chiral classes}\\
\hline
AIII & chGUE  &  $-$ \ \  $\pm$ &
$\text{U}(p+q)/\text{U}(p){\times}\text{U}(q)$ &
$\text{U}(p,q)/\text{U}(p){\times}\text{U}(q)$
& A$|$A & $\text{U}(n)$
\\
\hline
BDI & chGOE & $+$ \ \ $+$ &
$\text{SO}(p+q)/\text{SO}(p){\times}\text{SO}(q)$ &
$\text{SO}(p,q)/\text{SO}(p){\times}\text{SO}(q)$
& AI$|$AII &  $\text{U}(2n)/\text{Sp}(2n)$
\\
\hline
CII & chGSE & $+$ \ \ $-$ &
$\text{Sp}(2p+2q)/\text{Sp}(2p){\times}\text{Sp}(2q)$ &
$\text{Sp}(2p,2q)/\text{Sp}(2p){\times}\text{Sp}(2q)$
& AII$|$AI &  $\text{U}(n)/\text{O}(n)$
\\
\hline
\multicolumn{7}{l}{Bogoliubov - de Gennes classes}\\
\hline
C &  & $-$ \ \  $+$ &
$\text{Sp}(2N){\times}\text{Sp}(2N)/\text{Sp}(2N) \equiv \text{Sp}(2N)$
&  $\text{Sp}(2N,\mathbb{C})/\text{Sp}(2N)$  &
DIII$|$CI & $\text{Sp}(2n)/\text{U}(n)$
\\
\hline
CI &  &  $+$ \ \  $+$ & $\text{Sp}(2N)/\text{U}(N)$ & 
$\text{Sp}(2N,\mathbb{R})/\text{U}(N)$  &
D$|$C & $\text{Sp}(2n)$ 
\\
\hline
BD & &  $-$ \ \  $-$ &
$\text{SO}(N){\times}\text{SO}(N)/\text{SO}(N)\equiv \text{SO}(N)$ &
$\text{SO}(N,\mathbb{C})/\text{SO}(N)$  &
CI$|$DIII  &  $\text{O}(2n)/\text{U}(n)$
\\
\hline 
DIII & & $+$ \ \  $-$ &  $\text{SO}(2N)/\text{U}(N)$ & 
 $\text{SO}^*(2N)/\text{U}(N)$ & C$|$D  & $\text{O}(n)$ 
\\
\hline
\end{tabular}\hfill \\
\caption{Symmetry classification of disordered systems. First column:
  symbol for the symmetry class of the Hamiltonian. Second column:
  names of the corresponding RMT. Third column: presence (+) or
  absence ($-$) of the time-reversal (T) and spin-rotation (S)
  invariance. Fourth and fifth columns: families of the compact and
  non-compact symmetric spaces of the corresponding symmetry class. 
The Hamiltonians span the tangent space to these symmetric spaces.
Sixth column: symmetry class of the $\sigma$-model; the first symbol
corresponds to the non-compact (``bosonic'') and the second to the
compact (``fermionic'') sector of the base of the $\sigma$-model
manifold. The compact component ${\cal M}_F$ (which is particularly
important for theories with non-trivial topological properties) is
explicitly given in the last column.}   
\label{t4.1}
\end{table*}

This correspondence is summarized in Table~\ref{t4.1}, where the first
three rows correspond to the Wigner-Dyson classes, the next three to
the chiral classes (Sec.~\ref{s4.3}) and last four to the
Bogoliubov-de Gennes classes  (Sec.~\ref{s4.4}). The last two columns
of the table specify the symmetry of the corresponding $\sigma$-model. 
In the supersymmetric formulation, the base of the $\sigma$-model
target space $\cal{M}_B\times\cal{M}_F$ is the product of a 
non-compact symmetric space $\cal{M}_B$ corresponding to the bosonic
sector and a compact (``fermionic'') symmetric space $\cal{M}_F$.
(In the replica formulation, the space is $\cal{M}_B$ for bosonic
replicas or $\cal{M}_F$ for fermionic replicas, supplemented with the
replica limit $n\to 0$.)
The Cartan symbols for these symmetric spaces are given in the sixth
column, and the compact components  $\cal{M}_F$ are listed in the last
column. It should be stressed that the symmetry classes of
$\cal{M}_B$ and $\cal{M}_F$ are different from the symmetry class of
the ensemble (i.e. of the Hamiltonian) and in most cases are also
different from each other. 
Following the common convention, when we refer to a system as 
belonging to a particular class, 
we mean the symmetry class of the Hamiltonian. 

It is also worth emphasizing that the orthogonal groups appearing
in the expressions for  $\cal{M}_F$
are $\text{O}(N)$ rather than $\text{SO}(N)$. This difference
(which was irrelevant when we were discussing the symmetry of the
Hamiltonians, as it does not affect the tangent space) is
important here, since it influences topological properties of the
manifold. As we will detail in Sec.~\ref{s5},\ref{s6}, the
topology of the $\sigma$-model target space often affects the
localization properties of the theory in a crucial way.

\subsection{ Chiral classes}
\label{s4.3}

The Wigner-Dyson classes are the only allowed if one looks for a
symmetry that is translationally invariant in energy, i.e. is not
spoiled by adding a constant to the Hamiltonian. However, additional
discrete symmetries may arise at some particular value of energy
(which can be chosen to be zero without loss of generality), leading
to novel symmetry classes. As the vicinity of a special point in the
energy space governs the physics in many cases (i.e. the band center
in  lattice models at half filling, or zero energy in non-$s$-wave
superconductors), these ensembles  are of large interest. They can be
subdivided into two groups -- chiral and Bogoliubov - de Gennes
ensembles -- considered here and in Sec.~\ref{s4.4}, respectively.

The chiral ensembles appeared in both contexts of particle physics
and physics of disordered electronic systems about fifteen years
ago \cite{gade93,gade91,slevin93,verbaarschot93}. The corresponding
Hamiltonians have the form
\begin{equation}
\label{e4.5}
   { H } = \left( \begin{array}{cc}
             0           &    h \\
             h^{\dagger} &    0    
             \end{array}
             \right)\ ,
\end{equation}
i.e. they possess the symmetry
\be
\label{e4.6}
\tau_z H \tau_z = - H \ , 
\ee
where $\tau_z$ is the third Pauli matrix in a certain ``isospin''
space. In the condensed matter context, such ensembles arise, in
particular, when one considers a tight-binding model on a bipartite
lattice with randomness in hopping matrix elements only. In this case,
$H$ has the block structure (\ref{e4.5}) in the sublattice space.

In addition to the chiral symmetry, a system may possess 
time reversal symmetry and/or spin-rotation invariance. 
In full analogy with the
Wigner-Dyson classes, \ref{s4.1}, one gets therefore three chiral
classes (unitary, orthogonal, and symplectic). The corresponding
symmetric spaces, the Cartan notations for symmetry classes, and the
$\sigma$-model manifolds are given in the rows 4--6 of the
Table~\ref{t4.1}.

\subsection{ Bogoliubov - de Gennes classes}
\label{s4.4}

As was found in \textcite{altland97}, the Wigner-Dyson and chiral classes
do not exhaust all possible symmetries of disordered electronic
systems. The remaining four classes arise most naturally in
superconducting systems. The quasiparticle dynamics in such systems can be
described by the Bogoliubov-de Gennes Hamiltonian of the form
\begin{equation}
\label{e4.7}
\hat H = \sum_{\alpha\beta}^{N} h_{\alpha\beta} c^{\dagger}_{\alpha} c_{\beta} 
+ \frac{1}{2} \sum_{\alpha\beta}^{N} \left( \Delta_{\alpha\beta} 
c^\dagger_\alpha c^{\dagger}_\beta -
  \Delta_{\alpha\beta}^* c_\alpha c_\beta \right) 
\end{equation}
where $c^\dagger$ and $c$ are fermionic creation and annihilation operators,
and the $N\times N$ matrices $h$, $\Delta$ satisfy $h=h^{\dagger}$ and
$\Delta^T = -\Delta$, in view of hermiticity. 
Combining $c^{\dagger}_{\alpha},c_{\alpha}$ in a spinor
$\psi_{\alpha}^{\dagger}=(c_{\alpha}^{\dagger},c_{\alpha})$, one gets 
a matrix representation of the Hamiltonian,
$\hat H = \psi^\dagger H \psi$, where $H$
has the structure 
\begin{equation}
\label{e4.8}
      H = \left( \begin{array}{cc}
             h           &    \Delta\\
             -\Delta^*   &    -h^T    
             \end{array}
             \right), \qquad h = h^\dagger\ ,\ \ \Delta=-\Delta^T\ . 
\end{equation}
The minus signs in the definition of $H$ result form the fermionic
commutation relations between $c^{\dagger}$ and $c$. 
The Hamiltonian structure (\ref{e4.8}) corresponds to 
the condition 
\be
\label{e4.9}
H=-\tau_x H^T \tau_x 
\ee
(in addition to the Hermiticity $H=H^\dagger$), where $\tau_x$ is the
Pauli matrix in the particle-hole space.  Alternatively, one can
perform a unitary rotation of the basis, 
defining $\tilde{H}= g^\dagger H g$ with $g =
(1+i\tau_x)/\sqrt{2}$.  
In this basis, the defining condition of class D becomes $\tilde{H} =
- \tilde{H}^T$,  so that $\tilde{H}$ is pure imaginary.
The matrices $X=iH$ thus form the Lie algebra $\mathfrak{so}(2N)$,
corresponding to the Cartan class D. This symmetry class described
disordered superconducting systems in the absence of other symmetries.

Again, the symmetry class will be changed if the time reversal and/or
spin rotation invariance are present. The difference with respect to
the Wigner-Dyson and chiral classes is that now one gets four
different classes rather than three. This is because the spin-rotation
invariance has an impact even in the absence of time-reversal
invariance, since it combines  with the particle-hole symmetry in a
non-trivial 
way. Indeed, if the spin is conserved, the Hamiltonian has the form
  \begin{equation}
\label{e4.10}
    H = \sum_{ij}^{N} h_{ij}(c_{i\uparrow}^\dagger c_{j\uparrow} -
    c_{j\downarrow} c^\dagger_{i,\downarrow})
    + \sum_{ij}^{N} \Delta_{ij}
    c^{\dagger}_{i,\uparrow}c^{\dagger}_{j,\downarrow} 
    + \Delta^*_{ij} c_{i\downarrow} c_{j\uparrow}. 
    \end{equation}
where $h$ and $\Delta$ are $N{\times}N$ matrices satisfying
$h=h^\dagger$ and $\Delta=\Delta^T$.
Similar to (\ref{e4.8}), we can introduce the spinors 
$\psi_{i}^{\dagger} = (c_{i\uparrow}^\dagger, c_{i\downarrow})$ and
obtain the following matrix form of the Hamiltonian
\begin{equation}
\label{e4.11}
   { H } = \left( 
     \begin{array}{cc}
             { h } & { \Delta} \\
             { \Delta}^* & { -h}^{T}
             \end{array}
             \right), \qquad h = h^\dagger\ ,\ \ \Delta=\Delta^T\ .
\end{equation}
It exhibits a symmetry property
\begin{equation} 
\label{e4.12}
  H =   - \tau_{y} H^T \tau_y.
\end{equation}
The matrices $H=iX$ now form the Lie algebra $\mathfrak{sp}(2N)$,
which is the symmetry class C.

If the time reversal invariance is present, one gets two more classes
(CI and DIII). The symmetric spaces for the Hamiltonians and the
$\sigma$-models corresponding to the Bogoliubov--de Gennes classes are
given in the last four rows of the Table~\ref{t4.1}. 

\subsection{ Additional comments}
\label{s4.5}

i) In addition to Table~\ref{t4.1}, where the symmetry classes are
ordered according to the discrete symmetries of the underlying
physical systems, we include
Table~\ref{t4.2} with a more mathematical ordering. There, the first
row is formed by compact symmetric spaces that are groups and the rest
is a ``matrix'' of symmetric spaces $G/K$ arranged according to the type
(U, Sp, or O) of the groups $G$ and $K$. This ordering illustrates the
completeness of the classification scheme: all entries in the ``matrix''
are filled, except for two, as there is no symmetric spaces of the
O/Sp and Sp/O types. Also, we will see in Sec.~\ref{s6} that this
ordering is relevant to types of 2D critical behavior that the
corresponding systems may show. In particular, the first row are the
classes where different types of the QHE (IQHE, SQHE, and TQHE) take
place. The second row is formed by the classes allowing for the
Wess-Zumino type of criticality, while the third row are the systems
which allow for a $\mathbb{Z}_2$ topological term. The diagonal of the
``matrix'' is occupied by three chiral classes.

\begin{table*}
  \begin{tabular}{|cc|cc|cc|} \hline\hline 
${\mathrm U}(N)$  & A & 
$\mathrm{ Sp}(2N)$ & C &
$\mathrm{ SO}(N)$ & BD 
\\ \hline\hline   
${\mathrm U}(p+q)/{\mathrm U}(p){\times}{\mathrm U}(q)$ & AIII &  
${\mathrm Sp}(2N)/{\mathrm U}(N)$  & CI & 
${\mathrm SO}(2N)/{\mathrm U}(N)$ & DIII  
\\
\hline
${\mathrm U}(2N)/{\mathrm Sp}(2N)$ & AII &
$\mathrm{ Sp}(2p+2q)/{\mathrm Sp}(2p){\times}\mathrm{ Sp}(2q)$ & CII  &&
\\
\hline 
${\mathrm U}(N)/{\mathrm O}(N)$ & AI & &&
$\mathrm{ SO}(p+q)/{\mathrm SO}(p){\times}\mathrm{ SO}(q)$ & BDI 
\\ \hline\hline
\end{tabular}
\caption{Compact symmetric spaces arranged in a form of a ``matrix'',
  with the corresponding Cartan symbols. 
First row: U, Sp, and O groups. Second (third, fourth) row: symmetric
spaces G/K  (which are not groups themselves) with
K being the unitary (resp. symplectic, orthogonal) group.}
\label{t4.2}
\end{table*}

ii) 
Strictly speaking, one should distinguish between the orthogonal group 
$\mathrm{SO}(N)$ with even and odd $N$, which form different Cartan 
classes: $\mathrm{ SO}(2N)$ belongs to class D, while  $\mathrm{ SO}(2N{+}1)$
to class B. In the conventional situation of a disordered
superconductor, the matrix size is even  due to the
particle-hole space doubling, see Sec.~\ref{s4.4}. It was found, however,
that the class B can arise in $p$-wave vortices \cite{ivanov02,ivanov02a}. 
In the same sense, the class DIII should be
split in DIII-even and DIII-odd; the last one represented by the
symmetric space $\text{SO}(4N+2)/\text{U}(2N+1)$ can appear in
vortices in the presence of time-reversal symmetry.

\subsection{Perturbative $\beta$-functions for $\sigma$-models of
  different symmetry classes}
\label{s4.6}

Perturbative $\beta$-functions for $\sigma$-models on all the types of
symmetric spaces were in fact calculated \cite{hikami81,wegner88} 
long before the
physical significance of the chiral and Bogoliubov-de Gennes classes
has been fully appreciated. In Table~\ref{t4.3} we have collected the
results for all $\beta$-functions, $\beta(t)=-dt/d\ln L$, 
in $d=2+\epsilon$ dimensions up to four-loop order. Here $t$ is
the coupling constant inverse proportional to the dimensionless
conductance $g$, and the Anderson localization problem corresponds to the
replica limit $N=p=0$. The corresponding results for the Wigner-Dyson
classes have already been quoted in Sec.~\ref{s2.2.2}. For each
symmetry class of disordered electronic systems, the perturbative
$\beta$-function  can be
equivalently obtained from either compact or non-compact
$\sigma$-models on the appropriate manifolds. As an example, the
$\beta$-function for the Wigner-Dyson orthogonal class can be found 
by using the replica limit of the compact $\sigma$-model of the type CII or
of the non-compact $\sigma$-model of the type BDI.

It is seen that for
the classes A, AI, C, CI the $\beta$-function is negative in 2D in the
replica limit $N=p=0$ (at least, for small $t$). This indicates that normally
all states are localized in such systems in 2D. (This conclusion can
in fact be changed in the presence of topological or Wess-Zumino
terms, Sec.~\ref{s6.1}.) Above 2D, these systems undergo the Anderson
transition that can be studied within the $2+\epsilon$ expansion,
Sec.~\ref{s2.2.2}.  For the classes AIII, BDI, and CII (chiral
unitary, orthogonal and symplectic classes, respectively) the
$\beta$-function is exactly zero in 2D, implying a line of fixed
points, Sec.~\ref{s6.6}. Finally, in the classes AII, D, and DIII the
$\beta$-function is positive at small $t$, implying the existence of a
metal-insulator transition at strong coupling in 2D. The 2D
Anderson transitions in the Wigner-Dyson symplectic class AII and in the
Bogoliubov-de Gennes class D will be discussed in more detail in
Sec.~\ref{s6.2} and \ref{ss:tqhe}, respectively.

\begin{table*}[t]
\begin{tabular}{|c|c|c|c|c|c|c|c|}  \hline\hline
$\sigma$-model & compact & Ham. & $t$ & $t^2$ & $t^3$ & $t^4$ & $t^5$
\\ 
class & $\sigma$-model manifold & c $|$ n-c & &&&&
\\ \hline\hline 
AIII & $\mathrm{U}(N)/\mathrm{U}(p){\times}\mathrm{U}(N\!{-}\!p)$ & A
& $\epsilon$ & 
$-N$ & $-2(1\!{+}\!p(N\!{-}\!p))$ & $-\frac{1}{2}N(3p(N\!{-}\!p){+}7)$
& $c_4(N,p)$ \\\hline 
BDI & $\mathrm{O}(N)/\mathrm{O}(p){\times}\mathrm{O}(N\!{-}\!p)$ & AII$|$AI &
$\epsilon$ &  
$-(N\!{-}\!2)$ & $-(2p(N\!{-}\!p)\!{-}\!N)$ &
$-[\frac{3}{2}pN(N\!{-}\!p)-\frac{5}{4}N^2\!{+}
\!p(N\!{-}\!p)\!{+}\!\frac{1}{2}N]$ & $c_5(N,p)$  \\\hline 
CII &
$\mathrm{Sp}(2N)/\mathrm{Sp}(2p){\times}\mathrm{Sp}(2N\!{-}\!2p)$ &
AI$|$AII & 
$\epsilon$ & 
$\frac{1}{2}(N\!{-}\!2)$ & $-\frac{1}{4}(2p(N\!{-}\!p)\!{-}\!N)$&
$\frac{1}{8}[\frac{3}{2}pN(N\!{-}\!p)-\frac{5}{4}N^2\!{+}
\!p(N\!{-}\!p)\!{+}\!\frac{1}{2}N]$
& $\frac{1}{16}c_5(N,p)$  
\\\hline\hline
AI & $\mathrm{U}(N)/\mathrm{O}(N)$ & CII$|$BDI & $\epsilon$ & $-N$ &
$-N(\frac{1}{2}N+1)$ & $-N(\frac{3}{8}N^2+\frac{5}{4}N+1)$ &
$-{N\over 2}  c_2(-N/2)$  \\   \hline
AII & $\mathrm{U}(2N)/\mathrm{Sp}(2N)$ & BDI$|$CII & $\epsilon$ & $-2N$ &
$-2N(N-1)$ &  
$-N(3N^2-5N+2)$ & $Nc_2(N)$ \\\hline
CI & $\mathrm{Sp}(2N)/\mathrm{U}(N)$ & C$|$BD & $\epsilon$ & $-(N\!{+}\!1)$ &
$-\frac{1}{2}(N^2{+}3N{+}4)$ & $-\frac{1}{8}(3N^3{+}14N^2{+} 35N{+}28)$&
$c_3(N)$ \\\hline
DIII & $\mathrm{O}(2N)/\mathrm{U}(N)$ & BD$|$C & $\epsilon$ & $-(2N-2)$ &
$-(2N^2-6N+8)$ & $-(3N^3-14N^2+35N-28)$ & $c_3(-2N)$ \\\hline  
\hline 
A & $\mathrm{U}(N){\times}\mathrm{U}(N)/\mathrm{U}(N)$ & AIII & $\epsilon$ &  
$-N$ & $-\frac{1}{2}N^2$ & $-N^3(\frac{3}{8}+(\frac{19}{48} + a)N)$ & 
$aN^2(N\!{-}\!2)(N\!{+}\!2)$ \\\hline
C & $\mathrm{Sp}(2N){\times}\mathrm{Sp}(2N)/\mathrm{Sp}(2N)$  & CI$|$DIII
&$\epsilon$ & $-(N+1)$ & $-\frac{1}{2}(N+1)^2$ & $-\frac{3}{8}(N+1)^3$
& $-{N+1\over 8} c_1(-2N)$ \\  \hline
BD & $\mathrm{O}(N){\times}\mathrm{O}(N)/\mathrm{O}(N)$ & DIII$|$CI 
& $\epsilon$ & 
$-(N\!{-}\!2)$ & $ -\frac{1}{2}(N{-}2)^2 $ & $-\frac{3}{8}(N{-}2)^3$ &
$(N-2)c_1(N)$ 
\\\hline\hline
\end{tabular}
\caption{Perturbative $\beta$-functions, $\beta(t)= -d t/d\ln L$, up
  to the four-loop order for the
  $\sigma$-models in $d=2+\epsilon$ dimensions
with symmetric spaces as target manifolds 
\cite{hikami81,wegner88}. First column: Cartan symbol for the
$\sigma$-model symmetric space. Second column: compact $\sigma$-model
manifold. (The associated non-compact spaces can be found in
Table~\ref{t4.1}.) Third column: symmetry class of random Hamltonians
described by the replica version of the compact $\sigma$-model (c) and
of its non-compact counterpart (n-c). The last five columns give the
coefficients of the terms from $t$ to $t^5$ in the $\beta$-functions
for compact $\sigma$-models.   
The $\beta$-functions for the corresponding non-compact
$\sigma$-models  are obtained by the substitution
$\beta(t)\to-\beta(-t)$, i.e. by flipping the sign of the terms with
even powers of $t$.
Following notations are used: $a=\frac{3}{16}\zeta(3)$, 
$c_5(N,p)=-(\frac{p}{3}N^2(N-p)+5p^2(N-p)^2-\frac{5}{12}N^3
  -(\frac{23}{6}+8a)pN(N-p)+(-\frac{2}{3}+16a)p(N-p)
  +(\frac{7}{6}+16a)N^2+(\frac{1}{3}-64a)N+64a$);
   $c_4(N,p)=
   -(\frac{1}{3}pN^2(N\!{-}\!p)\!{+}\!5p^2(N\!{-}\!p)^2\!{+}
\!\frac{11}{6}N^2\!{+}\!  11p(N\!{-}\!p)\!{+}\!6)$;  
   $c_3(N)=-(\frac{19}{48}N^4+\frac{119}{48}N^3+
\frac{380}{48}N^2+\frac{578}{48}N+\frac{376}{48})$;   
   $c_2(N)= -(\frac{19}{3}N^3 - (\frac{43}{3}- 8a)N^2 + (9+8a)N
  -1)$ ;
   $c_1(N)= -[ (\frac{19}{48}\!{+}\!a)(N\!-\!2)^3\!{-}
\!a(N\!{-}\!3)(N\!{-}\!4)(N\!{+}\!2)]$
}
\label{t4.3}
\end{table*}

\section{\label{s5} Quasi-1D systems: Disordered wires}

Under usual conditions, electronic states in disordered wires are localized,
with localization length $\xi \sim Nl$, where $l$ is 
the mean free path and $N$ the number of conducting channels
\cite{berezinsky73,efetov83a}. This is, 
however, not the full story, and that is why we include a section
about quasi-1D systems in this review devoted to
localization-delocalization transitions and criticality. 
In fact, one route to delocalization and criticality in systems of one
1D geometry -- long-range $1/r$ hopping -- has already been discussed in
Sec.~\ref{s3}. In this section we consider possible types of
delocalization in disordered wires which are related to the symmetries of the
problem.  We will see later that in many cases there are close
connections between such phenomena in quasi-1D and 2D systems
(Sec.~\ref{s6}). 

Two approaches to quasi-1D disordered electronic 
systems have been developed. The first one is the supersymmetric
$\sigma$-model approach, Sec.~\ref{s2.2.1}. For the wire geometry, the
$\sigma$-model field $Q$ depends on the longitudinal coordinate
only. As a result, the problem becomes a kind of imaginary-time
quantum mechanics on the $\sigma$-model manifold, with the
longitudinal coordinate playing the role of the (imaginary) time.
This has allowed researchers to get a number of exact results for systems of
Wigner-Dyson symmetry classes, including the asymptotics of the
density-density correlation function and the value of the localization length 
\cite{efetov83a}, a detailed description of the wave function
statistics \cite{fyodorov94}, average conductance and its variance 
\cite{zirnbauer92,mirlin94c}. 
For reviews of results for statistical properties of
various quantities in disordered wires obtained within the
$\sigma$-model approach the reader is referred to
\textcite{efetov97,mirlin00}. 

The second approach is based on the description of a wire in terms of
its transfer-matrix $M$. The evolution of the corresponding
distribution ${\cal P}(M)$ with the wire length is described by the
Dorokhov-Mello-Pereyra-Kumar (DMPK) equation \cite{dorokhov82,mello88}. 
At variance with the $\sigma$-model approach, which allows to address
any observable, the DMPK approach is designed to study the transport
properties. On the other hand, the advantage of the DMPK approach is
that it allows one to study wires with arbitrary number $N$ of channels   
(not necessarily $N\gg 1$ as required for the derivation of the
diffusive $\sigma$-model)  and provides a very detailed information on
the whole distribution of transmission eigenvalues. The two approaches
are thus complementary; their equivalence for transport properties of
many-channel wires (when both of them are applicable) was explicitly
shown in \textcite{brouwer96}. A review of the results of the DMPK
method for Wigner-Dyson symmetry classes was given in
\textcite{beenakker97}. 

Peculiar transport properties of disordered wires of unconventional
symmetry classes have been mainly studied within the DMPK approach.
Below we briefly describe this method and present the key results. 
Whenever appropriate, we will also make contact to the results
obtained within the $\sigma$-model formalism.

\subsection{Transfer matrix and DMPK equations}
\label{s5.1}

In the transfer matrix approach, one imagines the wire attached to two
clean electrodes, where one can define asymptotic states. This allows
to formulate a scattering problem. The transfer matrix $M$ relates the
amplitudes in $N$ incoming and $N$ outgoing channels to the right of the wire
to the corresponding amplitudes on the left side of the wire,
\be
\label{e5.1}
      \left( \begin{array}{c}
               R^\text{out} \\
               R^\text{in}
             \end{array}
      \right)
       = 
       M
      \left( \begin{array}{c}
               L^\text{in} \\
               L^\text{out}
             \end{array}
      \right)
\ee
(In fact, for symmetry classes A, C, and BD with broken time-reversal
symmetry the number of incoming and outgoing channels can differ. 
We will discuss this peculiar situation in Sec.~\ref{s5.3}.)
The requirement of the current conservation
\be
\label{e5.2}
    |R^\text{out}|^2 - |R^\text{in}|^2 = |L^\text{in}|^2 - |L^\text{out}|^2  
\ee
implies that $M$ is an element of the pseudounitary group
$G=\text{U}(N,N)$. The transfer matrix can be presented in the form
(Cartan decomposition)
\be
\label{e5.3}
    M = \left(\begin{array}{cc}
              u & 0 \\
              0 & u'
              \end{array}
        \right) 
        \left(\begin{array}{cc}
              \cosh \hat{x} & \sinh \hat{x} \\
              \sinh \hat{x} & \cosh \hat{x} 
              \end{array}
        \right)
        \left(\begin{array}{cc}
              v & 0  \\
              0 & v'
              \end{array}
        \right) 
\ee     
where $\hat{x}=\text{diag}(x_1,\ldots,x_N)$ are ``radial coordinates'', while 
the left and right matrices (``angular coordinates'')
are the elements of $K={\mathrm U}(N){\times}{\mathrm U}(N)$. 
If the Hamiltonian of the system possesses some additional 
(time-reversal, spin-rotation, chiral, particle-hole) symmetries, the group
$G$ of the transfer matrices will  change correspondingly, with $K$
being the maximal compact subgroup of $G$. The coset spaces $G/K$
(which play the central role for the DMPK equations, see below) are
non-compact symmetric spaces; they are listed for all the symmetry
classes of the Hamiltonian in Table~\ref{t5.1}.   
The dimensionless conductance of the wire is expressed in terms of the radial
coordinates $x_n$ as  $G=d\sum_{n=1}^N T_n$, where $T_n=1/\cosh^2 x_n$
are transmission eigenvalues and $d$ is the degeneracy (1, 2, or 4)
depending on the symmetry class. 

When an additional thin slice with a transfer matrix $T$ is attached
to the wire, the new transfer matrix is obtained by simple
multiplication $M' = TM$. For a given distribution function of the
elementary transfer-matrices $T$, one then gets a stochastic process
on the space of transfer matrices. It is assumed in the derivation of
the DMPK equations that the distribution of $T$ is fully invariant
with respect to the subgroup $K$ (isotropy assumption), which
corresponds to a complete mixture of channels at each elementary
step. (This model assumption is justified by the fact that the mixing
of channels in  a realistic wire  takes place on a scale much shorter
than the localization length.) As a result, the stochastic process
develops in fact on the coset space $G/K$, describing a Brownian
motion on this manifold. In view of the rotational symmetry, the
distribution function depends on radial coordinates $x_i$ only. 
The corresponding evolution equation --- which is the DMPK equation
--- has the form
\be
\label{e5.4}
 \frac{d{\cal P}}{dL} = \frac{1}{2\ell\gamma} \sum_{i=1}^{N}
    \frac{\partial}{\partial x_i} J(x) 
\frac{\partial}{\partial x_i}J^{-1}(x) \ {\cal P}\ .
\ee
where the Jacobian  $J(x)$ of transformation to the radial
coordinates and the parameter $\gamma$ are specified below.

For the Wigner-Dyson and Bogoliubov-de Gennes classes the radial
coordinates $x_i$ satisfy\footnote{For the classes D and DIII the
  proper variation domain of $x_i$ is slightly different, 
 $|x_1| < x_2 < \ldots < x_N$  \cite{gruzberg05}.}
 $0< x_1 < x_2 < \ldots < x_N$ and the
Jacobian $J(x)$ is
\bea
\label{e5.5}
J(x) &=& \prod_{i<j}^{N}\ \prod_{\pm} |\sinh(x_i\pm x_j)|^{m_{o}}
\ \prod_{k}^{N}|\sinh 2x_k|^{m_{l}}\ \nonumber \\
&\times & \prod_{l}^{N}|\sinh x_l|^{m_{s}} \ . 
\eea
Here $m_{o}$, $m_{s}$, and $m_{l}$ are the multiplicities of the
ordinary, short, and long roots for the corresponding symmetric space
\cite{helgason78,caselle04}. The root multiplicities for all the
symmetry classes are listed in Table~\ref{t5.1}.
The first factor in Eq.~(\ref{e5.5}) is responsible for the repulsion
between eigenvalues $x_i$, analogous to the energy level repulsion in
RMT. The last two factors govern (when the corresponding
multiplicities are non-zero) the repulsion between an eigenvalue $x_i$
and its mirror $-x_i$, and the repulsion between $x_i$ and zero,
respectively.   
For the chiral classes, the variation domain of the coordinates $x_i$
does not restrict their sign, $x_1<x_2< \ldots <x_N$, and the Jacobian  
is
\be
\label{e5.6}
J(x) =  \prod_{i<j}^{N}\ |\sinh(x_i- x_j)|^{m_{\rm o}}\ .
\ee
(For these classes $m_{s} = m_{l} = 0$.) For the
Wigner-Dyson and the chiral classes the multiplicity  $m_{\rm o}$ of
the ordinary roots is the familiar parameter $\beta$ of the RMT, equal
to 1,2, and 4 for the orthogonal, unitary, and symplectic symmetry
classes, respectively. Further, the parameter $\gamma$ in the DMPK
equation is given by 
\be
\label{e5.7}
\gamma = \left\{ \begin{array}{ll}
m_{o}(N-1) + m_{l} + 1\ ,& \ \ \text{WD and BdG}\ ; \\
{1\over 2}[2+ m_{o}(N-1) ]\ ,& \ \ \text{chiral}\ .
\end{array}
\right.
\ee

In the short-wire regime, $L\ll\gamma\ell$, where the conductance is
large, $g\gg 1$, the DMPK equation yields Ohm's law for the
average conductance, the quasi-1D universal conductance
fluctuations, and the weak-localization corrections in the form of a
$1/g$ series. Our interest here is in the opposite, long-wire limit, 
$L\gg\gamma\ell$, where localization or critical properties of the
problem manifest themselves. This will be the subject of the remaining
part of Sec.~\ref{s5}. 
We also briefly mention that at the crossover scale,
$L\sim\gamma\ell$, an analytical treatment is most complicated. An
approximate scheme has, however, been developed in  
\textcite{muttalib99,muttalib02,muttalib03b} that allows one to understand
key properties of the conductance distribution in this regime.

\begin{table}
\begin{tabular}{|c|c|c|c|c|c|}
\hline
Ham. & transfer matrix & tr.matr. & $m_{o}$ & $m_{l}$ &
$m_{s}$  
\\
 class & symmetric space   & class & & & 
\\
\hline
\multicolumn{6}{l}{Wigner-Dyson classes}\\
\hline
A & $\text{U}(p+q)/\text{U}(p){\times}\text{U}(q)$ & AIII & 2 & 1 & $2|p-q|$
\\
\hline
AI & $\text{Sp}(2N,\mathbb{R})/\text{U}(N)$  & CI & 1 & 1 & 0
\\
\hline
AII &  $\text{SO}^*(2N)/\text{U}(N)$ \ \ 
$\begin{array}{l} $N$\ \text{even} \\
                  $N$\ \text{odd}
\end{array}$
& 
$\begin{array}{c} \text{DIII-e}\\
                  \text{DIII-o}
\end{array}$
& 4 & 1 & $\begin{array}{c} 0 \\ 4 \end{array}$
\\
\hline 
\multicolumn{6}{l}{chiral classes}\\
\hline
AIII &  $\text{GL}(N,\mathbb{C})/\text{U}(N)$ & A & 2 & 0 & 0
\\
\hline
BDI &  $\text{GL}(N,\mathbb{R})/\text{O}(N)$ & AI & 1 & 0 & 0
\\
\hline
CII &  $\text{U}^*(2N)/\text{Sp}(2N)$ & AII & 4 & 0 & 0 
\\
\hline
\multicolumn{6}{l}{Bogoliubov - de Gennes classes}\\
\hline
C &  $\text{Sp}(2p,2q)/\text{Sp}(2p){\times}\text{Sp}(2q)$ &
CII & 4 & 3 & $4|p-q|$
\\
\hline
CI &   $\text{Sp}(2N,\mathbb{C})/\text{Sp}(2N)$  & C & 2 & 2 & 0
\\
\hline
BD & $\text{O}(p,q)/\text{O}(p){\times}\text{O}(q)$ & BDI & 1 & 0 & $|p-q|$
\\
\hline 
DIII & $\text{SO}(N,\mathbb{C})/\text{SO}(N)$ 
\ \  $\begin{array}{l} $N$\ \text{even} \\
                  $N$\ \text{odd}
\end{array}$
& 
$\begin{array}{c} \text{D} \\
                  \text{B} 
\end{array}$
& 2 & 0 & $\begin{array}{c} 0 \\ 2 \end{array}$
\\
\hline
\end{tabular}\hfill \\
\caption{Transfer matrix spaces. First column: symmetry class of the
  Hamiltonian. Second and third columns: symmetric space for the
  transfer matrix and the corresponding Cartan symmetry class. Last
  three columns: multiplicities of the ordinary ($m_{o}$), long
  ($m_{l}$) , and short ($m_{s}$) roots.}     
\label{t5.1}
\end{table}

\subsection{Conventional localization in 1D geometry}
\label{s5.2}

We begin by considering the standard quasi-1D localization in
Wigner-Dyson classes ($m_l=1$, $m_o=\beta$, $m_s=0$). In the long-wire
limit the transmission eigenvalues satisfy the hierarchy $1\gg T_1 \gg
T_2 \gg \ldots$, i.e. the consecutive $x_i$ are separated by intervals
much larger than unity. One can thus approximate the hyperbolic sine
functions in Eq.~(\ref{e5.5}) by the exponentials. As a result, all the
variables in the DMPK equation (\ref{e5.4}) fully decouple; the resulting
Fokker-Planck equation for each of $x_k$ reads 
\be
\label{e5.8}
{d{\cal P}(x_k)\over dL} = \frac{1}{2\gamma\ell}\frac{\partial^2{\cal
      P}}{\partial x_k^2} - \frac{1}{\xi_k} \frac{\partial {\cal
      P}}{\partial x_k},  
\ee
where $\xi_k^{-1} = [1+\beta(k-1)]/\gamma\ell$. This is an equation of
the advection-diffusion type, with
$1/2\gamma\ell$ playing the role of the diffusion constant and $1/\xi_1$
of the drift velocity. The solution has a Gaussian form with 
$\langle x_k\rangle = L/\xi_k$ and ${\rm var}(x_k) = L/\gamma\ell$.
This immediately implies that the logarithm of the conductance 
$g\simeq d/\cosh^2x_1$ 
has a Gaussian distribution with the following average and variance
\cite{beenakker97}: 
\be
\label{e5.9}
-\langle \ln g \rangle = 2L/\gamma\ell\ ; \qquad {\rm var}(\ln g) =
4L/\gamma\ell\ .
\ee
On the side of atypically large $g$ this distribution is cut at
$g\sim 1$. The average conductance is straightforwardly found to be
determined by this cutoff (i.e. by rare events of $g\sim 1$), 
$-\ln \langle g \rangle = L/2\gamma l$. Defining the typical
and the average localization length via 
\be 
\label{e5.10}
-\langle \ln g \rangle = 2L/\xi_{\rm typ}\ ;\qquad
 -\ln \langle g \rangle = 2L/\xi_{\rm av}\ ,
\ee
we thus find
\be
\label{e5.11}
\xi_{\rm typ} = \gamma\ell \ ;\qquad \xi_{\rm av} = 4\gamma\ell\ .
\ee
These results are in full agreement with those obtained within the
$\sigma$-model formalism \cite{mirlin00}, which corresponds to $N\gg
1$, so that $\gamma = \beta N$.

A similar behavior is also found in the Bogoliubov--de Gennes classes
C and CI  \cite{brouwer00prl85}.  
The same derivation yields the equation (\ref{e5.8}) with 
$\xi_k^{-1} = [3+4(k-1)]/\gamma\ell$ for class C and 
$\xi_k^{-1} = [2+2(k-1)]/\gamma\ell$ for class CI. 
Thus,
\bea
\label{e5.12}
\hspace*{-1cm}& -\langle \ln g \rangle = 6L/\gamma\ell\ ; \ \  {\rm
  var}(\ln g) = 
4L/\gamma\ell &\ \ \  {\rm (C)}\ ; \\
\hspace*{-1cm} & -\langle \ln g \rangle = 4L/\gamma\ell\ ; \ \  {\rm
  var}(\ln g) = 
4L/\gamma\ell &\ \ \ {\rm (CI)}\ .
\label{e5.13}
\eea
Calculating the average conductance, one gets $-\ln \langle g \rangle
= 4L/\gamma l$ (class C) and  $-\ln \langle g \rangle
= 2L/\gamma l$ (class CI), so that
\bea
\label{e5.14}
& \xi_{\rm typ} = \gamma\ell/3 \ ;\qquad \xi_{\rm av} = \gamma\ell/2 &
\qquad {\rm (C)}\ ;\\
& \xi_{\rm typ} = \gamma\ell/2 \ ;\qquad \xi_{\rm av} = \gamma\ell &
\qquad {\rm (CI)}\ .
\label{e5.15}
\eea
A quasi-1D model of class C was also studied within the $\sigma$-model
formalism in \textcite{bundschuh98}. It was found there that $\xi_{\rm
  av}$ is 8 times shorter than for class A. This agrees with
the result of the DMPK approach, as is seen by comparison of 
$\xi_{\rm av}$ in  Eq.~(\ref{e5.14}) and  Eq.~(\ref{e5.11}).

\subsection{Types of delocalization in disordered wires}
\label{s5.2a}

Having discussed in Sec.~\ref{s5.2} how conventional localization
in disordered quasi-1D systems takes place, we now analyze how  
electrons in such a system can escape exponential localization. 
Table~\ref{t5.1}
is very useful in this respect, showing that there are two distinct
mechanisms. 

First, we see that in five classes 
(in all three chiral classes AIII, BDI, and CII, as well as in the
superconducting classes with broken spin-rotation invariance, BD and DIII)
the multiplicity $m_l$ of long roots in zero. This implies that there
is no repulsion between the eigenvalue $x_i$ and its mirror $-x_i$, so
that the smallest (by absolute value) $x_i$ can be close to zero. 
We will analyze the critical behavior that emerges in these classes in
Sec.~\ref{s5.4} and \ref{s5.5}.

The second type of delocalization is related to the multiplicity of short
roots $m_s$ in the Table~\ref{t5.1}. In the conventional situation it
is equal to zero. However, there are five classes (A, C, BD, AII, and
DIII), where it can be non-zero. A non-zero value of $m_s$ implies a
repulsion of $x_i$ from zero, which is an indicator of the existence
of exactly zero eigenvalues of the transfer-matrix. These zero
eigenvalues imply perfectly transmitting channels, yielding a finite
conductance in the limit of infinite system length. We will discuss
this class of systems in Sec.~\ref{s5.3}.

\subsection{Models with perfectly conducting channels}
\label{s5.3}

In this subsection, we consider the models with perfectly transmitting
channels, in which case $m_s\ne 0$, see Table~\ref{t5.1}. These models
are in turn subdivided in two types. 

In the classes A, C, and BD, the transfer-matrix spaces given in the
Table~\ref{t5.1} correspond to a model with $p$ channels propagating
to the left and $q$ channels propagating to the right. While in the
conventional situation $p=q$, symmetric spaces with $p\ne q$ are
allowed as well. It is not difficult to understand what is the physical
realization for these models, if one recalls that these are exactly
those classes whose 2D representatives show the quantum Hall effects, 
Sec.~\ref{s6}. When such a 2D system is on the quantum Hall plateau of
order $p$, there are $p$ edge channels that propagate on its
boundary. These channels are chiral in the sense that they can
propagate in one direction only. The edge of such a system represents
thus a wire of the corresponding symmetry class with $p$ modes
propagating in one direction and zero in the opposite. Since there is
no backscattering, the conductance of such a wire is identically equal
to $p$.  Let us consider
now parallel edges of two quantum Hall systems with counterpropagating
modes separated by a potential barrier. Assuming that there are $p$
modes in one edge and $q$ in the other and that they are coupled by
tunneling, we get a wire of the $(q,p)$ type from the corresponding
class. In the long length limit, $|p-q|$ modes will then remain perfectly
conducting, while the rest will be localized.   

The situation with the classes AII and DIII is much more
intricate. The systems of these classes may possess a single perfectly
conducting channel. As we explain below, this 
reflects an underlying  $\mathbb{Z}_2$ topological structure. 
A delocalization in the symplectic Wigner-Dyson class AII was obtained
for the first time within the $\sigma$-model formalism in
\textcite{zirnbauer92,mirlin94c}. It was found in these works that the
average conductance and its variance remain finite in the long-wire
limit, $\langle g\rangle \to 1/2$, ${\rm var}(g)\to 1/4$ due to a zero mode of
the corresponding transfer-operator. The physical significance of
these results was not understood at this stage. Further, it was shown
in \textcite{brouwer96} that the above zero mode is double-valued on the
$\sigma$-model manifold and thus does not contribute in the case of a
conventional wire with spin-orbit interaction. More recently, it was
understood, however, that a model of symplectic symmetry with a
perfectly conducting channel arises if one considers transport in
carbon nanotubes \cite{ando02,suzuura02}. The problem is described by
two species of Dirac fermions corresponding to two valleys in the
graphene spectrum. If the scatterers are of long-range character and
the intervalley scattering can be neglected, the problem acquires the
symplectic (AII) symmetry, with the sublattice space taking the role
of isospin (Sec.~\ref{s6.7.2}). 
Furthermore, in contrast to conventional wires of AII
symmetry, where the number of channels is even, there is just a single
channel here. (More precisely, its ``partner'' belongs to the other
node, and they ``do not talk to each other''.) As a result, the
channel remains perfectly transmitting independently of the length of
the wire. 

In subsequent works \cite{takane04b,takane04c,sakai05,caselle06}
quasi-1D systems of the AII symmetry class with odd number $N$ of 
channels were studied within the DMPK approach. 
It was found that for any odd $N$ a single perfectly transmitting
channel remains in the limit $L\gg \gamma\ell$. As emphasized in
\textcite{takane04a}, these results are in full agreement with the earlier
$\sigma$-model results of \textcite{zirnbauer92,mirlin94c}, if the latter
are complemented by the following interpretation. The Fourier expansion 
employed in \textcite{zirnbauer92,mirlin94c} contained eigenfunctions of
the Laplace operator in the $\sigma$-model manifold of two types --
with even and odd parity (the zero mode is of the odd-parity type). 
For systems with even (odd) number of channels one should keep only
even-parity (resp. odd-parity) eigenmodes and include an overall
factor of two.  The original result of
\textcite{zirnbauer92,mirlin94c} with $\langle g\rangle, \ \langle
g^2\rangle \to 1/2$ at $L\to \infty$ 
corresponds thus to an average over wires with
even and odd number  of channels.

In the problem with perfectly transmitting channels, one can determine
the localization length for the remaining modes by considering the
deviation $\delta g$ of the conductance from its $L\to\infty$
limit. A straightforward generalization of the consideration sketched
in Sec.~\ref{s5.2} yields \cite{caselle06}
\be
\label{e5.15a}
-\langle \ln \delta g \rangle = (2 m_l + m_s) L/\gamma\ell\ ,
\ee
so that $\xi_{\rm typ} = \gamma\ell / (m_l + m_s/2)$.
This implies for both the AII and DIII models with an odd number of
channels
\be
\label{e5.15b}
\xi_{\rm typ} = \gamma\ell/3\ ;\qquad 
\xi_{\rm av} = \gamma\ell/2\ .
\ee

The qualitative different behavior of class-AII (and DIII) wires with
even and odd number of channels is intimately connected with a
non-trivial topology of the corresponding $\sigma$-model manifold
(or, more specifically, of its compact component $\mathcal{M}_F$): the
first homotopy group $\pi_1(\mathcal{M}_F)$ is equal to
$\mathbb{Z}_2$. This enables a topological
$\theta$-term with $\theta$ equal to $0$ or $\pi$ \cite{ostrovsky07a}, 
in analogy
with the 2D situation, see Sec.~\ref{s6.1.5}, \ref{s6.2.5} for these
symmetry classes. The topological term with
$\theta = \pi$ is present if the number of channels is odd. 

Another realization of a symplectic-symmetry wire with 
an odd number of channels has
recently emerged in the context of the
quantum spin Hall (QSH) effect in systems of Dirac fermions with spin-orbit
coupling  \cite{kane05a,kane05b,bernevig06}. 
Such systems were found to possess two
distinct insulating phases (with a transition between them driven by
Rashba spin-orbit coupling 
strength), both having a gap in the bulk electron spectrum but
differing by the edge properties. The topological distinction between
the two insulating phases retains 
its validity in the presence of disorder
\cite{sheng05,onoda07,obuse07a,essin07}. 
While the normal insulating phase has no edge
states, the QSH insulator is characterized by a pair of mutually
time-reversed spin-polarized edge states penetrating the bulk gap.
These edge states in the QSH phase, which do not get localized by
disorder, represent a realization of the class-AII wire with a non-trivial
$\mathbb{Z}_2$ topology.

\subsection{Chiral classes}
\label{s5.4}

For the chiral classes, the  consideration analogous to that in
Sec.~\ref{s5.2} \cite{mudry99a,mudry00,brouwer00npb}
yields $\langle x_k\rangle$ = $(2k-1-N)\beta L/2\gamma\ell$. 
If the number of channels is even, then the smallest by absolute value
eigenvalues $x_k$ are separated by a large gap from zero:
$- \langle x_{N/2} \rangle =  \langle x_{N/2+1} \rangle =
L/2\gamma\ell$. Therefore, the exponential localization is preserved,
$-\langle \ln g \rangle = \beta L/\gamma\ell$, ${\rm var}(\ln g) =
4L/\gamma\ell$, yielding the localization lengths
\be
\label{e5.16}
\xi_{\rm typ} = 2\gamma\ell/\beta \ ;\qquad \xi_{\rm av} =
16\gamma\ell/\beta^2 \ .
\ee

On the other hand, if $N$ is odd, one of the eigenvalues is close to
zero, $\langle x_{(N+1)/2} \rangle = 0$. This leads to a completely
different behavior of the conductance,
\bea
\label{e5.17}
\hspace*{-1cm}
&&-\langle \ln g \rangle = \left({8L\over\pi\gamma\ell}\right)^{\!\!1/2};
\ \ 
{\rm var}(\ln g) = \left(4-{8\over\pi}\right){L\over\gamma\ell}\:;\\
\hspace*{-1cm}
&&\langle g\rangle = (2\gamma\ell /\pi L)^{1/2}\ ;
\qquad {\rm var}(g)  = (8\gamma\ell /9\pi L)^{1/2}.
\label{e5.18}
\eea 
It is seen from Eq.~(\ref{e5.17}) that while the typical conductance
decays in a stretched-exponential way, its fluctuations are very
strong, so that the probability to have $g\sim 1$ is small as
$L^{-1/2}$ only. This determines the very slow decay (\ref{e5.18}) of
the average conductance, which is even slower than the in classical Ohm's
law. 

The delocalization takes place for arbitrary odd $N$, including $N=1$.
The single-channel model with chiral-class disorder has been studied,
in its various incarnations, in a large number of works, starting form
the pioneering paper by \textcite{dyson53}. We list most salient
features characterizing (in addition to the above results for the
statistical properties of the conductance) this critical point.

{\it i) Localization length.} 
If the energy $E$ deviates from zero, the chiral symmetry is
broken, and the exponential localization establishes. One can thus ask
how the corresponding localization length diverges at $E\to 0$. It has
been found that one should distinguish between average and typical
observables (e.g. conductance) whose spatial dependence is governed by
two parametrically different lengths \cite{fisher95,balents97},
\be
\label{e5.19}
\xi_{\rm typ} \sim |\ln E|\ ;\qquad
\xi_{\rm av} \sim |\ln E|^2\ .
\ee

{\it ii) Staggering.}
An alternative way to drive the system out of criticality is to
introduce a staggering $M$ in the hopping strength
which opens a gap around zero energy in the
spectrum of a clean system. The corresponding localization lengths 
behave as follows \cite{fisher95,balents97,mathur97},
\be
\label{e5.20}
\xi_{\rm typ} \sim M^{-1}\ ;\qquad
\xi_{\rm av} \sim M^{-2}\ .
\ee

{\it iii) Wave function at criticality.} The Hamiltonian of a
single-chain problem can be written in a Dirac form \cite{balents97},
\be
\label{e5.21}
H=-i\sigma_z\partial_x + m(x)\sigma_y\ ,
\ee
where $m(x)=M +\tilde{m}(x)$
and $\tilde{m}(x)$ is the disorder (e.g. of the white-noise type). The
zero-energy eigenfunction can then be found explicitly,
\bea
\label{e5.22}
&& \Psi(x) = \left(
\begin{array}{c} 1 \\ \pm 1 \end{array}
\right)
{\psi_{\pm}(x) \over \left[\int dx\: \psi_\pm^2(x)\right]^{1/2}}, \nonumber\\
&& \psi_{\pm}(x) = \exp\left[\pm\int^x dx'\:m(x')\right]\:.
\eea 
The properties of this wave function were analyzed in
\textcite{balents97}. The following scaling of the spatial correlation
function of the moments of $\psi(x)$ at criticality ($M=0$) was found,
\be
\label{e5.23}
\langle |\psi(x)\psi(0)|^q\rangle \sim L^{-1} |x|^{-3/2}\ ,
\ee
for all $q>0$. This result can be interpreted in terms of the
following picture of wave functions at criticality
\cite{balents97}. The wave function 
is typically quasi-localized, showing a stretched-exponential
decay with respect to its principal maximum. 
However, with a probability $\sim x^{-3/2}$ it shows a secondary
maximum of a magnitude close to the primary one and separated from it
by a distance $x$.   

{\it iv) Density of states.} The DOS shows at criticality the Dyson
singularity \cite{dyson53,mckenzie96,titov01}, 
\be
\label{e5.24}
\rho(E) \sim 1/|E \ln^3 E|\ .
\ee
When the system is driven away from criticality by a non-zero
staggering parameter $M$, the singularity weakens and becomes
non-universal, $\rho(E)\sim |E|^{-1+\delta}$ with
$\delta>0$. 

Finally, it has been shown that a sufficiently strong 
staggering $M$ can also drive a system with even $N$ into a critical state
\cite{brouwer98}. More specifically, the staggering shifts all the variables
$x_k$ by a constant. With increasing $M$, the average values 
$\langle x_k\rangle$ consecutively cross zero; whenever this happens,
the system is at criticality. Therefore, whether $N$ is odd or even,
changing $M$ will drive the system through $N$ transition points.  

\subsection{Bogoliubov-de Gennes classes with broken spin-rotation
  invariance} 
\label{s5.5}

Analysis of the DMPK equation for the classes BD and DIII 
\cite{brouwer00prl85} leads to results identical to those obtained
for chiral classes, Eq.~(\ref{e5.17}), (\ref{e5.18}). Furthermore,
the DOS was found to show the Dyson singularity  (\ref{e5.24}), again
in full analogy with the chiral classes.

On the other hand, \textcite{motrunich01} studied 
certain single-channel models of the classes D and
DIII via a strong-disorder real-space RG. It was found
that generically these systems are in localized phases and the DOS
diverges in a power-law fashion with a non-universal exponent, 
$\rho(E)\sim E^{-1+\delta}$ with $\delta>0$. Only at
phase boundaries the system is critical and the DOS takes the Dyson
form (\ref{e5.24}). 

An apparent contradiction between the results of both papers was
resolved in \textcite{gruzberg05}. It was shown there that, generically,
quasi-1D systems of the classes BD and DIII are in a localized phase,
in agreement with \textcite{motrunich01}. The terms that drive the system
towards localization are usually neglected within the DMPK approach, as they
are irrelevant at the short-distance (diffusive) fixed point. It was
found, however, in \textcite{gruzberg05} that these terms 
become relevant at the long-distance (critical) fixed point and drive
the system away from it, into the localization fixed point. Only if
the disorder is fine tuned, the system is at the critical point. 
On the other hand, the length at which the crossover from criticality
to localization happens becomes exponentially large with increasing
number of channels $N$. Therefore, in the thick-wire limit, $N\gg 1$, 
the system is essentially at criticality. An analogous conclusion was
also reached in \textcite{brouwer03}.

It was also argued in   \textcite{motrunich01,gruzberg05} that critical
points with Dyson singularities of all five symmetry classes (AIII,
BDI, CII, BD, and DIII) belong to the same universality class. To
establish this remarkable ``superuniversality'', 
\textcite{gruzberg05} pointed out that all the
universal properties can be obtained from $N=1$ models and then
constructed mappings between single-channel models of all the
five classes.

\section{Criticality in 2D}
\label{s6}

\subsection{Mechanisms of criticality in 2D}
\label{s6.1}

As was discused in Sec.~\ref{s2.2.2}, conventional Anderson
transitions in the orthogonal and unitary symmetry classes take place
only if the dimensionality is $d>2$, whereas in 2D all states are
localized. It is, however, well understood by now that there is a rich
variety of mechanisms that lead to emergence of criticality in 2D
disordered systems. Such 2D critical points have been found to exist
for 9 out of 10 symmetry classes, namely, in all classes 
except for the orthogonal class AI.
A nice summary of possible types of 2D criticality was given in
\cite{fendley00}; we closely follow this work in our presentation in
this subsection. We now list and briefly describe the mechanisms for
the emergence of criticality; a detailed discussion of the
corresponding critical points will be given in the remaining
subsections of Sec.~\ref{s6}.

\subsubsection{Broken spin-rotation invariance: Metallic phase}
\label{s6.1.1}

We begin with the mechanism that has been already mentioned in
Sec.~\ref{s2.2.2} in the context of the Wigner-Dyson symplectic class
(AII). In this case the $\beta$-function  [(\ref{e2.20}) with
$\epsilon=0$] is positive for not too large $t$ (i.e. sufficiently
large conductance), so that the system is metallic ($t$ scales to zero
under RG). On the other hand, for strong disorder (low $t$) the system
is an insulator, as usual, i.e. $\beta(t)<0$. Thus, $\beta$-function
crosses zero at some $t_*$, which is a point of the Anderson
transition. Properties of this critical point will be discussed in
detail in Sec.~\ref{s6.2}.   

This mechanism  (positive $\beta$-function and, thus, metallic
phase at small $t$, with a transition at some $t_*$) is also realized in
two of Bogoliubov-de Gennes classes -- D and DIII, see
Table~\ref{t4.3}.  All these
classes correspond to systems with broken spin-rotation invariance.
The unconventional sign of the $\beta$-function in these classes,
indicating weak antilocalization (rather then localization), is
physically related to destructive interference of time reversed paths
for particles with spin $s=1/2$. 

\subsubsection{Chiral classes: Vanishing $\beta$-function}
\label{s6.1.2}

Another peculiarity of the perturbative $\beta$-function takes place
for three chiral classes -- AIII, BDI, ad CII.  Specifically, for
these classes $\beta(t)\equiv 0$ to all orders of the perturbation
theory, as was first discovered by Gade and Wegner
\cite{gade91,gade93}. 
As a result, the conductance is not renormalized at all,
serving as an exactly marginal coupling.  There is thus a line of
critical points for these models, labeled by the value of the
conductance. In fact, the $\sigma$-models for these classes contain an
additional term \cite{gade91,gade93} that does not affect the
absence of renormalization of the conductance but is crucial for the analysis
of the behavior of the DOS. A detailed discussion of the chiral
classes will be given in Sec.~\ref{s6.6}.

\subsubsection{Broken time-reversal invariance: Topological
$\theta$-term and quantum Hall criticality}
\label{s6.1.3}

For several classes, the $\sigma$-model action allows for inclusion of
a topological term, which is invisible to any order of the
perturbation theory. This is the case when the second homotopy group
$\pi_2$  of the $\sigma$-model manifold ${\cal M}$ (a group of homotopy
classes of maps of the sphere $S^2$ into ${\cal M}$) is 
non-trivial.\footnote{A pedagogical introduction of topological 
concepts in the 
context of condensed matter theory can be found in the recent
monograph by \textcite{altland06a}.}
From this point of view, only the compact sector ${\cal M}_F$
(originating from the fermionic part of the supervector field) of the
manifold base matters. There are five classes, for which 
$\pi_2({\cal M}_F)$ is non-trivial, namely A, C, D, AII, and CII.

For the classes A, C, D the homotopy group $\pi_2({\cal M}_F) = \mathbb{Z}$.
Therefore, the action $S[Q]$ may include the (imaginary) $\theta$-term,
\be
\label{e6.1}
iS_{\rm top}[Q] = i \theta N[Q]\ ,
\ee
where an integer $N[Q]$ is the winding number of the
field configuration $Q({\bf r})$. Without loss of generality, $\theta$
can be restricted to the interval $[0,2\pi]$, since the theory is
periodic in $\theta$ with the period $2\pi$. 

The topological term (\ref{e6.1}) breaks the time reversal invariance,
so it may only arise in the corresponding symmetry classes.
The by far most famous case is the Wigner-Dyson unitary class (A).
As was first understood by Pruisken \cite{pruisken84,pruisken87}, 
the $\sigma$-model of this class with the topological term
(\ref{e6.1}) describes the integer quantum Hall effect (IQHE), 
with the critical
point of the plateau transition corresponding to $\theta=\pi$. 
More recently, it was understood that  counterparts of the IQHE
exist also in the Bogoliubov-de Gennes classes with broken
time-reversal invariance -- classes C and D. They were called {\it spin} and
{\it thermal} quantum Hall effects (SQHE and TQHE), respectively. 
The criticality at the IQHE, SQHE, and TQHE transitions will be discussed
in detail in Sec.~\ref{s6.3}, \ref{s6.4}, and \ref{ss:tqhe},
respectively.

\subsubsection{ $\mathbb{Z}_2$ topological term}
\label{s6.1.4} 

For two classes, AII and CII, the second homotopy group is
$\pi_2({\cal M}_F) = \mathbb{Z}_2$. This allows for the $\theta$-term
but $\theta$ can only take the value $\theta=0$ and $\theta=\pi$.
It was shown very recently \cite{ostrovsky07a} that the $\sigma$-model
of the Wigner-Dyson symplectic class (AII) with a $\theta=\pi$
topological angle arises from a model of Dirac fermions with random
scalar potential, which describes, in particular, graphene with
long-range disorder. Like in the case of quantum-Hall systems, 
this topological term inhibits localization. Whether the model then flows
unavoidably into the ideal-metal fixed point or, else, there is
also a novel attractive fixed point is a matter of ongoing research. 
The reader is referred to Sec.~\ref{s6.2} for more detail.

\subsubsection{Wess-Zumino term}
\label{s6.1.5}

Finally, one more mechanism of emergence of criticalty is the
Wess-Zumino (WZ) term. It is known that this term may appear in
$\sigma$-models of the classes AIII, CI, and DIII. For these classes,
the compact component ${\cal M}_F$ of the manifold is the group 
$H\times H/ H = H$, where $H$ is $\text{U}(n)$, $\text{Sp}(2n)$, and
$\text{O}(2n)$, 
respectively. The corresponding theories are called ``principal chiral
models''. The WZ term has the following form:
\bea
\label{e6.2}
&& iS_{\rm WZ}(g) = {ik\over 24\pi} \int d^2r \int_0^1 ds \:
\epsilon_{\mu\nu\lambda} \nonumber \\
&& \times {\rm Str} (g^{-1}\partial_\mu g)
(g^{-1}\partial_\nu g)  (g^{-1}\partial_\lambda g),
\eea
where $k$ is an integer called the level of the WZW model. 
The definition (\ref{e6.2}) of the WZ term requires an extension of
the $\sigma$-model field $g({\bf r}) \equiv g(x,y)$ to the third
dimension, $0\le s \le 1$, such that $g({\bf r}, 0) = 1$ and 
$g({\bf r},1) = g({\bf r})$. Such an extension is always possible,
since the second homotopy group is trivial, $\pi_2(H)=0$, for all
the three classes. Further, the value of the WZ term does not depend
on the particular way the extension to the third dimension is
performed. (This becomes explicit when one calculates the variaton of
the WZ term: it is expressed in terms of   $g({\bf r})$ only.) More
precisely, there is the following topological ambiguity in the
definition of $S_{\rm WZ}(g)$. Since the third homotopy group is
non-trivial,  $\pi_3(H)=\mathbb{Z}$,  $S_{\rm WZ}(g)$ is defined up
to an arbitrary additive integer $n$ times $2\pi k$. 
This, however, does not affect any
observables, since simply adds the phase $nk \times 2\pi i$ to the
action. 

The WZ term arises when one bosonizes certain models of Dirac fermions
\cite{witten84} and is a manifestation of the chiral anomaly. 
In particular, a $\sigma$-model for a system of 
the AIII (chiral unitary) class  with the WZ term describes
Dirac fermions in a random vector potential. In this case
the $\sigma$-model coupling constant is truly marginal (as is
typical for chiral classes) and one finds a line of fixed
points. On the other hand, for the class CI there is a single fixed
point. The WZW models of these classes were encountered in the
course of study of dirty $d$-wave superconductors 
\cite{nersesyan95,altland02} and, most recently, in the
context of disordered graphene. We will discuss
critical properties of these models in Sec.~\ref{s6.7.3}.

\subsection{Symplectic Wigner-Dyson class (AII) }
\label{s6.2}

In metals with spin-orbit coupling the spin of a particle
is no longer conserved. The spin-up and spin-down channels
are coupled, and an electron needs to be represented as a
two component spinor. If the time-reversal symmetry is preserved, the system
belongs to the symplectic Wigner-Dyson class AII. The one-loop quantum
correction at large conductance $g$ takes then the form of weak
antilocalization, see Sec.~\ref{s2.2.2}, \ref{s6.1.1}. At lower $g$
the one-loop $\beta$-function is not sufficient anymore, and
higher-order terms lead to localization, with the Anderson transition
at some $g_*$. 
While the $\beta$-function has been calculated up to
the four-loop order, see Eq.~(\ref{e2.20}),  this does not help to get
quantitative predictions for critical properties. In particular,
an attempt to use the four-loop $\beta$-function to extract the
localization length exponent \cite{wegner88} yields 
$
    \nu {=} \frac{1}{5} \left(
      \frac{3}{4}\zeta(3)\right)^{1/3}{\approx} 0.193, 
$
which is an order of magnitude smaller than the numerical result (see
below) and even violates the Harris criterion $\nu\geq 2/d$  \cite{chayes86}.
This is not very surprising: the considered Anderson transition takes
place at strong coupling, $g_*\sim 1$, so that keeping just the first
few terms of the perturbative expansion is an uncontrolled
procedure. In this situation, numerical simulations are particularly
important. On their basis, a detailed quantitative picture of the
transition has been developed; the key findings  are summarized
below.

\subsubsection{Microscopic models}
\label{s6.2.1}

Most numerical studies employed a tight binding Hamiltonian. It is  defined
on a two dimensional square lattice with nearest neighbor coupling
\be
  H = \sum_{i,\sigma} \epsilon_i^{\phantom{\dagger}} 
           c^\dagger_{i,\sigma} c^{\phantom{\dagger}}_{i,\sigma}
    + \sum_{\langle i,j \rangle, \sigma, \sigma'} 
         V_{i,\sigma;j,\sigma'}^{\phantom{\dagger}}
         c^\dagger_{i,\sigma} c^{\phantom{\dagger}}_{j,\sigma'}. 
\label{e6.1.1}
\ee
Here, $c^\dagger_{i,\sigma}$ ($c^{\phantom{\dagger}}_{i,\sigma})$ denote
creation (annihilation) operators of an electron with spin $\sigma$
on site $i$.
The on-site energies $\epsilon_i$ are taken to be random numbers
drawn from the interval $[-W/2,W/2]$ with a homogeneous distribution. 
There exist various versions of the model, that differ by the choice of
the hopping matrix $V_{i,\sigma;j,\sigma'}$.

Most studies employ the Ando model \cite{ando89}
characterized by non-random hopping between next neighbors only, 
\be
  V_{i,\sigma;i+k,\sigma'} = 
    \left( V_0 \exp (i \theta_k \sigma_k) \right)_{\sigma,\sigma'}, \qquad
  k = x,y,
\label{e6.1.2}
\ee
where  $\sigma_x, \sigma_y$ are the Pauli matrices.
Conventionally, the spin-orbit energy scale is set to unity, $V_0{=}1$,
and the mixing angles take constant values $\theta_k{=}\pi/6$.
In the Evangelou-Ziman model \cite{evangelou87,evangelou95}, 
the components of
$V$ that are proportional to $\sigma_{x,y,z}$,
are chosen to be random with prefactors drawn independently from a
box distribution of a  width $V_0$. Recently, a third
variant was introduced \cite{asada02,asada04} -- 
the SU(2) model, in which the random matrix $\exp (i \theta_k
\sigma_k)$ is chosen to be uniformly distributed on the SU(2) group. 
This last model was found particularly
suitable for numerics, since finite size corrections appear
to be very small.

\subsubsection{Localization length exponent}
\label{s6.2.2}

The small magnitude of finite-size corrections in the SU(2) model has
allowed \textcite{asada02,asada04} to determine the
localization length exponent with a high precision, 
$\nu{=}2.746\pm0.009$. Very recently, a similar result (though with somewhat
higher uncertainty) was obtained \cite{markos06} 
in the context of the Ando model, $\nu = 2.8 \pm 0.04$.
\textcite{asada04} evaluated numerically the whole  $\beta$-function
and found that it has the expected shape, with a single
zero determining the transition point.

\subsubsection{Critical conductance}
\label{s6.2.3}

Since the transition takes place in the strong coupling regime, the
mesoscopic 
conductance fluctuations at criticallity are comparable to the mean value,
so that an ensemble of macroscopically identical coherent samples is
to be characterized by the whole distribution function ${\cal P}(g)$. 
Quite generally speaking, 
critical conductance distributions are scale-invariant but depend
on the shape of the sample (similar to the IPR distribution function,
Sec.~\ref{s2.3.5} and the level statistics, Sec.~\ref{s2.5}); a
review of numerical results has been given by \textcite{markos06b}. 
The distribution ${\cal P}(g)$ for a square sample at the
symplectic Anderson transition was determined in \textcite{ohtsuki04} for
the SU(2) model, see Fig. \ref{f11b}, 
and in \textcite{markos06} for the Ando model, with
essentially identical results. The average value  $\langle g \rangle$
was found to be  $\langle g\rangle{=}1.42\pm 0.005$ 
with a variance $\text{var} g {=} 0.36$  \cite{markos06}.

\subsubsection{Multifractal spectrum}
\label{s6.2.4}

The spectrum $\tau_q$ can be calculated for the 2D symplectic
transition with very good accuracy, because corrections to scaling
turn out to be extremely small \cite{asada04,mildenberger07}.
Therefore, many generic features of the critical wavefunction
statistics can be studied in great detail. 

\begin{figure}[t]
\includegraphics[width=0.9\columnwidth,clip]{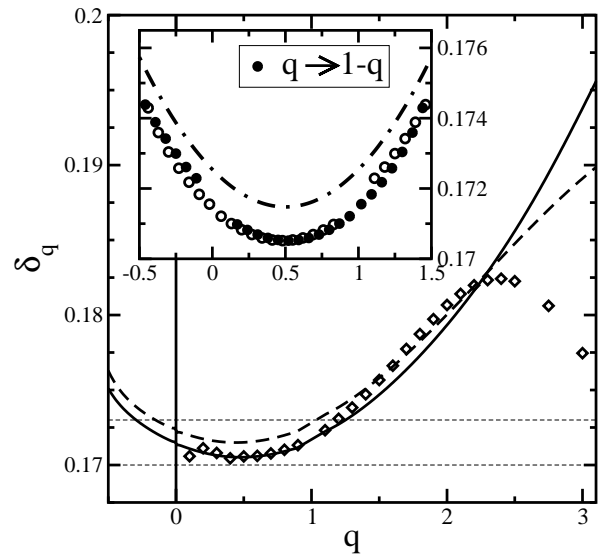}
\caption{Multifractal spectrum $\delta_{q}$  
 for the Ando model (dashed, $W_c{=}5.84$)
  and the SU(2) model (solid, $W_c{=}5.953$). To highlight deviations
  from parabolicity, reduced anomalous dimensions
  $\delta_q=\Delta_q/q(1-q)$ are plotted. Anomalous dimensions
  $\delta_q^{\rm typ}$ obtained from typical IPR are also shown
 ($\diamond$). Dashed lines indicate 
  the estimated error (2$\sigma$) in $\delta_0$. {\it Inset:} blow up of the
  solid line behavior near $q{=}{1\over 2}$ is represented by empty circles
  ($\circ$). Filled symbols ($\bullet$) show original trace
  after reflection at $q{=}{1\over 2}$. Dot-dashed line is a fit
  (offset: $10^{-3}$) $\delta_q = 0.1705 + 0.0043 (q-{1\over 2})^2$. 
  \cite{mildenberger07}} 
\label{fa}
\end{figure}

The results of a high-precision study of wave function multifractality 
at the symplectic Anderson transition \cite{mildenberger07} 
are summarized in Fig.~\ref{fa}. 
In order to highlight the non-trivial
features, the reduced anomalous dimensions
$\delta_q{=} \Delta_q/q(1{-}q)$ are depicted. The key observations are
as follows:

(i) One finds that $\delta_0 \equiv \alpha_0 - 2{=}0.172\pm 0.002$.
This result may be used as a check on the conformal invariance, which
imposes the exact condition \cite{janssen94a,janssen98}
\be
     \pi \delta_0 \Lambda_c = 1.
\label{e6.1.3}
\ee
Here $\Lambda_c{=}\xi_M / M$, where $\xi_M$ is the  localization
length in a quasi-1D strip of width $L$ at criticality.) 
With the above value for $\delta_0$ and with
$\Lambda_c{=}1.844\pm 0.002$ \cite{asada04},
the left-hand side of Eq.~(\ref{e6.1.3}) 
becomes indeed very close to unity: $0.996\pm 0.013$. 

(ii) The function $\delta_q$ fulfills the symmetry relation
Eq.(\ref{e2.32}), as demonstrated in the inset of Fig. \ref{fa}
where the data for the range $-0.5< q < 1.5$ are displayed.

(iii) $\delta_q$ has a small but non-zero curvature, implying that the
the multifractal spectrum is not parabolic.

(iv) The results are essentially identical for the SU(2) and Ando
models, confirming the universality of the transition.
An abrupt change in the behavior 
of the data for the typical IPR in Fig.~\ref{fa} at $q\simeq 2.5$
is related to the fact that at $q>q_+$, when the average IPR probes
the tail of its distribution function, the exponents $\tau_q$ and
$\tau_q^{\rm typ}$ start to differ, see Eq.~(\ref{e2.36}). In this range
of $q$ the statistical  uncertainty  in determination of $\tau_q$ also
increases, explaining some deviation between the data for  both
models. 

Very similar results for the multifractality spectrum were obtained in
\cite{obuse07}. In this work, the multifractality was also studied at
the boundary and at the corner of the system. It was found that the
multifractal exponents fulfill the relation (\ref{e2.48}), thus
providing a further strong evidence for the conformal invariance at this
critical point.     

\subsubsection{Symplectic-class theories with $\mathbb{Z}_2$ topology. }
\label{s6.2.5}

As was explained in Sec.~\ref{s6.1.4}, the $\sigma$-model of the
symplectic class allows for an inclusion of a topological term 
with $\theta=\pi$. A microscopic realizations of such a
non-trivial topology was for the first time identified in 
\textcite{ostrovsky07a} where the model of Dirac fermions in disordered
graphene was studied (see Sec.~\ref{s6.7}). 
It was found that for the case of long-range
impurities, when two valleys in the spectrum are
decoupled and the problem reduces to that of a single species of Dirac
fermions in random potential, the field theory is the class-AII
$\sigma$-model with $\theta=\pi$ topological term.  

The fermionic sector of the corresponding $\sigma$-mode manifold is 
$ \mathcal{M}_F = \text{O}(4n)/\text{O}(2n)\times \text{O}(2n)$. 
In fact, for the ``minimal'' supersymmetric $\sigma$-model ($n=1$) 
the second homotopy group is richer:
\begin{equation}
 \pi_2 \bigl[ \left. \mathcal{M}_F\right|_{n=1} \bigr]
  = \mathbb{Z} \times \mathbb{Z}, \qquad
 \pi_2 \bigl[ \left. \mathcal{M}_F\right|_{n\geq2} \bigr]
  = \mathbb{Z}_2.
 \label{pi2}
\end{equation}
For $n=1$ the compact sector of the model is the manifold  
$(\text{S}^2 \times \text{S}^2)/\mathbb{Z}_2$
(product of the ``diffuson'' and ``Cooperon'' 2-spheres divided by
$\mathbb Z_2$). 
Thus two topological invariants, $N_{1,2}[Q]$, counting the
covering of each sphere, emerge in accordance with Eq.\ (\ref{pi2}). The most
general topological term is $i S_{\rm top} = 
i\theta_1 N_1 + i\theta_2 N_2$. However, the time-reversal symmetry
requires that 
the action is invariant under interchanging the diffuson and
Cooperon spheres, which yields $\theta_1 = \theta_2 \equiv \theta$ where
$\theta$ is either $0$ or $\pi$. Hence only a
$\mathbb{Z}_2$ subgroup of the whole $\mathbb{Z} \times
\mathbb{Z}$ comes into play as expected: the phase diagram of the theory should
not depend on $n$.
For the Dirac fermion problem \cite{ostrovsky07a}, 
an explicit expression for $n=1$ topological
term can be written using $\mathbf{u} = T \nabla T^{-1}$ (where
the $\sigma$-model field $Q=T^{-1}\Lambda T$, see Sec.~\ref{s2.2.1}), 
\begin{equation*}
 i S_2[Q]
  = \frac{\epsilon_{\alpha\beta}}{8} \mathop{\mathrm{Str}} \bigl[
      (\Lambda \pm 1) \tau_2 u_\alpha u_\beta
    \bigr]
  \equiv i\pi(N_1[Q] + N_2[Q]),
\end{equation*}
yielding $\theta = \pi$. The non-trivial value of the topological angle
($\theta=\pi$) holds for higher $n$ as well. It implies that
all configurations $Q({\bf r})$ are subdivided into two topologically
distinct classes (``even'' and ``odd''); the former give positive and
the latter negative contribution to the $\sigma$-model partition
function. This was confirmed numerically in \textcite{ryu07}.
   
At large conductance $g$ the contribution of topologically non-trivial
configurations is exponentially small and can not affect the metallic
phase in any essential way. On the other hand, at small $g$ the
topological term  is expected to suppress localization, similarly to
its role in the quantum Hall effect (Sec.~\ref{s6.3}) and in the quasi-1D
symplectic model (Sec.~\ref{s5.3}). This leaves room to two
possibilities: 

(i) The $\beta$-function changes sign twice. This would mean that, 
in addition to the conventional repulsive fixed point of the
symplectic class,
a new attractive fixed point arises. This scenario  was proposed in
\textcite{ostrovsky07a}. Then, if the RG flow starts with a sufficiently
low conductivity, it ends up in this new critical point with a
universal conductivity of order unity.

(ii) The $\beta$- function remains positive everywhere. The RG flow
then necessarily leads the system into the ideal-metal fixed point
with infinite conductivity.

In view of the strong-coupling nature of the problem, numerical
simulations are needed to resolve this dilemma. Recent simulations
of disordered graphene \cite{nomura07,rycerz06,bardarson07,nomura07a} 
do confirm the suppression of localization in the symplectic class
with $\mathbb{Z}_2$ topology. While the results of
\textcite{nomura07,rycerz06} were consistent with the scenario (i),
with a critical  conductivity $\sim e^2/h$, most recent works
\cite{bardarson07,nomura07a} appear to favor the second scenario. 

It is worth reminding the reader of a different type of
$\mathbb{Z}_2$ topology in 2D systems of the symplectic symmetry
class (AII). It arises in the context of the quantum spin Hall (QSH) effect
and is related to a non-trivial first homotopy group,  
$\pi_1(\mathcal{M}_F)=\mathbb{Z}_2$.
This enables, in full similarity
to the 2D situation, a $\theta$-term with $\theta$ equal to $0$ or
$\pi$ also in 1D case, inducing a $\mathbb{Z}_2$ 
topological classification of edge states in QSH systems, see
Sec.~\ref{s5.3}. 
Therefore, these systems possess in addition to the metallic phase, two
distinct insulating phases, with different edge properties (normal
insulator and QSH insulator). An important question is whether there
is a direct, quantum-Hall-type transition between these two phases. 
Recent numerics \cite{onoda07,obuse07a,essin07} on some models of QSH
systems gives a negative answer: the insulating phases are found to be
everywhere separated by the metallic phase. 
It would be interesting to find out
whether such a direct transition is generically prohibited,
independently of the microscopic model. More activity in this
direction may be expected in near future.

\subsection{The integer quantum Hall effect }
\label{s6.3}

Our presentation in this section complements the reviews
\cite{huckestein95,kramer05}. 

\subsubsection{Pruisken's $\sigma$-model}
\label{s6.3.1}

As was discovered in \textcite{klitzing80}, the Hall conductivity
$\sigma_{xy}$ of a 2D electron gas in a strong transverse magnetic
field develops plateaus at values quantized in units of $e^2/h$. While
the physics of the Hall plateau is fairly well understood by now, the
theory of the quantum critical points separating the plateaus -- the
quantum Hall transition -- remains a challenging issue. 

From the field-theoretical point of view, the IQHE is described by the
$\sigma$-model (\ref{e2.12}) with a topological term (\ref{e6.1}).
It was first derived by Pruisken in the replica formalism
\cite{levine83,pruisken84,pruisken87}; a supersymmetric generalization
was obtained in \textcite{weidenmueller87}. The action of the model reads
\be
\label{iqhe1}
S[Q] = {1\over 8} {\rm Str} [ -\sigma_{xx} (\nabla Q)^2 + 2
\sigma_{xy} Q \nabla_x Q \nabla_y Q ], 
\ee
where $\sigma_{xx}$, $\sigma_{xy}$ are dimensionless
conductivities. The Hall conductivity $\sigma_{xy}$ is related to the
topological angle as $\sigma_{xy} = \theta/2\pi$. 
There is strong evidence that the corresponding two-parameter 
flow diagram has the form shown in Fig.~\ref{iqhe-f1}, as proposed in 
\textcite{khmelnitskii83,pruisken85,pruisken87}.
The fixed point at $\theta = (2n+1)\pi$
describes the QH transition. While the theory (\ref{iqhe1}) is highly
important for understanding of qualitative features of the problem, it
allows to make only rough predictions for  parameters of the critical
behavior \cite{pruisken04}.
This is because a controllable calculation in this framework 
can only be performed at
weak coupling, $\sigma_{xx}\gg 1$, while the fixed points are at
strong coupling, $\sigma_{xx}\sim 1$. In this situation, numerical
simulations are particularly important; their results will be reviewed
in Sec.~\ref{s6.3.5}--\ref{s6.3.7}.

\begin{figure}[btp]
    \begin{center}\leavevmode
\includegraphics[width=0.8\linewidth]{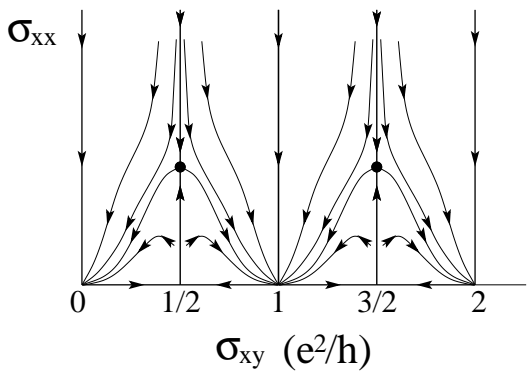}
    \end{center}
      \caption{Two-parameter flow diagram of the Pruisken
        $\sigma$-model, as first proposed in \cite{khmelnitskii83}.}
      \label{iqhe-f1}
\end{figure}

\subsubsection{Further analytical approaches}
\label{s6.3.2}

A great deal of effort has been invested by many researchers in order
to attack the problem of the QH transition from the analytical side. 
In addition to the Pruisken model, Sec.~\ref{s6.3.1}, several other
analytical frameworks have been used. While this activity has not led
to an ultimate success in the quantitative description of  critical
behavior,  a variety of important connections between the
models has been established. In particular, it has been shown that 
the $\sigma$-model (\ref{iqhe1}) is also obtained as a continuum limit
of the Chalker-Coddington network described in
Sec.\ref{s6.3.4} \cite{zirnbauer97}. Further, either of these two models can be
mapped onto a quantum antiferromagnetic superspin chain 
\cite{zirnbauer94,lee94,zirnbauer97,kondev97,marston99}.
Unfortunately, attempts to find an integrable deformation of this spin
chain have failed. A further approach to QH criticality is based
on the model of Dirac fermions; it will be reviewed in Sec.~\ref{s6.7.2}.

\subsubsection{Quest for conformal field theory}
\label{s6.3.3}

Another line of activity is the search for a conformal field theory of
the QH transition. The guiding principle is related to the fact that a
relative 
of the Pruisken's model, the O(3) $\sigma$-model with $\theta=\pi$
topological term, describing a 1D quantum antiferromagnet with
half-integer spin, flows under renormalization to a SU(2) WZW model.
This means that the target space -- which is the 2-sphere O(3)/O(2) =
SU(2)/U(1)=S$^2$ for the O(3) $\sigma$-model -- is promoted to the
group SU(2) (isomorphic to the 3-sphere S$^3$) at criticality.  
The idea is thus to identify the corresponding critical theory for
the QH problem, with a hope that it is of the WZW type and is
solvable by the methods of the conformal field theory. 
Such a proposal was made in \textcite{zirnbauer99}, along with a detailed
analysis of constraints on the sought fixed-point theory. The
target space of the theory conjectured by Zirnbauer 
is a real form 
of the complex supergroup PSL($2|2$). Its base ${\cal M}_F \times
{\cal M}_B$ is a product of the 3-sphere    ${\cal M}_F = \text{SU}(2)
=\text{S}^3$ 
and the 3-hyperboloid $\text{SL}(2,\mathbb{C})/\text{SU}(2) =
\text{H}^3$.  A model of the 
same type was also proposed in \cite{bhaseen00} and most recently
in \cite{tsvelik07}. The proposed theories have the form of the WZW
model, see Sec.~\ref{s6.1.5},
\begin{equation}
  \label{iqhe2}
        S[g] = \frac{1}{8\pi t} \int \! d^2x {\rm Str} \partial_\mu
        g^{-1} \partial_\mu g + iS_{\rm WZ}[g],
\end{equation}
where $iS_{\rm WZ}$ is the WZ term (\ref{e6.2}). The peculiarity of
the WZW models on the considered manifold is that they are critical at
any value of the coupling constant $t$ and level $k\in
\mathbb{N}$.   
While \cite{zirnbauer99} argues for $k=1$, \cite{bhaseen00} considers
the model with Kac-Moody symmetry, $k=1/t$. The later
condition restricts $1/t$ to be integer but facilitates the
analysis of the model. Very recently, it was proposed \cite{tsvelik07}
that the later model with $k=8$ may be the required fixed-point
theory.  

Both variants of the theory make a prediction for the statistics
of critical eigenfunctions. Specifically, it is found that the
multifractality spectrum is exactly parabolic, 
\bea
\label{iqhe2a}
&&\hspace*{-1.3cm} \Delta_q = \gamma q (1-q)\ ; \\
&&\hspace*{-1.3cm} f(\alpha) = 2 - (\alpha - \alpha_0)^2 / 4(\alpha_0 -2)\:;
\ \ \  \alpha_0 = 2+\gamma\: ,
\label{iqhe2b}
\eea
with 
$\alpha_0-2 = 4t$ in the case of \cite{zirnbauer99} and
$\alpha_0-2 = 2t$ for \cite{bhaseen00}. This prediction of
parabolicity of $\Delta_q$ and $f(\alpha)$ indeed agrees with numerical
simulations, Sec.~\ref{s6.3.7}, supporting this type of models. However,
many questions related to the above conjectures remain open. In
particular, if there is a whole line of fixed points (parametrized by
$t$), then is there universality at the QH transition? If yes,
how is it established? From the numerical point of view, there is no
indication of non-universality at present.

\subsubsection{Chalker-Coddington network}
\label{s6.3.4}

The Chalker-Coddington network (CCN) model  was introduced in
\textcite{chalker88} as an effective description of the IQHE in a smooth
random potential. In brief, the model is motivated in the following
way. One considers electrons in a Landau level broadened by a
potential with large correlation length. The electrons then drift
along equipotential lines and tunnel between the lines near saddle
points of the random potential. When the energy is sufficiently close to
the band center (classical percolation threshold), the tunneling
probability becomes $\sim 1$, and a random network with directed links
is formed. At each node of the network two incoming and two ongoing
links meet. 
In the CCN model, this geometrically random structure 
is replaced by a regular square
network, as shown in Fig.~\ref{iqhe-f2}; the disorder is accounted for
by random phases associated with all links. A state $\Psi$ of the network is
defined by its amplitudes on the edges of the network. Originally
\cite{chalker88} the network was characterized by a transfer
matrix, as appropriate for finite-size scaling analysis of the
localization length. Later \cite{klesse95} an equivalent description in
terms of a scattering matrix was introduced. Each
realization of the network is determined by a unitary operator 
${\cal U}={\cal U}_N{\cal U}_E$ acting on states $\Psi$ and modelling
the evolution of the state in a time step. Here ${\cal U}_N$ is an
operator describing the unitary scattering at nodes with amplitudes
$\pm\cos\theta$, $\pm\sin\theta$, as shown in Fig.~\ref{iqhe-f2}. The
second factor, ${\cal U}_E$, is a diagonal operator with
random elements $e^{i\phi_e}$ on all edges $e$ of the
network. In the simplest formulation of the model, the angle $\theta$
is the same for all nodes, and the phases $\phi_e$ are independent
random variables distributed uniformly over
$[0;\:2\pi]$. Changing the parameter $\theta$ allows one to drive the
system through the IQH transition, with the critical point at
$\cos^2\theta=1/2$.

\begin{figure}[btp]
    \begin{center}\leavevmode
 \includegraphics[width=0.7\linewidth]{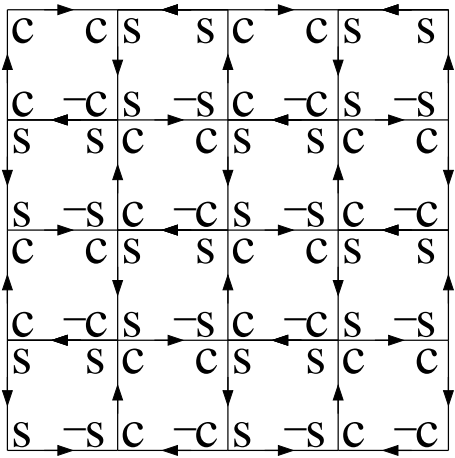}
    \end{center}
      \caption{Chalker-Coddington network model of the IQH transition
        \cite{chalker88}. The symbols $\pm c, \ \pm s$ denote the
        components $\pm \cos\theta$, $\pm \sin\theta$ of scattering
        matrices at the nodes of the network.}
      \label{iqhe-f2}
\end{figure}

The CCN model has been extensively used for numerical simulations of the
IQH transition point; it turned out to be particularly well suited for
the analysis of statistical properties of energy levels and wave
functions. Key results of computer simulations are reviewed below.

The model has been generalized to other symmetry classes. This is most
naturally done for the counterparts of the IQHE in the superconducting
classes C and D, namely, SQHE and TQHE. In these classes the symmetric
spaces of the Hamiltonian (Table~\ref{t4.1}) are the groups Sp(N) and O(N),
and the required modification amounts to a replacement of the
factors $e^{i\phi_e}\in U(1)$ by the elements of Sp(2)=SU(2) for the
SQHE (in this case the amplitudes are spin doublets) and of O(1) for
the TQHE, see Sec.~\ref{s6.4}, \ref{ss:tqhe} 
for more detail. For several other symmetry
classes non-directed generalizations of the CCN
have been constructed and used to study the corresponding critical
behavior; specifically, this has been done for the chiral classes
(AIII, BDI, CII) \cite{bocquet02}, Sec.~\ref{s6.6},  and for the symplectic
class AII \cite{merkt98,obuse07a} considered in Sec.~\ref{s6.2}. 

A further important aspect of the CCN model and its generalizations 
is that they can serve as a starting point for analytical work. We
have already mentioned established equivalences between the CCN and
other IQH models (Pruisken model and superspin chain) in
Sec.~\ref{s6.3.2}. Further, a connection with the models of disordered
Dirac fermions has been established \cite{ho96}. The network model of
the SQHE has led to a number of exact analytical results, Sec.~\ref{s6.4}.

\subsubsection{Localization length exponent}
\label{s6.3.5}

Several microscopic models have been used to study numerically the
critical properties at the IQH transition. This includes tight-binding
models, Landau-space models where the problem is projected on one or
several Landau levels, and the CCN models, Sec.~\ref{s6.3.4}. For a
more detailed review of the models the reader is referred to
\textcite{huckestein95}. The first high-precision determination of the
localization length exponent $\nu$ has been achieved in the framework
of the Landau-space model
\cite{huckestein90,huckestein92,huckestein95}, with the result $\nu =
2.35 \pm 0.03$. Results of later simulations on different models are all in
agreement with this value, thus favoring the universality of the IQH
critical behavior. At present, the precise value of the leading
irrelevant scaling index, $y$, is known with much less accuracy. 
While several authors find values close to $y{=}0.4$,
e.~g. \cite{huckestein95,evers98a}, in some cases values up to 
$y{=}0.6$ have been reported \cite{kramer05}.


 \subsubsection{Critical conductivity and conductance distribution}
\label{s6.3.6}

In a number of works, the critical conductivity was found numerically
in the range 0.5 -- 0.6. More specifically, the results are:
from 
$0.50 \pm 0.03$ to $0.55 \pm 0.05$ for different types of disorder 
\cite{huo93}; 
$0.50 \pm 0.02$ \cite{gammel94}; 
$0.58 \pm 0.03$ \cite{schweitzer05}.
Results in higher Landay levels are consistent with these values to
the extent that the critical regime (of system sizes) could be reached
\cite{gammel98}.

Several authors also studied the conductance distribution ${\cal P}(g)$
of a square
sample with periodic boundary conditions in the
transverse direction; see Sec.~\ref{s6.2.3} for a qualitative
discussion of  ${\cal P}(g)$ at criticality.
As expected, a scale invariant distribution was found, see
Fig. \ref{f11b}, with
the average $\langle g\rangle = 0.58 \pm 0.03$ and the variance
${\rm var}(g) = 0.081 \pm 0.005$ \cite{wang96}; similar results were
obtained in \textcite{cho97a}, in \textcite{ohtsuki04,kramer05} where the
average found to be $\langle g \rangle = 0.57 \pm 0.02$, 
as well as in \textcite{schweitzer05}; 
the latter work yields $\langle g \rangle = 0.60 \pm 0.02$.

\begin{figure}[btp]
    \begin{center}\leavevmode
      \includegraphics[width=0.65\linewidth, angle=270]{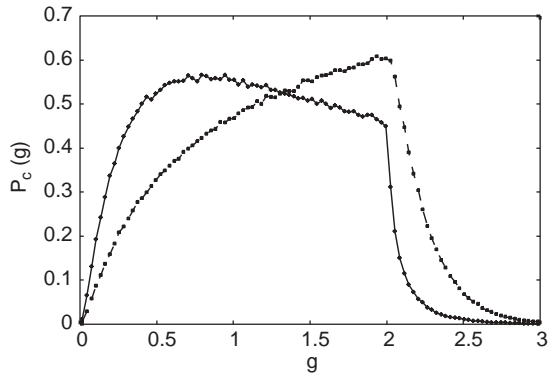}
    \end{center}
      \caption{Distribution function of the conductance at
        the IQH (solid) and the 2d symplectic (dashed) critical point
        using periodic boundary conditions and a square sample geometry
        \cite{ohtsuki04}}. 
      \label{f11b}
\end{figure}

 \subsubsection{Wave function multifractality}
\label{s6.3.7}

A high-precision evaluation of the multifractal spectrum at
the IQH transition was carried out in \textcite{evers01} for  the CCN
model of a size $L\times L$ with $L$ ranging from 16 to 1280.
Figure \ref{f12} shows results for the $f(\alpha)$ spectrum. It is
seen that after extrapolation to the thermodynamic limit, $L\to
\infty$, the $f(\alpha)$ spectrum is well described by the
parabolic form
(\ref{iqhe2b}) with $\alpha_0-2 = 0.262 \pm 0.003$.  
One observes that deviations
from parabolicity -- if they exist -- are too small to 
be resolved in this plot. In Fig.~\ref{f13} reduced anomalous
dimensions $\Delta_q/q(1-q)$ are plotted. This quantity is constant
(equal to $\alpha_0-2$) for an exactly parabolic spectrum,
Eq.~(\ref{iqhe2a}).  
The observed deviations from the constant are very small ($\sim 1\%$)
and within the error bars. Since there is no small parameter in the
problem, it seems that an accidental closeness of the spectrum to
a parabola with such a high accuracy is very improbable. So, the
numerical results are in favor of exact parabolicity of the
multifractal spectrum, thus supporting the possibility of a conformal 
theory of the IQH critical point of the type discussed in
Sec.~\ref{s6.3.3}.  

It is worth mentioning that earlier studies of the IQH multifractality
[the references can be found in \cite{evers01}] gave considerably
different results, showing, in particular, strong deviations from
parabolicity. As was shown in  \textcite{evers01}, earlier numerics
suffered strongly from the absence of ensemble averaging (its
role was explained in Sec.~\ref{s2.3.5}) and from finite-size
effects. Importance of a careful analysis of the latter
is illustrated in  Fig.~\ref{f12}, where also data for finite sizes of
a system are included.

\begin{figure}[btp]
    \begin{center}\leavevmode
      \includegraphics[width=0.8\linewidth]{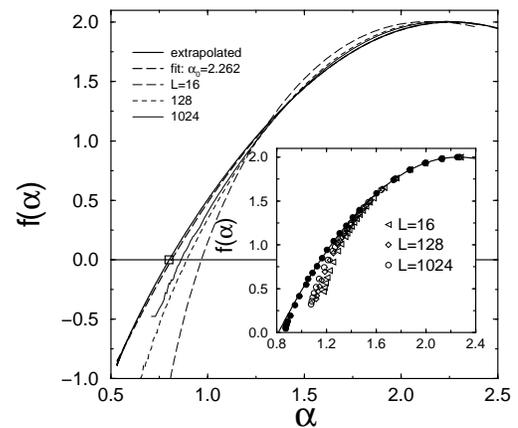}
    \end{center}
      \caption{Multifractal spectrum at the IQH
      transition. Fat solid line: numerical results for $f(\alpha)$
      obtained from scaling of average IPR after extrapolation to
      $L\to\infty$. Datas for several finite values of $L=16, 128,
      1024$ are also shown.  
      Dashed line: parabolic approximation, 
      Eq.~(\ref{iqhe2a}). Inset: data points from typical IPR
      from systems with sizes $L=16,128,1024$ (open symbols) 
      and extrapolation to
      infinite system size (full circles); data from average
      IPR are shown by solid line. \cite{evers01}}
      \label{f12}
\end{figure}

\begin{figure}[btp]
    \begin{center}\leavevmode
      \includegraphics[width=0.8\linewidth]{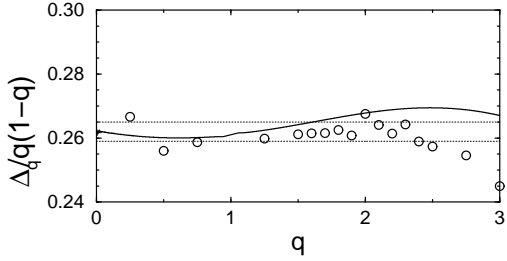}
    \end{center}
      \caption{Anomalous multifractal dimensions $\Delta_q$ [divided
        by $q(1-q)$] at the 
        IQHE critical point, as obtained from extrapolation of the
        average inverse participation ratio (IPR).
        In the case of exact parabolicity, the plotted quantity should
        be constant. 
        The dotted lines indicate the error bars obtained for
        $\alpha_0-2$. The circles give the $\Delta^{\rm typ}_q/q(1-q)$ 
        as obtained from the typical IPR. As explained in
        Sec.~\ref{s2.3.5}, $\Delta^{\rm typ}_q=\Delta_q$ for $q<q_+$;
        for IQHE with parabolic spectrum (\ref{iqhe2a}),
        (\ref{iqhe2b}) we find $q_+ = [2/(\alpha_0-2)]^{1/2} \simeq
        2.76$. Thus, the last two data points for 
        $\Delta^{\rm typ}_q/q(1-q)$ are in the range $q>q_+$,
        which explains their downward deviations from  
        $\Delta_q/q(1-q)$. 
         \cite{evers01}} 
      \label{f13}
\end{figure}

A high-precision evaluation of multifractality allows to test the
conformal invariance of the problem via Eq.~(\ref{e6.1.3}). The parameter
$\Lambda_c\equiv \xi_M/M$  was found to be $\Lambda_c = 1.22 \pm 0.01$
\cite{evers98a,evers01}, implying, in combination with the above value
of $\alpha_0-2$, that (\ref{e6.1.3}) is perfectly fulfilled and thus
confirming the expectation that the IQH critical theory is conformally
invariant.

 \subsubsection{Statistics of the two-point conductance}
\label{s6.3.8}

Closely related to wavefunction multifractality is the statistics of
two-point conductances, Sec.~\ref{s2.3.8}. 
In \textcite{klesse00} a relation between statistical properties of the
wave functions and two-point conductances was derived, 
\begin{eqnarray}
\label{iqhe3}
2\pi \rho \langle y_m f(y_m/y_l) \rangle = 
\langle F(T_{lm} \rangle\:; \nonumber
\\
F(T) = \int_{0}^{2\pi} \frac{d\phi}{2\pi}
f\left(T^{-1}|1-e^{i\phi}\sqrt{1-T}|^2\right)\:. 
\end{eqnarray}
Here $f(x)$ is an arbitrary function,  
$y_l=|\Psi_l|^2$,  $y_m=|\Psi_m|^2$ are wave function
intensities at two links $l$ and $m$ for an eigenstate $\Psi$ of a
closed network, and $T_{l,m}$ is the two-point
conductance defined for an open network with the edges $l$ and $m$ cut
and attached to two terminals. This result was used in \textcite{evers01}
to derive the the relation (\ref{e2.52}) between the corresponding
critical exponents. The parabolic spectrum for wave function
multifractality, Eq.~(\ref{iqhe2a}), found for
IQHE translates thus into parabolic spectrum of exponents $X_q$ for
the two-point conductance,
\be
\label{iqhe4}
X_{q\le 1/2}  = X_t q (1-q)\ ; \qquad X_{q\ge 1/2} = X_t/4\ ,
\ee
with
$X_t = 2(\alpha_0-2) = 0.524\pm 0.006$. 
In an earlier work \cite{janssen99} an explicit expression for the
distribution of the two-point conductance on the CCN was derived under the
assumption of the parabolic law (\ref{iqhe4}) for $X_q$. In
\textcite{klesse00} this result was used to test the conformal invariance
of the theory, utilizing a numerical analysis of the moments of the two-point
conductance in the quasi-1D (cylinder) geometry.  
The result of \textcite{janssen99} for the distribution of the conductance
$T$ between the points $(0,0)$ and $(x,y)$ in this geometry reads
\bea
\label{iqhe5}
&& \vspace*{-1.5cm}{\cal P}(T) = 
{2\pi^{-1/2}\zeta^{-X_t/4} \over T^2(X_t\ln\zeta)^{3/2}}
\int_{{\rm arcosh} {1\over\sqrt{T}}}^\infty {e^{-t^2/X_t\ln\zeta}\:t\: dt\over
\sqrt{\cosh^2 t - T^{-1}}}\:,\nonumber \\
&& \vspace*{-1.5cm}
\zeta = (W / \pi a) |\sinh [\pi(x+iy) / W]|\:,
\eea
where $a$ is the non-universal microscopic scale that sets the length
unit. The numerically determined moments of $T$ were in perfect
agreement with this formula, thus supporting the conformal invariance. 
The best fits yielded the values $X_t = 0.54\pm 0.01$ from the
analysis of the 
moments $\langle T^{n+1/2}\rangle $ and  $X_t = 0.57 \pm 0.05$ from
the analysis of the typical conductance, $\langle \ln T \rangle$. 
These values are consistent with the above result $X_t=0.524\pm 0.006$. 

 \subsubsection{Classical percolation vs. quantum Hall effect }
\label{s6.3.9}

If the disorder correlation length is large, there is an intermediate,
parametrically broad, range of energies where the physics is
dominated by classical percolation. On corresponding length
scales, tunneling between the percolating contours is exponentially
small. This range of energies was studied in \textcite{milnikov88}
where the scaling of the localization length with an exponent 
$\tilde{\nu}=\nu_{\rm perc}+1=7/3$ was found ($\nu_{\rm perc}=4/3$ is the
correlation length index of the 2D percolation problem).
In this regime quantum interference effects play no role.
When energy approaches still closer the critical point, the
probability of tunneling between the contours ceases to be small
and the quantum interference becomes crucially important -- the system
enters the true critical regime of the quantum Hall transition. It is
this latter regime that is described by the CCN model. While the
``quasiclassical QHE'' exponent $\tilde{\nu}=7/3$ is remarkably close to the
numerical value of the true QH exponent $\nu=2.35\pm 0.03$, this
coincidence is apparently fully accidental, as the physics of the 
true QH critical regime and the intermediate quasiclassical regime of
\textcite{milnikov88} is completely different.

It is worth mentioning that the physics of the intermediate,
quasiclassical regime,  where the physics is dominated by the vicinity
of the percolation fixed point
\cite{milnikov88,kratzer94,evers94,klesse95,gammel96,evers98a}
is interesting in its own right. 
In particular, quasiclassical time evolution generates a long time
tail in the velocity correlation function 
$\langle v_x(t)v_x(0)\rangle\sim t^{-2}$, which leaves $\sigma_{\rm
  xx}(\omega)$
with a nonanalytic $\omega$-dependence inside an intermediate
(quasiclassical) frequency window.
The corresponding 
results may be relevant to
experiments if the latter are performed on structures with smooth
disorder, in which case the true QH criticality may in fact be totally 
unobservable for realistic temperatures.

 \subsubsection{Experiment vs. theory. Interaction effects}
\label{s6.3.10}

\paragraph{Experimental results.}
The IQH transition and the associated critical properties have been
studied in numerous experiments. All the basic features --- the
existence of phase transitions with critical values $\sigma_{xy} =
n+1/2$ and $\sigma_{xx}$ of the order of unity, as well as the power-law
scaling behavior --- are in agreement with the theoretical
expectations. The situation with the values of critical exponents is
not so simple, as we are going to discuss. To do this, we will have to
touch the question of interaction effects, which is left out in the
rest of this review, except for Sec.~\ref{s7.1}.

Experiments yield the following results for the critical
exponents. First, the index $\nu$ of the localization length is found
to be $\nu=2.3\pm 0.1$ \cite{koch91}; this value was confirmed more
recently in \textcite{hohls01,hohls02}. Second, the width of the critical
region (peak in $\sigma_{xx}$ and plateau transition in $\sigma_{xy}$)
scales with the temperature $T$ as $\Delta B \propto T^\kappa$ with
$\kappa = 0.42\pm 0.04$ \cite{wei88}. While different values of
$\kappa$ were obtained in other works, it was emphasized in
\textcite{van-schaijk00} that this results from macroscopic
inhomogeneities that complicate observation of the true IQH critical
behavior with $\kappa\simeq 0.42$. 
More recent work of the same group \cite{pruisken06a,deVisser06} favors 
again $\kappa{=}0.56\pm0.02$, however. On the other hand, the impact
of density inhomogeneities was reconsidered by \textcite{li05}. It 
was found there, 
that for short-range disorder,
when the true IQH criticality can be achieved, 
$\kappa = 0.42\pm 0.01$,  
whereas the larger value $\kappa{=}0.58$ was ascribed to impurity
clustering. 
Finally, the frequency scaling of the transition width was
found to be $\Delta B \sim \omega^\zeta$, with $\zeta=0.41\pm 0.04$
\cite{engel93}. A more recent work \cite{hohls02a} yields a result
consistent with this value, but with somewhat larger uncertainty,
$\zeta = 0.5\pm 0.1$. To summarize, the experiments yield $\nu$ that
agrees with its numerical value ($2.35\pm 0.03$), as well as the
dynamical exponents $z_T
\equiv 1/\kappa\nu \simeq 1$ and $z \equiv 1/\zeta\nu \simeq 1$.  
The remarkable agreement in the value of $\nu$ is in fact surprising,
in view of the electron-electron interaction. 

\paragraph{Finite-range interaction.}
We consider first the case of a finite-range interaction 
$v({\bf r}-{\bf r'})$, following
\textcite{lee96,wang00}. In this case the interaction is
irrelevant. Indeed consider the Hartree-Fock interaction between the
(close in energy) states $\alpha$ and $\beta$, normalized to the level
spacing,
\bea
\label{iqhe6}
&& \lambda = \rho L^2 \int d^2 r d^2 r'\langle 
[|\psi_\alpha({\bf r})|^2 |\psi_\beta({\bf r'})|^2 \nonumber \\
&&\ \ - \psi_\alpha({\bf r})\psi_\alpha^*({\bf r'})  
  \psi_\beta({\bf r'})\psi_\beta^*({\bf r'})] v({\bf r}-{\bf r'})\rangle.     
\eea
 In the $\sigma$-model language, see Sec.~\ref{s2.2.1}, the scaling of
 (\ref{iqhe6}) with the system size $L$ is governed by the scaling
 dimension  of the operator 
\be
\label{iqhe7}
Q_{11}^{\rm bb}({\bf r})Q_{22}^{\rm bb}({\bf r}) - 
Q_{12}^{\rm bb}({\bf r})Q_{21}^{\rm bb}({\bf r})\ .
\ee
While each of the two terms in (\ref{iqhe7}) is relevant in the RG
sense, having a dominant negative scaling dimension $\Delta_2$ (which governs
the IPR scaling, Sec.~\ref{s2.3.1},\ref{s6.3.7}), the difference
(\ref{iqhe7}) is RG-irrelevant \cite{wegner80}, with a scaling
dimension $x_2>0$. The numerical value of $x_2$ at the IQH critical
point was estimated to be $x_2 = 0.66\pm 0.04$ \cite{lee96}. With
increasing $L$ the interaction (\ref{iqhe6}) scales as $\lambda\propto
L^{-x_2}$, so that the fixed point is unaffected by it. This
implies that the critical index $\nu$ of the localization length and
the multifractality spectrum $\Delta_q$ remain the same as in
the non-interacting problem. This is also true for the dynamical
exponent $z$ governing the destruction of localization by finite
frequency: the corresponding scaling variable is $\omega\xi^z$ with
$z=2$. (In general, for a non-interacting transition with finite DOS in
$d$ dimensions the scaling variable is $\omega/q^d$
\cite{wegner76}). The interaction 
cannot be fully discarded, however (it is said to be {\it dangerously
irrelevant}),  as the conductivity at finite $T$ would be zero without
the interaction-induced dephasing. The dephasing rate
scales as $\tau_{\phi}^{-1} \propto T^p$ with $p = 1+2x_2/z$
\cite{wang00}. Thus, the dephasing length is $L_\phi \propto
\tau_\phi^{1/2}   \propto T^{-1/z_T}$ with $z_T = 2/p = 2z/(2x_2+z)$,
yielding $z_T \simeq 1.2$ with the above estimate for $x_2$. Therefore, for
a system with a metallic gate (which screens the interaction) one expects
the exponents $\nu \simeq 2.35$, $z=2$, and $z_T\simeq 1.2$. 

\paragraph{Coulomb interaction.}
The situation with $1/r$ Coulomb interaction (as in typical
experiments) is much less clear. In this case the interaction is
RG-relevant and drives the system to a novel fixed point
\cite{lee96,baranov02}. While the conductivity and the screened
compressibility  $\partial n/\partial\mu$ remain finite at the
transition \cite{finkelstein90,belitz94}, much less is known
theoretically about other critical properties.  
Several authors \cite{polyakov98,wang02}
argued that charging effects analogous to those responsible for the
linear Coulomb gap in the tunneling DOS of the insulator will lead to
$z=z_T=1$ (which is the natural scaling dimension of the $1/r$
interaction).  The status of this argumentation is unclear, however, 
for the following reasons:

(i) These arguments essentially identify 
$z$ and $z_T$ with a dynamical exponent ($z_3$ in notations of
\textcite{belitz94}) governing the scaling of the density response
function. It is known, however, that in metals the plasmon pole
governed by this exponent, $z_3=1$, determines the interaction-induced
quantum correction to
tunneling DOS but not to conductivity \cite{altshuler84}, in view of
gauge invariance \cite{finkelstein94}. The conductivity correction is
governed by the conventional diffusion pole in the {\it irreducible}
density-density response function, which corresponds to the dynamical
exponent $z_3^{\rm irr}=2$.

(ii) Each Goldstone mode (or, equivalently, conserved quantity) is in
general characterized by a dynamical exponent. For the QH transition
this implies that, in addition to the exponents related to the
particle number conservation -- $z_3=1$ and its irreducible
counterpart, $z_3^{\rm irr} = 2$, -- there is another exponent
associated with the energy conservation. This latter exponent, 
controlling the renormalization of the frequency term in the 
$\sigma$-model, is denoted by $\zeta$ in
\textcite{finkelstein90}, $z_1$ in \textcite{belitz94}, and by
$2+\gamma^*$ in \textcite{baranov02}. It is believed to govern the frequency 
scaling of the conductivity at the critical point of the Anderson
transition in a system with Coulomb interaction and broken
spin-rotation invariance in $2+\epsilon$ dimensions 
\cite{finkelstein90,belitz94}. One may thus expect that this dynamical
exponent plays a central role at the quantum Hall 
transition as well \cite{baranov02,burmistrov06}.    
 
The
problem of the index $\nu$ of the localization length is also far from
being solved. While it was found that $\nu$ is equal to its
non-interacting value within the Hartree-Fock theory \cite{yang95}, it
is not clear whether this should be applicable to the true fixed point
in the problem with Coulomb interaction. To summarize, in our view, the
theoretical problem of the critical behavior in the presence of
Coulomb interaction remains open. In particular, it remains to be seen
whether  the remarkable agreement of $\nu$ with its non-interacting
value as well as $z=z_T=1$ --- as suggested by experiments  --- are
indeed exact properties of the interacting problem.

\subsection{Spin Quantum Hall Effect  (Class C)}
\label{s6.4}

\subsubsection{Physical realization}
\label{s6.4.1}

The SQHE is a counterpart of the IQHE in superconductors with
broken time-reversal but preserved spin-rotation invariance
\cite{kagalovsky99,gruzberg99,senthil99}. The class-C Hamiltonian
satisfies the  symmetry (\ref{e4.12}) 
(with $\tau_y$ the Pauli matrix in
the particle-hole space) and has the block structure (\ref{e4.11}).
Several possible physical realization of SQHE systems have been
proposed: (i) a $d$-wave superconductor with complex $d_{x^2-y^2} +
id_{xy}$ pairing \cite{kagalovsky99,senthil99} that was
conjectured for high-$T_c$ superconductors \cite{balatsky98,laughlin}; 
(ii) granular superconducting film in a magnetic field
\cite{kagalovsky99}; (iii) a state of composite fermions at filling
fraction $\nu=5/2$  with $d$-wave pairing \cite{read00}, as proposed
in \textcite{haldane88}.  

Similar to the case of the IQHE, the key signature of the 
SQHE is the quantization of the appropriate Hall
conductance. Specifically, while the quasiparticle number is not
conserved, the spin is, so that the relevant quantity is the  spin
Hall conductivity 
$\sigma_{xy}^{\rm s}$. It describes the transverse 
spin current induced in response to a gradient of the Zeeman field,  
\be
\label{sqhe1}
    j_{x}^{z} = \sigma_{xy}^{\rm s} 
[- \partial B^z(y) / \partial y ]
\ee
In the IQHE the step $\Delta\sigma_{xy}$ between the quantized values
of the Hall conductivity is $e^2/h$ per spin orientation. 
In the case of the SQHE, the elementary charge $e$ is
replaced by $\hbar/2$, yielding \cite{senthil99}  
\be
\label{sqhe2}
\Delta \sigma_{xy}^{\rm s} = 2n \cdot \hbar/8\pi\ ,
\ee
where the factor 2 accounts for the spin and $n$ for the valley
degeneracy. In particular, for the case of Dirac fermions in a $d+id$
superconductor ($n=2$), one finds two SQH phases with
quantized values
\be
\label{sqhe3}
\sigma_{xy}^{\rm s} = \pm \hbar/4\pi\ .
\ee
In the presence of disorder, these two phases become Anderson
insulators separated by the SQH transition. The field theory of this
problem is analogous to Pruisken's theory of the
IQHE (Sec.~\ref{s6.3.1}) 
--- it is a $\sigma$-model of the class C with the topological
term (\ref{e6.1}}), possessing a critical point at $\theta=(2n+1)\pi$ 
\cite{senthil98,senthil99,read00,altland02}.
The corresponding flow diagram is expected to have qualitatively the
same form as for the IQHE, Fig.~\ref{iqhe-f1}. 
The critical behavior at the SQH
transition is analyzed below. In the presence of a Zeeman
term, the spin-rotation symmetry is broken, and the system crosses
over to the symmetry class A of the IQHE transition
\cite{senthil98,kagalovsky99,senthil99,cui04}.   

\subsubsection{Mapping to percolation}
\label{s6.4.2}

The network model for the SQHE \cite{kagalovsky99} is the SU(2) version of the
Chalker-Coddington IQHE network, see Sec.~\ref{s6.3.4}  and
Fig.~\ref{iqhe-f2}.  The directed
links of the network carry doublets of complex fluxes representing
propagation of spin-1/2 particle. The
scattering at each node is spin-independent and defined in the same
way as for the IQHE.  Each
realization of the network is characterized by a set of random
$2\times 2$ spin matrices $U_e$ associated with all edges $e$ of the
network. In view of (\ref{e4.12}), the evolution operator of the
network ${\cal U}$ satisfies the symmetry
${\cal U}=\tau_y{\cal U}^*\tau_y$, implying that $U_e\in {\rm
SU(2)}$. Starting with this network model, it turns out to be possible
to establish a remarkable property of the SQH transition: some physical
observables and critical indices can be calculated exactly via mapping
to classical percolation. This was shown for the
DOS and the conductance  in \textcite{gruzberg99} via the supersymmetry; an
alternative derivation of these results was presented in
\textcite{beamond02}. In \textcite{mirlin02} the mapping was extended  to all
two- and three-point correlation functions describing, in particular,
the wave function statistics. It was also shown there
that the mapping breaks down for generic $n$-point correlation functions with
$n > 3$. 

We briefly sketch the idea of the approach of
\textcite{beamond02,mirlin02}. The
primary objects are Green functions on the network,
\be
\label{sqhe4}
G(e',e;z)=\langle e'|(1-z{\cal U})^{-1}|e\rangle\ ;
\ee
for $z=\exp i(\epsilon\pm i\eta)$ 
they have a meaning of retarded ($G_R$, $|z|<1$)
and advanced ($G_A$, $|z|>1$) Green functions at energy $\epsilon$.  
The Green function is straightforwardly
represented in the form of a sum over paths
\be
\label{sqhe5}
G(e,e';z)=\sum_{{\rm paths}\ e'\to e} \ldots \cdot zU_{e_j}s_j \cdot 
zU_{e_{j+1}}s_{j+1}\cdot \ldots,
\ee
where $s_j$ is the corresponding matrix element ($\cos\theta$,
$\sin\theta$, or $-\sin\theta$) of the $S$-matrix between the edges $e_j$
and $e_{j+1}$. Equation (\ref{sqhe5}) generates a convergent expansion
in powers of $z$ when $|z|<1$; otherwise the identity 
\be
\label{sqhe6}
G^\dagger(e,e';z)={\bf 1}\cdot\delta_{ee'}-G(e',e;(z^*)^{-1})
\ee
is to be used. 
As shown below, each of the sums over paths obtained by
substituting (\ref{sqhe5}), (\ref{sqhe6}) in products of $n\le 3$
Green functions can be reduced after disorder averaging to a sum over
classical paths (hulls) in  the percolation problem. This remarkable
reduction crucially relies on the following two statements:

 1. Only paths visiting each edge of the network either 0 or 2 times are
 to be taken into account; contributions of all the remaining paths sum up
 to zero.
 
\begin{figure}
\centerline{
\includegraphics[width=0.9\columnwidth,clip]{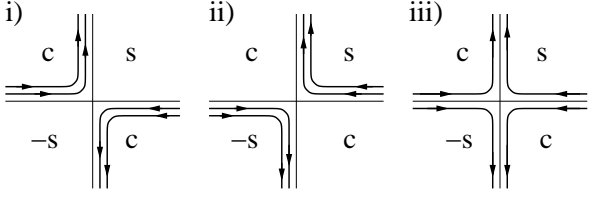}
}
\caption{Possible configurations of paths passing four times through a
network node. The symbols $c$ and $\pm s$ denote the elements
$\cos\theta$, $\pm\sin\theta$ of the $S$-matrix at the node. The
statement 2 in Sec.~\ref{s6.4.2} allows one to get rid of the quantum
interference contribution (iii) and to associate with the
contributions (i) and (ii) the  weights $\cos^2\theta$ and
$\sin^2\theta$, respectively, yielding a mapping to 
the classical percolation.} 
\label{four-paths}
\end{figure}

 2. Using the statement 1, it is easy to see that each node may be
 visited 0, 2, or 4 times. The second statement concerns the nodes visited
 four times. As illustrated in Fig.~\ref{four-paths}, 
there are three possibilities 
 how this may happen; the corresponding contributions have weights (i)
 $\cos^4\theta$, (ii) $\sin^4\theta$, and (iii) $-\sin^2\theta\cos^2\theta$
 from the scattering matrix at this node. The statement is that one can
 equivalently take into account only the contributions (i) and (ii) with
 the weights $\cos^2\theta$ and $\sin^2\theta$, respectively.

After application of the statement 2 to all nodes, the network
is reduced to a weighted sum over all its possible decompositions
in a set of closed loops (such that each edge belongs to exactly one
loop). These loops can be viewed as hulls of the bond percolation
problem. As a result, the correlation functions (averaged products of
Green functions) are expressed in terms of classical sums over the
percolation hulls. The results obtained in this way are listed below.

\subsubsection{Density of states and localization length}
\label{s6.4.3}

The result for the average of a single Green function reads
($|z| < 1$) \cite{beamond02}:
\bea
\label{sqhe7}
&& \langle {\rm Tr} G(e,e,z) \rangle = 2 - \sum_{N>0} P(e;N)\ z^{2N}
\ , \\
&&  \langle {\rm Tr} G(e,e,z^{-1}) \rangle = \sum_{N>0} P(e;N)\
z^{2N}\ ,
\label{sqhe8}
\eea
where $P(e,N)$ is the probability that the edge $e$ belongs to a loop
of the length $N$. (In the bulk of a large system, $L\to\infty$, 
or for periodic boundary conditions, this probability does
not depend on $e$, $P(e,N) = P(N)$.) This yields the DOS
\be
\label{sqhe8a}
\rho(E) = (1/2\pi) [ 1 - \sum_{N>0} P(N)
\cos(2N E)]\ . 
\ee 
In the insulating phases ($t \equiv \cos^2\theta \ne 1/2$) this yields 
\be
\label{sqhe8b}
\rho(E) \simeq \pi^{-1}\langle N^2 \rangle E^2, 
\ee
which is the expected behavior of DOS in the class C. 
On approaching the percolation transition point, $t=1/2$, the characteristic 
diameter of largest loops diverge,
\be
\label{sqhe9}
\xi\sim |t-1/2|^{-\nu}\ ,\qquad \nu=4/3.
\ee 
At the critical point ($t=1/2$)
\be
\label{sqhe10}
P(N)\sim N^{-2/d_h} = N^{-8/7},
\ee
where $d_h{=}7/4$ is the fractal dimension of the percolation hull
\cite{saleur87,isichenko92}, yielding \cite{gruzberg99,beamond02} 
\be
\label{sqhe11}
\rho(E) \sim |E|^{1/7}\ .
\ee 
The characteristic length of loops contributing to (\ref{sqhe11}) is
$N_E \sim E^{-1}$, yielding their characteristic ``diameter'' 
$\xi_E \sim |E|^{-1/d_h} = |E|^{-4/7}$, which is
the localization length of the states with energy $E$. 
The percolation hull exponent 
$d_h$ plays therefore a role of the dynamical exponent for the SQH
transition.

\subsubsection{Conductance} 
\label{s6.4.4}

To define the two-terminal conductance $g$, one opens the system by
cutting two subsets of links and attaching them to
two reservoirs.  The average dimensionless conductance is given by
\be
\label{sqhe12}
\langle g \rangle = 2 \sum_{e\in 1_{\rm out};\ e'\in 2_{\rm in}}
P(e,e')\ ,
\ee
where $P(e,e')$ is the probability that a path incident from the
second reservoir on the link $e'$ escapes to the first contact via
the link $e$, and the sum goes over all such links.
Equation (\ref{sqhe12}) was used in \cite{cardy00} to
calculate the critical conductivity by determining the conductance of
a wide sample ($W\gg L$), $\langle g \rangle L/W = \sqrt{3}/2$. A very
close result was earlier obtained by numerical evaluation of the Kubo
formula, $\sigma_c = 2 \times (0.45\pm 0.01)$ \cite{evers97}. It is worth
stressing that, despite of the vanishing density of states, the critical
conductance is finite. From the point of view of the Einstein
relation,  $\sigma = h \rho D$, this results from a mutual
cancellation of the percolation exponents \cite{cohen91,evers94},    
$\rho(E) \sim N_E /\xi_E ^{2} \sim
E^{2/d_h-1} = E^{1/7}$ and 
$D(E) \sim \xi_E ^2/N_E\sim E^{1-2/d_h}
=  E^{-1/7}$. 

For the average two-point conductance (Sec.~\ref{s2.3.8}),
Eq.~(\ref{sqhe12}) yields \cite{gruzberg99} 
\be
\label{sqhe13}
\langle g(e,e')\rangle = 2 P(e,e') \sim r^{-1/2}\ ,
\ee
where $r\gg 1$ is the distance between $e$ and $e'$.

\subsubsection{Higher correlation functions and multifractality}
\label{s6.4.5}

The results presented in this subsection were obtained in \textcite{mirlin02}.
To determine the fractal dimension $\Delta_2$ governing the scaling of
two-point correlations of wave functions (\ref{e2.25}), (\ref{e2.26}),
one considers the correlation functions  
\bea
\label{sqhe-e6}
{\cal D}(e',e;\gamma)&=&(2\pi)^{-2}
\langle{\rm Tr}[G(e',e;z)-G(e',e;z^{-1})] \nonumber \\
&\times & [G(e,e';z)-G(e,e';z^{-1})]\rangle, \\
\label{sqhe-e7}
\tilde{D}(e',e;\gamma)&=&(2\pi)^{-2}
\langle{\rm Tr}[G(e,e;z)-G(e,e;z^{-1})] \nonumber\\
&\times & {\rm Tr}[G(e',e';z)-G(e',e';z^{-1})]\rangle,
\eea
with a real $z=e^{-\gamma}<1$ and $\gamma \ll 1$ playing a role of the
level broadening. The mapping to percolation yields for the
averaged products of two Green functions entering (\ref{sqhe-e6}), 
\bea
\langle{\rm Tr} G(e',e;z)G(e,e';z)\rangle & = &
\langle{\rm Tr} G(e',e;z^{-1})G(e,e';z^{-1})\rangle \nonumber \\ 
&=& -2\sum _N P(e',e;N)z^{2N}, \label{sqhe-e8} \\
\langle{\rm Tr} G(e',e;z) G(e,e';z^{-1})\rangle & = &
-2\sum_N P_1(e',e;N)z^{2N}, \label{sqhe-e9} 
\eea
where $P(e',e;N)$ and $P_1(e',e;N)$ are probabilities that the edges
$e$ and $e'$ belong to the same loop of the length $N$ (resp. with the
length $N$ of the part corresponding to the motion from $e$ to
$e'$). According to the classical percolation theory, $P$ and $P_1$
scale as
\be
P(e',e,N),\ P_1(e',e,N) \sim N^{-8/7}r^{-1/4}\ , \qquad r\lesssim N^{4/7}
\label{e10}
\ee
and fall off exponentially fast at $r\gg N^{4/7}$, where $r$ is the distance
between $e$ and $e'$. This yields for the correlation functions in
(\ref{sqhe-e8}) and (\ref{sqhe-e9}) (which we abbreviate as 
$\langle G_R G_R\rangle$, $\langle G_A G_A\rangle$, 
$\langle G_R G_A\rangle$) 
\bea
&& \langle G_R G_R\rangle = \langle G_A G_A\rangle \simeq
\langle G_R G_A\rangle \sim r^{-1/2}, \nonumber\\
&& \hspace*{3cm} r\ll \xi_\gamma\equiv 
\gamma^{-4/7}
\label{sqhe-e12}
\eea
in full agreement with the scaling argument of \cite{gruzberg99}. 
However, these leading order terms cancel in (\ref{sqhe-e6}), and  
the result is 
non-zero due to the factors $z^{2N}$ only, implying that relevant $N$
are $N\sim \gamma^{-1}$. As a consequence, $\langle(G_R-G_A)(G_R-G_A)\rangle$
scales differently compared to (\ref{sqhe-e12}),
\bea
&& {\cal D}(e',e;\gamma) =  {1\over\pi^2}\sum_N
[P(r,N)-P_1(r,N)](1-e^{-2N\gamma}) \nonumber\\
&& \qquad \sim  P(r,\gamma^{-1})\gamma^{-1}\sim
(\xi_\gamma r)^{-1/4},\qquad r\lesssim \xi_\gamma.
\label{sqhe-e13}
\eea
The analysis of the correlation function (\ref{sqhe-e7}) yields
similar results. One thus finds the scaling of  two-point wave
function correlations for $r\lesssim \xi_E$,   
\begin{eqnarray}
\label{sqhe14}
\left.\begin{array}{r}
L^{4}\langle\psi^*_{i\alpha}(e)\psi_{j\alpha}(e)
\psi_{i\beta}(e')\psi^*_{j\beta}(e')\rangle  
\\
L^4 \langle |\psi_{i\alpha}(e)|^2 | \psi_{j\beta}|^2 \rangle
      \end{array}
      \right\}
\sim \left(\frac{\xi_{E}}{r}\right)^{1\over 4}.
\end{eqnarray}
with $\alpha,\beta$ labeling the spin indices.
This  implies that the fractal exponent $\Delta_2 \equiv -\eta$ is
\be
\label{sqhe-e39a}
\Delta_2= - 1/4,
\ee
at variance with what one might naively expect
from the $r^{-1/2}$ scaling of the diffusion propagator 
$\langle G_RG_A\rangle$,  Eq.~(\ref{sqhe-e12}). 
An analogous
calculation for three-point correlation functions yields
$\Delta_3=-3/4$. For correlation functions of higher orders 
(determining, in particular, the exponents $\Delta_q$ with $q>3$) the
mapping to percolation breaks down. 

The point $q=3$ deserves special attention. It satisfies the relation
\be
\label{sqhe15}
\Gamma(q) \equiv qx_\rho + \Delta_q = 0\ ,
\ee
where $x_\rho=1/4$ is the scaling dimension of DOS defined by
$\rho(E)\sim \xi_E^{-x_\rho}$. It separates two regimes
with different scaling of correlation functions. For smaller $q$, when
$\Gamma(q)>0$, the correlation functions 
\bea
\label{sqhe-e56}
&& \Pi^{(q)}_{s_1\ldots
s_q}(e_1,\ldots,e_q;E_1,\ldots,E_q)
\nonumber \\
&& =\langle{\rm Tr}G_{s_1}(e_1,e_2;e^{iE_1})
\ldots G_{s_q}(e_q,e_1;e^{iE_q})\rangle,
\eea
where $s_j=R$ or $A$, show the scaling
\be
\label{sqhe-e59}
\Pi^{(q)}_{s_1\ldots
s_q}(e_1,\ldots,e_q;E_1,\ldots,E_q)\sim r^{-qx_\rho}
\ee
and are, to the leading approximation, independent of the indices
$s_i$. However, when one calculates the wavefunction correlations, 
\bea
\label{sqhe-e55}
&& {\cal D}^{(q)}(e_1,\ldots,e_q;E_1,\ldots,E_q)
=(2\pi)^{-q} \nonumber \\
&&\times \langle{\rm Tr}[(G_R-G_A)(e_1,e_2;e^{iE_1}) 
 (G_R-G_A)(e_2,e_3;e^{iE_2}) \nonumber\\ 
&& \times \ldots \times (G_R-G_A)(e_q,e_1;e^{iE_q})]\rangle,
\eea
these leading-order terms cancel, yielding the multifractal behavior,
\be
\label{sqhe-e57}
{\cal D}^{(q)}(e_1,\ldots,e_q;E_1,\ldots,E_q)\sim
(r/\xi_E)^{\Delta_q} \xi_E^{-qx_\rho}, \qquad r\lesssim
\xi_E.  
\ee
On the other hand, for $q>3$, when $\Gamma(q)$ is negative, the
correlation functions 
 $\Pi^{(q)}_{s_1,\ldots,s_q}$ start to depend in
a singular way on the infrared cutoff ($\xi_E$) and 
scale in the same way as ${\cal D}^{(q)}$,
Eq.~(\ref{sqhe-e57}) (with a numerical prefactor depending on indices
$s_i$), similarly to the conventional Anderson localization
transition.

It is instructive to analyze this situation within the
field-theoretical approach to the wave-function multifractality
\cite{wegner80,wegner85,duplantier91,mudry96,bhaseen00,bernard02}.
In the renormalization-group language, $\Gamma(q)$ defined by
Eq.~(\ref{sqhe15}) are scaling dimensions of operators of the type 
${\cal O}^{(q)}\sim \psi_{s_1}\psi^\dagger_{s'_1}\ldots
\psi_{s_q}\psi^\dagger_{s'_q} $, where $\psi,\psi^\dagger$ are
electronic fields. Averaged products of Green functions are expressed
as correlation functions of the corresponding operators ${\cal
O}^{(q)}$; in particular, (\ref{sqhe-e56}) takes the form
\be
\label{sqhe-e60}
\Pi^{(q)}_{s_1\ldots s_q} \sim \langle {\rm Tr} 
{\cal O}^{(1)}_{s_1s_2}(e_2) {\cal O}^{(1)}_{s_2s_3}(e_3) \ldots     
{\cal O}^{(1)}_{s_qs_1}(e_1)  \rangle.
\ee
To calculate the scaling behavior of such correlation functions, one
applies the operator product expansion (OPE)
\cite{wegner85,duplantier91,mudry96}. Generically, the identity
operator will be among those generated by  the OPE. Moreover, under
the condition $\Gamma(q)>0$ it will be the most
relevant operator and will dominate the expansion, leading to the gap
scaling $\Pi^{(q)}\sim r^{-q\Gamma(1)}$, in agreement with
(\ref{sqhe-e59}). On the other hand, if $\Gamma(q)<0$, the operator 
${\cal O}^{(q)}$ will give a dominant contribution to OPE, leading to a
multifractal type of scaling, $\Pi^{(q)}\propto 
r^{-q\Gamma(1)}(r/\xi_E)^{\Gamma(q)}$, as in
Eq.~(\ref{sqhe-e57}). What is, however, non-trivial from this point of
view, is that the scaling of the wave function correlator (\ref{sqhe-e55})
has the multifractal form (\ref{sqhe-e57}) independently of the sign of
$\Gamma(q)$. This means that in the regime $\Gamma(q)>0$ the leading
(gap scaling) terms (\ref{sqhe-e59}) cancel in the particular combination
of the functions $\Pi^{(q)}$ corresponding to ${\cal D}^{(q)}$, and
subleading terms determine the result (\ref{sqhe-e57}).

A related analysis can also be performed for the moments of the
two-point conductance, Eq.~(\ref{e2.51}). The corresponding exponents
$X_q$ are found to be linked to the wavefunction multifractal indices
via (\ref{e2.53}).

\subsubsection{Numerical results}
\label{s6.4.6}

The numerical simulations of the SQHE network \cite{mirlin02} have
allowed to confirm the analytical predictions
(Sec.~\ref{s6.4.3}--\ref{s6.4.5}) as well as to
determine some physical quantities that are not known analytically, --
most notably, the whole spectrum of multifractality. We present a
brief summary of the numerical results.

\begin{figure}
\centerline{
\includegraphics[width=0.8\columnwidth,clip]{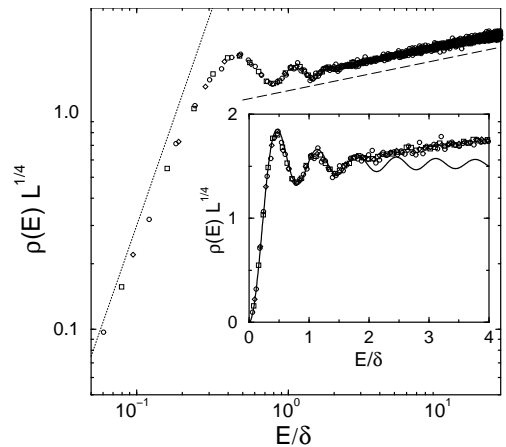}}
\caption{Scaling plot of the density of states for system sizes
$L=16 (\diamond), 32 (\Box), 96 (\circ)$.
Dashed and dotted lines indicate power laws
(dashed: $E^{1/7}$, dotted: $E^{2}$),
$\delta=1/2\pi L^{7/4}$ denotes the level spacing at $E=0$.
{\it Inset:} same data on a linear scale and the RMT result 
(solid curve). \protect\cite{evers02}   
} 
\label{f:SDoS}
\end{figure}

In Fig.~\ref{f:SDoS} the numerically calculated
DOS $\rho(E)$ for different system sizes $L$ is displayed. 
After a proper rescaling all data collapse onto a single curve. 
 The scale invariance of
$\rho(E)$ at criticality is reminiscent of the analogous
property of the level statistics at the conventional Anderson or QH
transition, Sec.~\ref{s2.5}. 
At $E\gg\delta$ the critical DOS scales as $|E|^{1/7}$,
in agreement with the analytical
prediction (\ref{sqhe11}). On the other hand, at
$E\sim\delta$ one observes an oscillatory structure
qualitatively analogous to the RMT behavior  
for the class C \cite{altland97}.

\begin{figure}[btp]
    \begin{center}\leavevmode
      \includegraphics[width=0.85\linewidth]{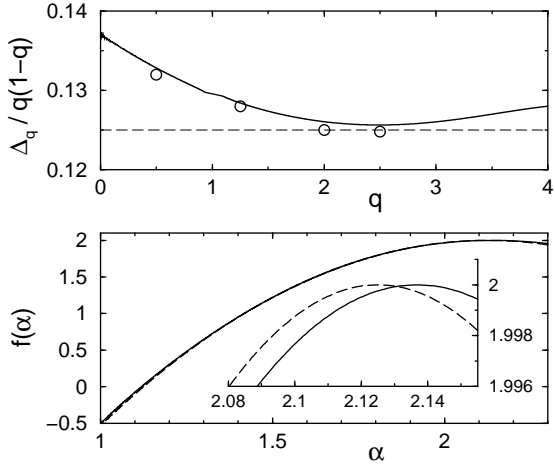}
    \end{center}
      \caption{{\it Upper panel:} anomalous dimensions $\Delta_q$
        (solid line) and $\Delta_q^{\rm typ}$ (circles)
        at the SQHE critical point. Numerical results agree
        very well with analytical findings
        $\Delta_2=-1/4$ and $\Delta_3=-3/4$. 
Data are presented in the form $\Delta_q/q(1-q)$ emphasizing deviations 
from exact parabolicity (dashed line).
{\it Lower panel:} singularity spectrum $f(\alpha)$
        (solid) and the parabolic approximation (dashed). {\it Inset:}
        magnified view of the apex region. \protect\cite{mirlin02} }
      \label{f:sqhe-mf}
\end{figure}

The anomalous multifractal dimensions $\Delta_q$ [divided by $q(1-q)$] 
are shown by a solid line in the
upper panel of Fig.~\ref{f:sqhe-mf}. 
They have been obtained from the scaling of 
the average IPRs.  According to the analytical calculations
(Sec.~\ref{s6.4.5}),  
$\Delta_q/q(1-q)$ is equal to 1/8 for both $q=2$ and $q=3$; this value
is marked by the dashed line in the figure. The
numerical results agree perfectly well with the analytical findings at
$q=2$ and $q=3$.   Furthermore, the parabolic dependence may serve as a
numerically good approximation,
\be
\label{sqhe-num3}
\Delta_q\simeq {q(1-q) / 8}. \qquad {\rm (approximate!)}
\ee
Nevertheless, Eq.~(\ref{sqhe-num3}) is not exact: 
at $0 < q <2$ the numerically found $\Delta_q$ show clear
deviations from exact parabolicity (\ref{sqhe-num3}), 
which are of the order of $10\%$
near $q=0$. In particular, the deviation of the limiting value 
$\Delta_q/q(1-q)|_{q\to 0}=0.137\pm 0.003$ from 1/8 well exceeds the
estimated numerical uncertainty. 
The lower panel of Fig.~\ref{f:sqhe-mf} depicts the singularity spectrum
$f(\alpha)$. The dashed line represents the parabolic approximation
Eq.~(\ref{e2.31}) with $\alpha_0-2=1/8$,
corresponding to (\ref{sqhe-num3}).
 The deviation of $\alpha_0-2=0.137\pm 0.003$ from 1/8 highlighted in
 the inset corresponds  to non-parabolicity
of $\tau_q$ discussed above. The numerical results for the
multifractality at the SQH transition rule out conjectures of critical
field theories that predict exactly parabolic spectra
\cite{bernard01,bernard02}.  

\begin{figure}
\centerline{
\includegraphics[width=0.8\columnwidth,clip]{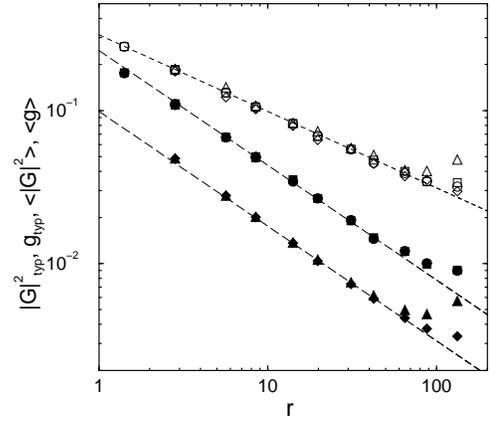}}
\caption{Scaling of the two-point conductance
with distance $r$ between the contacts: 
average value (empty symbols), $\langle g\rangle$,
and typical value (filled symbols),
$g_{{\rm typ}}=\exp \langle \ln g\rangle$,
in systems of sizes $L=128 (\Box)$ and $L=196 (\circ)$.
Also shown is scaling of the two-point Green function, $\langle
|G|^2 \rangle$ and 
$|G|^2_{{\rm typ}}=\exp \langle \ln |G|^2 \rangle$ 
($L=128 (\bigtriangleup), L=196 (\diamond)$). The
lines correspond to the $r^{-1/2}$ (dotted) and $r^{-3/4}$ (dashed)
power laws. Deviations from power-law
scaling at large values of $r$ are due to the finite system
size. \protect\cite{mirlin02} } 
\label{sqhe-fig8}
\end{figure}

In Fig.~\ref{sqhe-fig8}
the scaling of the average $\langle g\rangle$ and the typical
$g_{\rm typ}=\exp\langle\ln g\rangle$ values of the two-point
conductance is shown, along with analogous quantities
$\langle |G|^2\rangle$ and 
$|G|^2_{\rm typ}=\exp\langle\ln|G|^2 \rangle$ for a closed system,
$|G|^2\equiv-{\rm Tr}G(e',e;1)G(e,e';1)$. For the average values, 
$\langle g\rangle$ and $\langle |G|^2\rangle$, the numerics fully
confirm the theoretical results (\ref{sqhe13}), (\ref{sqhe-e9}) predicting
that 
both quantities scale as $r^{-1/2}$ and, moreover, are equal to each
other. A non-trivial character of the equality $\langle
g\rangle=\langle |G|^2\rangle$ is well illustrated by the data for
typical quantities: $g_{\rm typ}$ and $|G|^2_{\rm typ}$ are not
equal. Nevertheless, they are found to share a common scaling: 
$g_{\rm typ}, |G|^2_{\rm typ}\sim r^{-X_t}$, confirming the analytical
expectations. Furthermore, the numerically obtained value of the
exponent, $X_t\simeq 3/4$, is in agreement with the theoretical
prediction based on the relation (\ref{e2.53}), $X_t = 2x_\rho +
2(\alpha_0-2) \simeq 0.774$.

\subsection{\label{ss:tqhe} Thermal quantum Hall effect (class  D)}

\subsubsection{Physical realizations and general considerations}
\label{sss:tqhe-1}

Systems belonging to the class D are disordered superconductors where
both time reversal and spin rotation symmetries are broken. The
corresponding Hamiltonian has the structure described in
Sec.~\ref{s4.4}, see Eqs.~(\ref{e4.8}), (\ref{e4.9}) and text below
them. 
Possible physical realizations of this symmetry class include: (i)
$d$-wave superconductors with strong spin-orbit scattering, (ii)
$p$-wave paired states of spinless or spin-polarized fermions,
e.g. paired states of composite fermions \cite{read00}; (iii) triplet
odd-parity ($p$- or $f$-wave) superconductors, like
$\mathrm{SrRu}_3\mathrm{O}_4$ \cite{nelson04}; (iv) type-II
superconductors in a strong magnetic field in the presence of
spin-orbit scattering impurities \cite{senthil00}.  
While neither the quasiparticle number nor the spin are conserved for
this symmetry class, one still can speak about thermal transport. The TQHE
corresponds to the quantization  of the ratio $\kappa_{xy}/T$ of the
thermal Hall conductance to the temperature 
in units of $\pi^2 k^2_{B}/6h$ \cite{senthil00}. 

For a combination of reasons, class-D systems show particularly rich
behavior from the point of view of localization, quantum phases, and phase
transition. First of all, class D allows for two mechanisms of
2D criticality, Sec.~\ref{s6.1}: (i) a topological $\theta$-term
associated with a quantum-Hall-type transition and (ii) a metallic
phase, in view of broken spin-rotation invariance. Thus, generically,
three phases are possible: metal, insulator, and quantized Hall
conductor. A further striking feature of class D is that 
the type of disorder affects crucially the phase diagram. 
At the level of the $\sigma$-model,
the reason is believed to be that the relevant target space has 
two disconnected pieces, and that, depending on the choice of the
underlying microscopic model, it may or may not be necessary to
consider configurations containing
domain walls on which the $\sigma$-model field jumps between the two 
components \cite{bocquet00,chalker00,read01,gruzberg05}.
In the following we mainly concentrate on the
Cho-Fisher (CF) network model \cite{cho97} of the TQHE, which 
is generic in the sense that it displays all
three possible phases.
Other models of disorder will be briefly discussed in the end of
Sec.~\ref{sss:tqhe2}.

\subsubsection{Network model and phase diagram}
\label{sss:tqhe2}

To obtain a disordered network model of class D, one can start from the
ordered network, Fig.~\ref{iqhe-f2}, and then allow for independent
fluctuations of the node parameters $\theta_i$ with some distribution
function ${\cal P}(\theta)$. The CF model corresponds to the
choice  
\begin{equation}
\label{tqhe-e2}
{\cal P}(\theta) = (1-p)\delta(\theta-\theta_0) +
{p\over 2}\delta(\theta+\theta_0) + 
{p\over 2}\delta(\theta+\theta_0-\pi),
\end{equation}
implying that disorder is introduced as 
isolated defects by making the change
$\theta\to -\theta$ or $\theta\to \pi-\theta$,
for a subset of nodes randomly distributed with a concentration $p$. 
This amounts to flipping signs of either both
$\sin\theta$  or both $\cos\theta$ associated with such a node. 
This procedure can be viewed as the insertion of 
two additional half-flux lines into two plaquettes adjacent to the node
and belonging to the same sublattice.
Note that the vortex pair appears with equal probability on the C- or
S-sublattice. It is this feature that distinguishes the CF
model from the random bond Ising model (RBIM) \cite{cho97,merz01}, 
which is obtained if all the additional vortices are placed
on the same sublattice.

\begin{figure}[tb]
  \includegraphics[width=0.8\columnwidth]{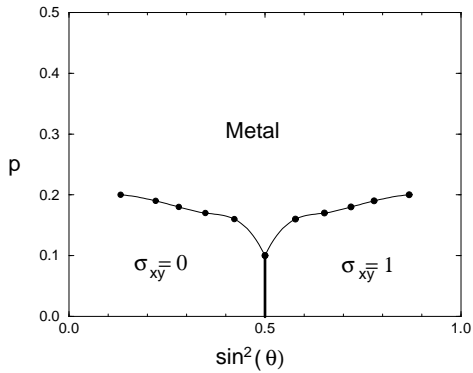}
\caption{Phase diagram of the Cho-Fisher model as obtained in
  \cite{chalker00} from transfer matrix calculations.
  The plane is spanned by the parameters $\sin^2 \theta$, the
  interplaquette tunneling probability, 
and $p$, the concentration of vortex disorder: these control the
short-distance values of the conductivity components $\sigma_{xy}$ and  
$\sigma_{\rm  xx}$ respectively.}
\label{tqhe-f1}
\end{figure}

The phase diagram of the CF model was established in \textcite{chalker00}, and
it was found that all the three expected phases are indeed present, 
Fig.~\ref{tqhe-f1}. The DOS in these phases and at transitions between
them was studied in \textcite{mildenberger07b}, the results will be
presented in Sec.~\ref{sss:tqhe3}, \ref{sss:tqhe4}.  
It was also checked in \textcite{mildenberger07b}
that the CF model is indeed generic: the same behavior is obtained for
a model with a Gaussian distribution ${\cal P}(\theta)$. 

We briefly discuss now two other disorder models, with properties
qualitatively different from the CF model:

(i) A fermionic version of the $\pm J$ RBIM is described by a disordered
network model with \cite{read01,chalker00,merz01}
\begin{equation}
\label{tqhe-e3}
{\cal P}(\theta) = (1-p)\delta(\theta-\theta_0) +
p\delta(\theta+\theta_0).
\end{equation}
This implies that all pairs of vortices are inserted in the same
sublattice. It has been shown analytically
\cite{read01} and verified numerically \cite{chalker00} 
that the metallic phase is absent in the RBIM. Two phases with
localized states (separated by the TQHE transition)
correspond to the paramagnetic and ferromagnetic phases in the Ising
spin language. The self-dual state of the disorder-free network ($p=0$,
$\sin^2\theta=1/2$) maps onto the critical point of the clean Ising
model.  

(ii) The $O(1)$ model is obtained if one includes in the regular
network factors (-1) for propagation along some links, randomly
selected with a concentration $p$. The crucial feature of such a
defect (that distinguishes it from the randomness in the nodal
parameter such as the vortex pairs in the CF model and the
RBIM) is that it cannot be ``switched off'' by any continuous
transformation. Therefore, such a defect has the topological character
of a vortex. It was found \cite{read01,chalker00} that such
topological defects destroy completely the localization, so that the
$O(1)$ model is always in the metallic phase. From the $\sigma$-model
perspective, it was shown that the effect of vortices is in
suppression of the second (disconnected) component of the target space
\cite{bocquet00}.

\subsubsection{Thermal metal}
\label{sss:tqhe3}

The metallic phase can be treated analytically by using the
$\sigma$-model approach. The corresponding RG analysis yields
\begin{equation}
\label{tqhe-e4}
  dt / d \ln L = - t^2 \ ,
  \end{equation}
where $t$ is the running coupling constant inversely proportional to the
dimensionless conductivity, $t=1/\pi g$. The infrared behavior of the
system is governed by
the perfect-metal fixed point, $t\to 0$. Specifically, the
conductance increases logarithmically with the system size, 
$g(L) = g_0 +(1/\pi) \ln L/\ell_0$ ($\ell_0$ is the mean free path), 
justifying the perturbative RG. The RG equation for the second coupling
constant $\varepsilon$, whose bare value is given by the energy $E$, reads
\begin{equation}
\label{tqhe-e5}
  d\varepsilon/ d \ln L = (2+t)\varepsilon \ ,
  \end{equation}
leading to a logarithmic increase of DOS \cite{senthil00,bocquet00},
\begin{equation}
 \rho(E) = \rho_0 + \frac{1}{4\pi^2D}\ln
 \frac{D}{|E| \ell_0^2}\,,
\label{tqhe-e6}
\end{equation}
where $D$ is the diffusion constant (remaining unrenormalized to this
order), $g_0 = 2\pi\rho_0 D$.

\begin{figure}[tb!p]
\includegraphics[width=0.8\columnwidth,clip]{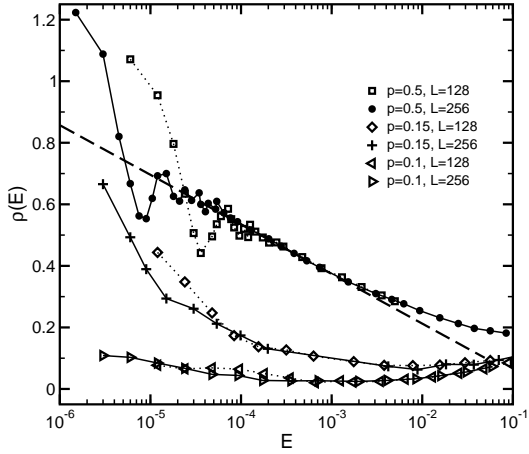}
\caption{Low-energy DOS in the metallic phase.
Parameters (upper curves): $p{=}0.5$, $\alpha{=}\pi/4$,
system sizes $L{=}128$ (squares) and $L{=}256$ (full circles).
The straight dashed line represents the logarithmic asymptotics.
For lowest energies the RMT oscillations are clearly visible; they can
be collapsed on a single curve as shown in Fig.~\ref{tqhe-f3}. For
comparison, the results for $p=0.15$ and $p=0.1$ are also shown;
the latter point is close to the expected
boundary of the metallic phase, see Fig.~\ref{tqhe-f1}. It is seen
that when the system approaches the phase boundary, the
logarithmic increase of the DOS disappears and the RMT
oscillations get damped. \cite{mildenberger07b}.
}
\label{tqhe-f2}
\end{figure}

Numerical results for the DOS in the
metallic phase are shown in Fig.~\ref{tqhe-f2}. The data
exhibit a logarithmic increase of the DOS over almost
three decades in $E$ for the larger system size, $L{=}256$.
It is worth stressing that the increase continues to be of logarithmic form
even though the renormalized DOS at small energies becomes
much larger than its bare (large-$E$) value $\rho_0{\simeq} 0.1$.
This is a signature of the fact that the RG flow is towards weak
coupling, so that the one-loop result (\ref{tqhe-e6}) is valid down to
arbitrarily low energies in the thermodynamic limit.

\begin{figure}[tb!p]
\includegraphics[width=0.8\columnwidth,clip]{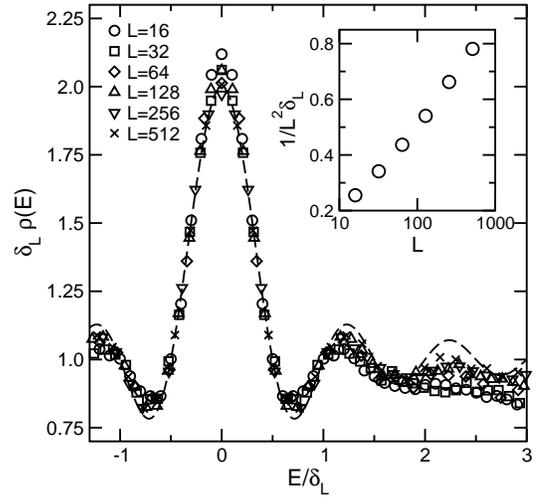}
\caption{Renormalized DOS at maximal disorder $p{=}0.5$ and
on the symmetry line $\sin^2\alpha {=}1/2$ for different
system sizes vs. the energy measured in units of the level spacing $\delta_L$.
The RMT result, Eq.~(\ref{tqhe-e7}), 
is plotted as a dashed line for comparison. {\it Inset:} 
logarithmic dependence of $1/L^2\delta_L$ on the system size $L$,
consistent with the data of Fig.~\ref{tqhe-f2}. \cite{mildenberger07b}. }
\label{tqhe-f3}
\end{figure}

At the smallest energies, pronounced oscillations in the
DOS are observed. 
These are RMT oscillations due to finite system size and serve as
another indication of the fact that we are dealing with a metallic
phase. The RMT origin of these oscillations is demonstrated in
Fig.~\ref{tqhe-f3}, where these parts of DOS curves are replotted,
with the energy rescaled to the mean
level spacing $\delta_L$ at lowest energy for the corresponding system
size. The data collapse on a single curve, which shows that the
renormalized level spacing 
\begin{equation}
\label{tqhe-e6a}
\delta_L=\frac{1}{L^2\rho(E_{\rm Th})} = 
{1\over L^2\rho_0[1+t_0\ln (L/\ell_0)]},
\end{equation}
with $E_{\rm Th}$ being the Thouless energy, 
is indeed the only relevant energy scale in the regime $E \lesssim E_{\rm
Th}$ where the RMT is applicable. 
As further seen in Fig.~\ref{tqhe-f3}, the obtained curve 
agrees with the RMT prediction,
\begin{equation}
\label{tqhe-e7}
 \rho(E)={1\over L^2 \delta_L}\left[ 1+ {\sin(2\pi E/\delta_L)\over 
2\pi E/\delta_L}\right],
\end{equation}
 up to $E/\delta_L {\sim} 1.5$--$ 2$; for larger
energies the oscillations are strongly suppressed. This is fully
consistent with the exponential vanishing of the RMT oscillations
beyond the Thouless energy, see the review \cite{mirlin00}.
 With increasing system size, the ratio
$E_{\rm Th}/\delta_L$ increases (though only logarithmically), so that
the RMT range includes progressively more oscillation
periods. This tendency is clearly seen in Fig.~\ref{tqhe-f3}.

\subsubsection{Localized phases and TQH transition}
\label{sss:tqhe4}

\begin{figure}[tb!p]
\includegraphics[width=0.8\columnwidth,clip]{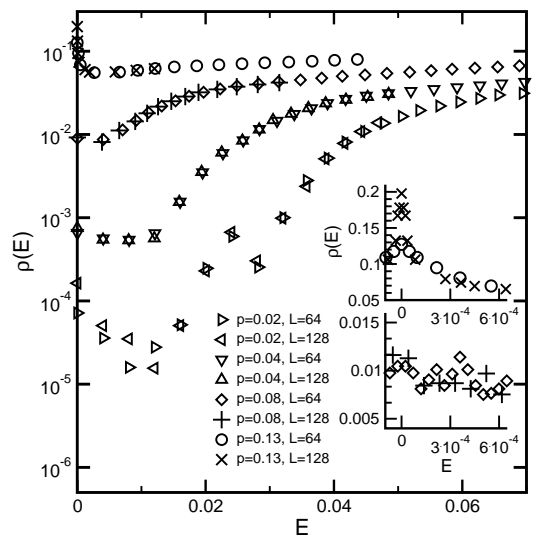}
\caption{DOS near $E=0$ for disorder values $p=0.13$, 0.08, 0.04,
  and 0.02 and for two system
sizes $L=64,\ 128$ at fixed interplaquette coupling $\sin^2(\theta)=0.579$. 
The DOS diverges logarithmically as $E\to 0$ in the metallic phase ($p=0.13$)
and remains finite in the localized phase (other values of $p$).
The results for the lowest impurity concentration, $p=0.02$, show an
oscillatory feature induced by the band structure of the clean system, as well
strong scatter in the data at the lowest energies, which is due 
to insufficient ensemble averaging. {\it Upper inset:} Low-energy peak at
$p=0.13$; its amplitude 
increases with $L$, in agreement with  Sec.~\ref{s3.1}. {\it Lower inset:}
Low-energy DOS at $p=0.08$. No peak at $E\to0$ is detected; $\rho(E\to
0)$ is a constant independent on $L$, indicating that the system is in
the insulating phase. Statistical noise in the lower inset is more
pronounced than in the upper one due to the smallness of the DOS.
\cite{mildenberger07b}.}
\label{tqhe-f4}
\end{figure}

An analytical approach to the problem  alternative to the $\sigma$-model 
is based on the model of Dirac fermions with random mass, 
in the spirit of the analysis of the Ising model 
in \textcite{dotsenko83a}. Being perturbative in the
disorder strength, this approach is appropriate for the description of
the localized phases and the transition between them.   
The disorder-free system has a transition, driven by tuning 
a uniform mass through zero, which in the CF model
lies at $p{=}0$, $\sin^2\theta{=}1/2$ and corresponds to the clean
Ising transition. 
In the vicinity of the clean fixed point representing this transition
the disorder strength $g_M$ is marginally
irrelevant. This implies 
for the critical DOS a logarithmic correction term of the form
\cite{bocquet00,mildenberger07b} 
\begin{equation}
\rho(E) = \frac{|E|}{2\pi}\left(1 + \frac{2g_M}{\pi}\ln
\frac{1}{|E|} \right).
\label{tqhe-e8}
\end{equation}
Slightly away from the critical value $\theta=\pi/4$, the system is in
a localized phase with a large localization length
$\xi\propto|\theta-\pi/4|^{-\nu}$. As the RG flow is towards the clean
Ising fixed point, the corresponding index should be the same as in
the Ising model, $\nu=1$. The behavior of the DOS in the localized
phases can be also understood using the Dirac fermion RG
\cite{mildenberger07b}. Specifically,
for energies that are not too small, behavior will
be the same as at criticality, Eq.~(\ref{tqhe-e8}). 
However, for smallest energies,
it is the localization length $\xi$ (rather than $E$) that will
terminate the RG process. In this sense, the role of $\xi$ is fully
analogous to that of finite system size $L$ at criticality. 
This implies that $\rho(E)$ saturates at the value
\begin{equation}
\label{tqhe-e9}
\rho(E) \sim 
{\ell_0\over \xi}
\left(1+2{g_M\over\pi}\ln{\xi\over\ell_0}\right)^{1/2}, \qquad
  E\lesssim E_\xi,
\end{equation}
The energy $E_\xi$ at which the saturation takes place is 
\begin{equation}
\label{tqhe-e10}
E_\xi \sim {\ell_0\over \xi}
\left(1+2{g_M\over\pi}\ln{\xi\over\ell_0}\right)^{-1/2}.
\end{equation}
The Anderson transition from the metallic to one of the
localized phases should therefore be signalled by a transition from 
logarithmically diverging to finite $\rho(E\to 0)$. This has been
verified by numerical simulations in \textcite{mildenberger07b},
Fig.~\ref{tqhe-f4}.

A brief comment on the regions of
localized phases where the interplaquette coupling is very weak
($\sin^2\theta$ close to zero or to unity), is in order here. As shown
recently \cite{mildenberger05}, 
in this situation the DOS of the RBIM acquires a
non-universal power-law singularity, $|E|^{1/z-1}$ with $z>1$
associated with Griffiths strings \cite{motrunich01,motrunich02}. 
The same mechanism of the formation of divergent DOS in these parts of
the localized phases is expected to be operative in the
CF model as well. 

\begin{figure}[tb!p]
\includegraphics[width=0.8\columnwidth,clip]{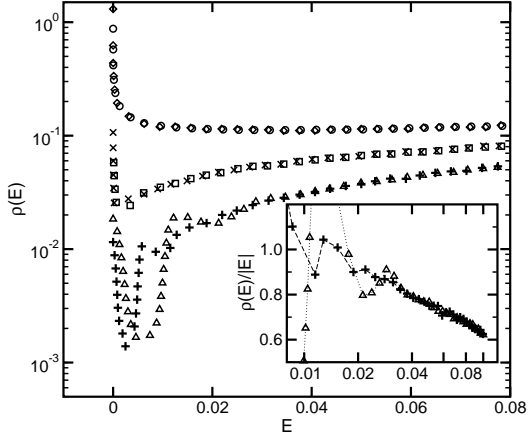}
\caption{DOS at low energy on the self-dual line $\sin^2(\alpha)=0.5$ 
for disorder concentrations
$p=0.2$ ($\circ$, $\diamond$), 0.1 ($\Box$, x), and  0.05 ($\triangle$, +),
where in each case the first symbol is for $L=128$ and the second is
for $L=256$. 
{\it Inset:} $\rho(E)/|E|$ at $p{=}0.05$ on a log-linear scale.
The logarithmic correction is clearly observed, in agreement with 
Eq.~(\ref{tqhe-e8}). \cite{mildenberger07b}. }
\label{tqhe-f5}
\end{figure}

We turn now to numerical results \cite{mildenberger07b} on the DOS at the
line $\sin^2\theta=1/2$, where the TQH phase boundary is located. They
are shown in Fig.~\ref{tqhe-f5}. While at sufficiently large $p$
($p\gtrsim 0.1$) the DOS shows a logarithmic increase characteristic
for the metallic phase, for lower $p$ the DOS behavior agrees with
Eq.~(\ref{tqhe-e8}), as expected for the TQH transition. This is
demonstrated in the inset of  Fig.~\ref{tqhe-f5}, where $\rho(E)/|E|$
as a function of $\log|E|$ is plotted for $p=0.05$. 

While at moderately low
$E$ the DOS at the TQH transition line 
is in good agreement with the Dirac-fermion RG, Eq.~(\ref{tqhe-e8}),
the results for the lowest energies obtained in \textcite{mildenberger07b}
constitute a puzzle. Specifically, it was found that the DOS saturates at
a very low energy scale and even shows a weak upturn. The reason for
this behavior  is not understood at present; several possible
scenarios were proposed in  \textcite{mildenberger07b}: (i) 
the position $p_T$ of the tricritical point 
T is in fact not $p_T\simeq 0.1$ as in Fig.~\ref{tqhe-f1} but rather
considerably smaller, $p_T<0.05$; (ii) in addition
to the tricritical point $p_T$ there is a second, repulsive fixed
point on the TQH transition line $\sin^2\theta=1/2$ , at some
$p_N<p_T$. This point would then act as a ``flow splitter'' which is similar
to the role of the Nishimori point in the RBIM; 
(iii) the RG treatment of the 
theory of Dirac fermions with Gaussian random mass is in fact insufficient,
and some effects -- possibly of non-perturbative origin --
eventually drive the system away from the clean Ising fixed point. 
The clarification of this important issue remains a subject for future
research.

\subsection{Chiral classes (AIII, CII, BDI)}
\label{s6.6}

We have seen in Sec.~\ref{s5.4} that quasi-1D systems of chiral
symmetry show a criticality accompanied by a very slow ($L^{-1/2}$) decay of
the average conductance and by a Dyson-type singularity
($1/|E\ln^3 E|$) in the DOS. As we discuss in the present
subsection, a similar type of criticality takes place also in two
dimensions.

\subsubsection{Gade-Wegner $\sigma$-model}
\label{s6.6.1}

In their pioneering works, \textcite{gade91,gade93} derived
$\sigma$-models for systems of the chiral classes and performed their
RG analysis at and near two dimensions. They used the fermionic
replicas, so that the models are defined on the spaces 
${\mathrm U}(N)$, ${\mathrm U}(2N)/{\mathrm Sp}(2N)$, and  
${\mathrm U}(N)/{\mathrm O}(N)$ for the chiral unitary, orthogonal and
symplectic classes, respectively. As usual, the supersymmetric
generalization (see Table~\ref{t4.1} for the symmetry classes of the
corresponding models) is equivalent to the replica version on the level
of the perturbation theory. For definiteness, we consider the chiral unitary
class (AIII), the results for the other two are very similar. The action of the
$\sigma$-model has the form 
\bea
\label{e6.6.1}
&& \hspace*{-0.6cm}
S[Q]= \int d^2r \left\{ {1\over 16\pi t}{\rm Tr} \nabla Q^{-1}\nabla Q - 
{1\over 64\pi c}[{\rm Tr} (Q^{-1}\nabla Q)]^2 \right.\nonumber \\
&& \left. \qquad +i{\pi\rho_0\over 4}\varepsilon{\rm Tr} (Q+Q^{-1})
\right\},
\eea
where $Q\in {\mathrm U}(N)$, $t$ and $c$ are two coupling constants,
the first of which is related to the conductivity in the usual way,
$t^{-1}=2\pi\sigma_{xx}$, $\varepsilon$ is the running coupling
whose bare value is the energy $E$ (which breaks the chiral
symmetry), and $\rho_0$ the bare density of states. 
A special feature of the $\sigma$-model for chiral classes is
the existence of the second term (known as Gade term), which governs
fluctuations of ${\rm det}\,Q$. 
This $\sigma$-model was later obtained and analyzed in a 
number of works
\cite{guruswamy99,fukui99,altland99,fabrizio00}.

The one-loop RG equations in $2+\epsilon$ dimensions read
\bea
\label{e6.6.2}
&& -dt/\ln L = \epsilon t \qquad \text{(exact!)}\ ; \\
&& -dc/ d\ln L = \epsilon c + 2 c^2\ ;
\label{e6.6.3}
\\
&& d\ln\varepsilon/d\ln L = (2+\epsilon) + t^2/2c\ .
\label{e6.6.4}
\eea
It is of central importance that in 2D ($\epsilon=0$) the
$\beta$-function (\ref{e6.6.2}) is identically zero \cite{gade91},
i.e. the coupling
constant $t$ (and thus, the conductivity) is not renormalized.
If one considers a system at non-zero (but small) energy, the
chiral symmetry is broken down to that of the unitary Wigner-Dyson
class (A), and 
the localization should happen at a sufficiently large scale. To find
the scaling of the localization length with energy, 
one performs the RG transformation until the running
energy $\varepsilon(L)$ ceases to be small (i.e. reaches a
characteristic scale $\Delta$ of the problem; typically, the
bandwidth). Beyond this scale, the RG takes the form characteristic
for class A, driving the system towards the localized regime.  
The localization length $\xi(E)$ is thus given by the crossover length
$L_c(E)$, up to an energy-independent factor $\sim\exp{(-1/4t^2)}$.
Further, the DOS is given by 
\be
\label{e6.6.5}
\rho(E) \sim 1/E L_c^2(E). 
\ee
Integration
of Eqs.~(\ref{e6.6.3}), (\ref{e6.6.4}) (with $\epsilon=0$) yields
\be
\label{e6.6.6}
\ln L_c(E)\simeq {1\over 2t^2}\left(\sqrt{B^2+4t^2|\ln (E/\Delta)|}-B\right)\ ,
\ee
where $B=2+t^2/c_0$ and $c_0$ is the bare value of the coupling $c$. 
For asymptotically low energies, $|\ln (E/\Delta)| \gg B^2/4t^2$, this
reduces to
\bea
\label{e6.6.7}
&& \xi(E)\propto \exp\left(t^{-1}|\ln (E/\Delta)|^{1/2}\right)\ , \\
&& \rho(E) \propto E^{-1} \exp\left(-2t^{-1}|\ln (E/\Delta)|^{1/2}\right)\ .
\label{e6.6.8}
\eea
The last formula is the 2D counterpart of the Dyson singularity.
On the other hand, at intermediate energies,   $|\ln (E/\Delta)| \ll
B^2/4t^2$, the behavior is of the power-law type with 
a non-universal exponent,
\be
\label{e6.6.9}
\xi(E)\propto E^{-1/B}\ ;\qquad \rho(E)\propto E^{-1+2/B}\ .
\ee

It is worth emphasizing that the bare coupling $c_0$ is of the order
of unity even for a system with large conductance ($t\ll 1$). This
means that $B\simeq 2$ and the asymptotic behavior (\ref{e6.6.7}),
(\ref{e6.6.8}) 
establishes at exponentially low energies only, $\ln (\Delta/E) \ll
1/t^2$. 
For this reason it is very difficult to reach the true asymptotics 
in numerical simulations, whereras the intermediate power law 
regime (\ref{e6.6.9}) can be studied very well, 
see e.g. \textcite{markos07}. 

Like in the 1D geometry, one can drive the system away from criticality by
introducing a staggering in hopping strength (``dimerization'') in a
lattice model. The DOS in the localized phase was predicted to show a
power-law behavior, $\rho(E) \sim E^{-1+2/z}$, governed by the Griffiths
mechanism, with a non-universal
dynamical exponent $z>0$  \cite{motrunich02}. In the work
\cite{bocquet02} network models for all the chiral classes were
constructed. A numerical study of the class-AIII network model 
confirmed the existence of the critical (Gade-Wegner) and
the localized (Griffiths-type) phases. 
It was found that the DOS exhibits a non-universal
$E^{-1+2/z}$ power-law behavior in the localized phase and that $z$
asymptotically tends to infinity with decreasing energy in the
critical phase, in agreement with analytical predictions.

\subsubsection{Dirac fermions approach. Strong-coupling effects}
\label{s6.6.2}

The Gade-Wegner prediction of a critical state with a diverging DOS of
the form (\ref{e6.6.8}) can also be reproduced by starting form a model
of disordered Dirac fermions \cite{guruswamy99}. Specifically, 
a model of fermions on a bipartite lattice with $\pi$ flux and random
real hopping \cite{hatsugai97}, which belongs to the class BDI, 
is described by a two-flavor model of Dirac fermions subject to a
random vector potential (coupling $g_A$) and a chiral random mass
($g_m$). This is a particular type of two-flavor disordered Dirac fermion 
models considered in Sec.~\ref{s6.7.4}; in the notations used there
$g_A=\gamma_\bot/2$, $g_m=\beta_z/2$. For this model, the exact
$\beta$-functions can be found  \cite{guruswamy99}
to all orders in $g_{m}$:
\be
\label{e6.6.10}
\frac{dg_{A}}{d\ln L} = \frac{2 g_m^2}{(1+2g_m)^2}, \qquad 
\frac{d g_m}{d\ln L} = 0\ .
\ee 
Further, the dynamical scaling function was found to be
\be
\label{e6.6.11}
{d\ln \varepsilon \over d\ln L} \equiv z  
= 1 + {2g_A \over (1+2g_m)^2} + 2g_m + O(g_m^2)\ .
\ee
According to Eqs.~(\ref{e6.6.10}), $g_m$ is not
renormalized, while $g_A$ grows logarithmically with $L$. Assuming for
simplicity weak disorder, $g_m\ll 1$, we have
\be
\label{e6.6.12}
g_A(L) \simeq g_{A}^{(0)} + 4 g_m^2 \ln L\ .
\ee
For sufficiently large $L$ the $\ln L$ term dominates. Substituting it
in Eq.~(\ref{e6.6.11}), one gets
\be
\label{e6.6.13}
\ln \varepsilon(L) \simeq 2 g_m^2 \ln^2 L\ .
\ee
This reproduces Eqs.~(\ref{e6.6.7}), (\ref{e6.6.8}) for the localization
length and the DOS with $t \to g_m\sqrt{2}$. 

As was shown in \textcite{motrunich02}, this result is, however, not fully
correct, for the following reason. It is known that the multifractal
spectrum of Dirac
fermions in a random vector potential undergoes a transition from a
weak-disorder to strong-disorder phase at $g_A=1$ (``freezing
transition''), see Sec.~\ref{s6.7.3}. 
As shown in \textcite{motrunich02,horovitz02}, this transition is
accompanied by a change of the behavior of the dynamical exponent
\be
\label{e6.6.14}
 z = \left\{\begin{array}{cc}
           1 +2 g_A\ , & \qquad g_A < 1\ ; \\
           4\sqrt{g_A} - 1\ , & \qquad g_A > 1\ . 
           \end{array}
           \right.
\ee
In the presence of the second coupling ($g_m$), $g_A(L)$ will flow
according to (\ref{e6.6.12}), and $z$ will develop following
Eq.~(\ref{e6.6.14}).  
While the first line of (\ref{e6.6.14}) agrees with Eq.~(\ref{e6.6.11}),
it is only valid at short distances, $g_A(L) < 1$ (which corresponds
to the ballistic regime). At sufficiently long (diffusive) scale,
where the $\sigma$-model is applicable, $g_A$ will become larger than
unity. Using the second line of Eq.~(\ref{e6.6.14}) instead of
Eq.~(\ref{e6.6.11}) yields 
\be
\label{e6.6.15}
\rho(E) \sim E^{-1} \exp\left\{-{1\over 2}(3g_m^{-1}|\ln
  (E/\Delta)|)^{2/3}\right\}\ . 
\ee
Therefore, the exponent $1/2$ in (\ref{e6.6.7}) and in the second
(subleading) factor in (\ref{e6.6.8}) is replaced by $2/3$. 

This result was also rederived in the framework of the RG approach.
The key point is that, when the running coupling $g_A$ ceases to be
weak, it is not sufficient anymore to characterize the disorder by the
lowest-order cumulants. Instead,  
one should take into account an infinite number of couplings.  
Such a functional RG method was developed in \textcite{carpentier99} in
the context of a related random XY model. For the problem 
of Dirac fermions with random mass and vector potential this program
was carried out in \textcite{mudry03}: the results confirm the
conclusion of \textcite{motrunich02}, Eq.~(\ref{e6.6.8}). Analogous results
were also obtained in \textcite{yamada03}.

\subsection{Disordered Dirac Hamiltonians}
\label{s6.7}

Localization and criticality in models of 2D Dirac fermions subjected
to various types of disorder have been studied in a large number of
papers and in a variety of contexts, including the
random bond Ising model \cite{dotsenko83a}, the quantum Hall effect
\cite{ludwig94}, dirty superconductors with unconventional pairing
\cite{nersesyan95,bocquet00,altland02}, and some lattice models with
chiral symmetry \cite{guruswamy99}. Recently, this class of problems
has attracted a great deal of attention in connection with its
application to graphene, see, in particular, 
\textcite{mccann06,khveshchenko06,aleiner06,altland06,ostrovsky06,ostrovsky07a,ostrovsky07b}.  

In the presence of different types of randomness, Dirac Hamiltonians 
realize all ten symmetry classes of disordered systems; see
\cite{bernard01b} for a detailed symmetry classification. Furthermore,
in many cases the Dirac character of fermions induces non-trivial
topological properties ($\theta$-term or WZ term) of the
corresponding field theory ($\sigma$-model). In Sec.~\ref{s6.7.1} we
review
the classification of disorder in  a two-flavor model of Dirac
fermions describing the low-energy physics of graphene, 
the RG treatment of the problem, and types of criticality.
The emergent critical theories will be discussed in
Sec.~\ref{s6.7.2}--\ref{s6.7.4}. In Sec.~\ref{s6.7.5} we briefly
discuss  a four-node Dirac fermion model appropriate for a description of 
dirty $d$-wave superconductors.

\subsubsection{Disordered two-node Dirac Hamiltonians: Symmetries of 
  disorder, renormalization group, and types of criticality.}
\label{s6.7.1}

The presentation below largely follows 
the papers \cite{ostrovsky06,ostrovsky07b}. We concentrate on a two-flavor
model, which is in particular relevant to the description of
electronic properties of graphene. Graphene is a semimetal; 
its valence and conduction bands touch each other in two conical
points $K$ and $K'$ of the Brillouin zone. In the
vicinity of these points the electrons behave as massless relativistic
(Dirac-like) particles. Therefore,  
the effective tight-binding low-energy Hamiltonian 
of clean graphene is a $4 \times 4$ matrix
operating in the $AB$ space of the two sublattices and in the
$K$--$K'$ space of 
the valleys. We introduce the four-component
wave function
\begin{equation}
 \Psi
  = \{\phi_{AK}, \phi_{BK}, \phi_{BK'}, \phi_{AK'}\}^T\ ,
 \label{4D}
\end{equation}
In this representation the Hamiltonian has the form
\begin{equation}
 H = v_0 \tau_3 \bm{\sigma}\mathbf{k}.
 \label{ham}
\end{equation}
Here $\tau_3$ is the third Pauli matrix in the $K$--$K'$ space, $\bm{\sigma}
= \{\sigma_1, \sigma_2\}$ the two-dimensional vector of Pauli matrices in
the $AB$ space, and $v_0$ the velocity  ($v_0 \simeq 10^8$ cm/s in graphene). 
It is worth emphasizing that the Dirac form of the Hamiltonian (\ref{ham}) does
 not rely on the tight-binding approximation but is protected by the
 symmetry of the honeycomb lattice which has two atoms in a unit cell. 

Let us analyze the symmetries of the clean 
Hamiltonian (\ref{ham}) in the $AB$ and $KK'$ spaces. First,  
there exists an SU(2) symmetry group in the
space of the valleys , with the generators \cite{mccann06}
\begin{equation}
 \Lambda_x
  = \sigma_3 \tau_1,
 \qquad
 \Lambda_y
  = \sigma_3 \tau_2,
 \qquad
 \Lambda_z
  = \sigma_0 \tau_3\ ,
 \label{Lambda-matrices}
\end{equation}
all of which commute with the Hamiltonian. 
Second, there are two more relevant symmetries of the clean Hamiltonian,
namely,
time inversion operation ($T_0$) and chiral symmetry ($C_0$). 
Combining $T_0$, $C_0$, and isospin rotations $\Lambda_{0,x,y,z}$,  
one can construct twelve symmetry operations, out of which four
(denoted as $T_\mu$) are of
time-reversal type, four ($C_\mu$) of chiral type, and four ($CT_\mu$) 
of Bogoliubov-de Gennes type:
\begin{eqnarray*}
 && T_0: \ A \mapsto \sigma_1 \tau_1 A^T \sigma_1 \tau_1, \qquad
 C_0: \ A \mapsto -\sigma_3 \tau_0 A \sigma_3 \tau_0, \\
 && \qquad\qquad CT_0:\  A
  \mapsto -\sigma_2 \tau_1 A^T \sigma_2 \tau_1, \\
 && T_x:\  A \mapsto \sigma_2 \tau_0 A^T \sigma_2 \tau_0, \qquad
 C_x:\ A \mapsto -\sigma_0 \tau_1 A \sigma_0 \tau_1, \\ 
&& \qquad\qquad CT_x:\  A \mapsto -\sigma_1 \tau_0 A^T \sigma_1 \tau_0, \\
&&  T_y:\  A \mapsto \sigma_2 \tau_3 A^T \sigma_2 \tau_3, \qquad
 C_y:\  A \mapsto -\sigma_0 \tau_2 A \sigma_0 \tau_2, \\ 
&& \qquad\qquad CT_y:\  A \mapsto -\sigma_1 \tau_3 A^T \sigma_1 \tau_3, \\
&& T_z:\  A \mapsto \sigma_1 \tau_2 A^T \sigma_1 \tau_2, \qquad
 C_z:\ A \mapsto -\sigma_3 \tau_3 A \sigma_3 \tau_3, \\ 
&& \qquad\qquad CT_z:\  A \mapsto -\sigma_2 \tau_2 A^T \sigma_2 \tau_2.
\end{eqnarray*}
It is worth recalling that the $C$ and $CT$ symmetries apply to the
Dirac point 
($E=0$), i.e. to undoped graphene, and get broken by a
non-zero energy $E$. We
will assume the average isotropy of the disordered graphene, 
which implies that $\Lambda_x$ and $\Lambda_y$
symmetries of the Hamiltonian are present or absent
simultaneously. They are thus
combined into a single notation $\Lambda_\perp$; the same applies to
$T_\perp$ and $C_\perp$. In Table \ref{Tab:sym} all possible
matrix structures of disorder along with their symmetries are listed.

\begin{table*}
\begin{center}
\begin{tabular}{c@{\ }cccc@{\,}cc@{\,}c@{\,}cc@{\,}c@{\,}cc@{\,}c@{\,}c}
\hline\hline
 structure\  & \ coupling\ \ \  &\ AE\ &\ GLL\ \ &
 \makebox[0.7cm]{$\Lambda_\perp$} & \makebox[0.7cm]{$\Lambda_z$} &
 \makebox[0.7cm]{$T_0$} & \makebox[0.7cm]{$T_\perp$} & \makebox[0.7cm]{$T_z$} &
 \makebox[0.7cm]{$C_0$} & \makebox[0.7cm]{$C_\perp$} & \makebox[0.7cm]{$C_z$} &
 \makebox[0.7cm]{$CT_0$} & \makebox[0.7cm]{$CT_\perp$} & \makebox[0.7cm]{$CT_z$}
\\ \hline
 $\sigma_0 \tau_0$ & $\alpha_0$ & \ $\gamma_0/2\pi v^2$\  & &
 $+$ & $+$ &
 $+$ & $+$ & $+$ &
 $-$ & $-$ & $-$ &
 $-$ & $-$ & $-$
\\
 $\sigma_{\{1,2\}} \tau_{\{1,2\}}$ & $\beta_\perp$ & $2\beta_\perp/\pi v^2$
 & &
 $-$ & $-$ &
 $+$ & $-$ & $-$ &
 $+$ & $-$ & $-$ &
 $+$ & $-$ & $-$
\\
 $\sigma_{1,2} \tau_0$ & $\gamma_\perp$ & $\gamma_\perp/\pi v^2$ & $2g_A$ &
 $-$ & $+$ &
 $+$ & $-$ & $+$ &
 $+$ & $-$ & $+$ &
 $+$ & $-$ & $+$
\\
 $\sigma_0 \tau_{1,2}$ & $\beta_z$ & $\beta_z/\pi v^2$ & $2g_m$ &
 $-$ & $-$ &
 $+$ & $-$ & $-$ &
 $-$ & $-$ & $+$ &
 $-$ & $-$ & $+$
\\
 $\sigma_3 \tau_3$ & $\gamma_z$ & $\gamma_z/2\pi v^2$ & &
 $-$ & $+$ &
 $+$ & $-$ & $+$ &
 $-$ & $+$ & $-$ &
 $-$ & $+$ & $-$
\\ \hline 
 $\sigma_3 \tau_{1,2}$ & $\beta_0$ & & $2 g_\mu$ &
 $-$ & $-$ &
 $-$ & $-$ & $+$ &
 $-$ & $-$ & $+$ &
 $+$ & $-$ & $-$
\\
 $\sigma_0 \tau_3$ & $\gamma_0$ & & &
 $-$ & $+$ &
 $-$ & $+$ & $-$ &
 $-$ & $+$ & $-$ &
 $+$ & $-$ & $+$
\\
 $\sigma_{1,2} \tau_3$ & $\alpha_\perp$ & & $2g_{A'}$ &
 $+$ & $+$ &
 $-$ & $-$ & $-$ &
 $+$ & $+$ & $+$ &
 $-$ & $-$ & $-$
\\
 $\sigma_3 \tau_0$ & $\alpha_z$ & & & 
 $+$ & $+$ &
 $-$ & $-$ & $-$ &
 $-$ & $-$ & $-$ &
 $+$ & $+$ & $+$
\\ \hline\hline
\end{tabular}
\end{center}
\caption{Disorder symmetries in graphene. 
The first five rows
of the table represent disorders preserving the time reversal symmetry
$T_0$;  the last four ---  violating $T_0$. 
First column: structure of disorder in the sublattice ($\sigma_\mu$)
and valley ($\tau_\nu$) spaces. Second column: disorder strength
according to the notations of \cite{ostrovsky06,ostrovsky07b} used in
this review. Third (AE) and fourth (GLL) columns: disorders considered
in \cite{aleiner06} and \cite{guruswamy99} and notations used
there. The remaining columns indicate which symmetries of the clean
Hamiltonian are preserved by disorder.  \cite{ostrovsky06}.}
\label{Tab:sym}
\end{table*}

To derive the field theory of the problem, one introduces a superfield
$\psi({\bf r})$, see Sec.~\ref{s2.2.1} (or, alternatively, its replica
counterpart). After the disorder averaging, 
the action containing all possible disorder structures from Table
\ref{Tab:sym} reads ($\bar\psi = \psi^\dagger\Lambda$)
\begin{equation}
\begin{split}
 S[\psi]
  &=\int d^2 r \biggl\{
      i \bar\psi \bigl(
        \varepsilon + i v_0 \tau_3 \bm{\sigma} \nabla - i0 \Lambda
      \bigr) \psi \\
 & \hspace*{-0.8cm}     +\pi v_0^2 \Bigl\{
        \alpha_0 (\bar\psi \sigma_0 \tau_0 \psi)^2 
        +\frac{\alpha_\perp}{2} \left[
          (\bar\psi \sigma_1 \tau_3 \psi)^2
          +(\bar\psi \sigma_2 \tau_3 \psi)^2
        \right] \\
&  \hspace*{-0.8cm}      +\alpha_z (\bar\psi \sigma_3 \tau_0 \psi)^2 
        +\frac{\beta_0}{2} \left[
          (\bar\psi \sigma_3 \tau_1 \psi)^2
          +(\bar\psi \sigma_3 \tau_2 \psi)^2
        \right] \\
& \hspace*{-0.8cm}       +\frac{\beta_\perp}{4} \left[
          (\bar\psi \sigma_1 \tau_1 \psi)^2
          +(\bar\psi \sigma_1 \tau_2 \psi)^2
          +(\bar\psi \sigma_2 \tau_1 \psi)^2
          +(\bar\psi \sigma_2 \tau_2 \psi)^2
        \right] \\
& \hspace*{-0.8cm}     +\frac{\beta_z}{2} \left[
          (\bar\psi \sigma_0 \tau_1 \psi)^2
          +(\bar\psi \sigma_0 \tau_2 \psi)^2
        \right] 
        +\gamma_0 (\bar\psi \sigma_0 \tau_3 \psi)^2 \\
& \hspace*{-0.8cm}       +\frac{\gamma_\perp}{2} \left[
          (\bar\psi \sigma_1 \tau_0 \psi)^2
          +(\bar\psi \sigma_2 \tau_0 \psi)^2
        \right] 
        +\gamma_z (\bar\psi \sigma_3 \tau_3 \psi)^2
      \Bigr\}
    \biggr\}.
\end{split}
\end{equation}


A complete set of one-loop perturbative RG equations  for nine disorder
amplitudes has the form \cite{ostrovsky06}
\begin{eqnarray}
 \frac{d \alpha_0}{d \ln L}
  &=& 2 \alpha_0 (\alpha_0 + \beta_0 + \gamma_0 + \alpha_\perp + \beta_\perp
     + \gamma_\perp  \nonumber \\
&+& \alpha_z + \beta_z + \gamma_z)
 2 \alpha_\perp \alpha_z
     + \beta_\perp \beta_z + 2\gamma_\perp \gamma_z; \nonumber \\
 \frac{d \alpha_\perp}{d \ln L}
  &=& 2 (2 \alpha_0 \alpha_z + \beta_0 \beta_z + 2 \gamma_0 \gamma_z);
\nonumber \\
 \frac{d \alpha_z}{d \ln L}
  &=& -2 \alpha_z (\alpha_0 + \beta_0 + \gamma_0 - \alpha_\perp - \beta_\perp
     - \gamma_\perp \nonumber \\
&+& \alpha_z + \beta_z + \gamma_z) + 2 \alpha_0 \alpha_\perp
     + \beta_0 \beta_\perp + 2 \gamma_0 \gamma_\perp; \nonumber \\
 \frac{d \beta_0}{d \ln L}
  &=& 2 [\beta_0 (\alpha_0 - \gamma_0 + \alpha_\perp + \alpha_z -
  \gamma_z) \nonumber \\
&+& \alpha_\perp \beta_z + \alpha_z \beta_\perp + \beta_\perp
     \gamma_0]; \nonumber  \\
 \frac{d \beta_\perp}{d \ln L}
  &=& 4 (\alpha_0 \beta_z + \alpha_z \beta_0 + \beta_0 \gamma_0
     + \beta_\perp \gamma_\perp + \beta_z \gamma_z); \nonumber \\
 \frac{d \beta_z}{d \ln L}
  &=& 2 [-\beta_z (\alpha_0 - \gamma_0 - \alpha_\perp + \alpha_z -
  \gamma_z) \nonumber \\
   &+& \alpha_0 \beta_\perp + \alpha_\perp \beta_0 + \beta_\perp
   \gamma_z]; \nonumber\\ 
 \frac{d \gamma_0}{d \ln L}
  &=& 2 \gamma_0 (\alpha_0 - \beta_0 + \gamma_0 + \alpha_\perp - \beta_\perp
     + \gamma_\perp \nonumber \\
&+& \alpha_z - \beta_z + \gamma_z) + 2 \alpha_\perp \gamma_z
     + 2 \alpha_z \gamma_\perp + \beta_0 \beta_\perp; \nonumber \\
 \frac{d \gamma_\perp}{d \ln L}
  &=& 4 \alpha_0 \gamma_z + 4 \alpha_z \gamma_0 + \beta_0^2 + \beta_\perp^2
     + \beta_z^2; \nonumber \\
 \frac{d \gamma_z}{d \ln L}
  &=& -2 \gamma_z (\alpha_0 - \alpha_\perp + \alpha_z - \beta_0 + \beta_\perp
     - \beta_z + \gamma_0 \nonumber \\
&-& \gamma_\perp + \gamma_z) + 2 \alpha_0 \gamma_\perp
     + 2 \alpha_\perp \gamma_0 + \beta_\perp \beta_z. 
\label{e6.7.1}
\end{eqnarray}
The RG equation for the running energy $\varepsilon$ (whose bare value
is $E$) reads
\bea
 \frac{d \ln \varepsilon}{d \ln L}
  &=  1 + \alpha_0 + \beta_0 + \gamma_0 + \alpha_\perp + \beta_\perp
     + \gamma_\perp \nonumber \\
& + \alpha_z + \beta_z + \gamma_z\ .
\label{e6.7.2}
\eea

If all types of disorder are present (i.e. no symmetries is
preserved), the RG flow is towards the
conventional localization fixed point (unitary Wigner-Dyson class A).
If the only preserved symmetry is the time reversal ($T_0$), again the
conventional localization (orthogonal Wigner-Dyson class AI) takes
place \cite{aleiner06}.  A non-trivial situation occurs if either (i) one
of chiral symmetries is preserved or (ii) the valleys remain
decoupled. In the Table \ref{Tab:result} we list those
situations when the symmetry 
prevents the localization and leads to criticality and non-zero conductivity at
$E=0$ (in the case of decoupled nodes  -- also at nonzero
$E$). Models with decoupled nodes will be analyzed in 
Sec.~\ref{s6.7.2}, and models with a chiral symmetry in
Sec.~\ref{s6.7.3} ($C_0$-chirality) and Sec.~\ref{s6.7.4}
($C_z$-chirality).

\begin{table}
\begin{center}
\begin{tabular}{cccc}
\hline\hline
Symmetries              & Class          & Criticality &
Conductivity \\
\hline
$C_z$, $T_0$            & BDI            & Gade &
$\approx 4e^2/\pi h$ \\
$C_z$                   & AIII           & Gade &
$\approx 4e^2/\pi h$ \\
$C_z$, $T_z$            & CII            & Gade &
$\approx 4e^2/\pi h$ \\
\hline
$C_0$, $T_0$            & CI             & WZW &
$4e^2/\pi h$ \\
$C_0$                   & AIII           & WZW &
$4e^2/\pi h$ \\
\hline
$\Lambda_z$, $C_0$      & $2 \times$AIII & WZW &
$4e^2/\pi h$ \\
\hline
$\Lambda_z$, $T_\perp$  & $2 \times$AII  & $\theta = \pi$ &
$4\sigma_{Sp}^{**}$ or $\infty$ (?) \\
$\Lambda_z$, $CT_\perp$ & $2 \times$D    & $\theta = \pi$ &
$4e^2/\pi h$ \\
$\Lambda_z$             & $2 \times$A    & $\theta = \pi$ &
$4\sigma_U^*$ \\
\hline\hline
\end{tabular}
\end{center}
\caption{Possible types of disorder in graphene leading to
  criticality.\protect\footnote{A further possible mechanism of
    delocalization is 
generation of the metallic phase due to broken spin- (or isospin-) rotation
invariance, Sec.~\ref{s6.1.1}. If the only preserved symmetry is
$T_z$, the system is in class AII, while if only $CT_0$ or $CT_z$
invariance is
preserved the system belongs to class D. (No topological term arises in
these situations.) In both these classes, the
system flows towards a perfect-metal fixed point if the bare
conductivity is above the localization threshold.} 
 First column: preserved symmetries. 
Second column: symmetry class.  Third
column: type of criticality. The first three row correspond to $C_z$
chiral symmetry leading to Gade-Wegner-type criticality, 
Sec.~\protect\ref{s6.7.4}. The next three rows contain models with $C_0$
chiral symmetry (random gauge fields), 
inducing a WZ term in the $\sigma$-model action,
Sec.~\protect\ref{s6.7.3}. The last four rows correspond to the case
of decoupled valleys (long-range disorder), see
Sec.~\protect\ref{s6.7.2}; from top to  bottom: random  
vector potential, scalar potential, mass, and any of their
combinations. For these models (except for the random vector
potential, which is $C_0$ chiral at the same time) the $\sigma$-model
acquires a topological term with $\theta=\pi$. Adapted from
\cite{ostrovsky07b}. }
\label{Tab:result}
\end{table}

\subsubsection{Decoupled nodes: Disordered single-flavor Dirac
  fermions and quantum-Hall-type criticality}
\label{s6.7.2}

If the disorder is of long-range character,  the valley
mixing is absent due to the lack of scattering with large momentum
transfer. The couplings that do not mix the valleys are 
$\alpha_0$, $\alpha_\perp$, $\alpha_z$, $\gamma_0$, $\gamma_\perp$, and
$\gamma_z$. For each of the nodes, the system can then be described 
in terms of a single-flavor Dirac Hamiltonian,
\begin{equation}
 H
  = v_0 [\bm{\sigma}\mathbf{k} + \sigma_\mu V_\mu(\mathbf{r})].
 \label{ham1}
\end{equation}
Here disorder includes random scalar ($V_0$) and vector ($V_{1,2}$) potentials
and random mass ($V_3$). The corresponding couplings are
$g_V = \alpha_0+\gamma_0$, $2g_A = \alpha_\perp + \gamma_\perp$, and
$\tilde{g}_m = \alpha_z +\gamma_z$ \footnote{(i) The tilde in
  $\tilde{g}_m$ serves
to distinguish it  from the ``chiral random mass'' coupling $g_m$,
see Table~\ref{Tab:sym} and Sec.~\ref{s6.6.2}, \ref{s6.7.4}; (ii) our
couplings are related to those of \textcite{ludwig94} via 
$g_A=\Delta_A/2\pi$, $g_V=\Delta_V/2\pi$, $\tilde{g}_m=\Delta_m/2\pi$.}.
The clean single-valley Hamiltonian (\ref{ham1}) obeys
the 
effective time-reversal invariance $H = \sigma_2 H^T\sigma_2$. This symmetry
($T_\perp$) is not the physical time-reversal symmetry ($T_0$): the latter
interchanges the nodes and is of no significance in the absence of inter-node
scattering. 
The RG equations have the form
\bea
d g_A / d\ln L & = & 2 g_V \tilde{g}_m\ ;
\label{e6.7.3}\\
d g_V / d\ln L & = & 2 (2g_A+g_V) (g_V+\tilde{g}_m)\ ;
\label{e6.7.4}\\
d \tilde{g}_m / d\ln L & = & 2 (2g_A-\tilde{g}_m) (g_V+ \tilde{g}_m)\ ;
\label{e6.7.5}\\
d \ln\varepsilon / d\ln L & = & 1 + 2g_A +\tilde{g}_m + g_V\ .
\label{e6.7.6}
\eea

Remarkably, single-flavor Dirac fermions are never in the conventional
localized phase! More specifically,
depending on which of the disorders are present, four different types
of criticality take place:

(i) The only disorder is the random vector potential ($g_A$). This is
a special case of the symmetry class AIII. This problem is exactly
solvable and has been studied in a great detail; we will discuss it in
Sec.~\ref{s6.7.3}. 

(ii) Only random mass ($\tilde{g}_m$) is present. The system belongs then to
class D. The random-mass disorder is marginally irrelevant, and the
system flows under RG towards the clean fixed point. This problem has
been analyzed in Sec.~\ref{sss:tqhe3}.

(iii) The only disorder is random scalar potential ($g_V$). The system
is then in the Wigner-Dyson symplectic (AII) symmetry class. As was
found in \textcite{ostrovsky07a}, the corresponding $\sigma$-model
contains a topological term with $\theta=\pi$, which leads to
delocalization. This model has been discussed in Sec.~\ref{s6.2.5}.

(iv) At least two types of randomness are present. This is the same as to
say that all of them are present, as the third one will be generated
by RG, see Eqs.~(\ref{e6.7.3})--(\ref{e6.7.5}). The model belongs to the
Wigner-Dyson unitary class A, and it was argued in
\textcite{ludwig94} that it flows into the IQH transition
fixed point. This is confirmed by the derivation of the corresponding
$\sigma$-model \cite{altland02,ostrovsky07a,ostrovsky07b}, which contains a
topological term with $\theta=\pi$, i.e. is nothing but 
the Pruisken $\sigma$-model at criticality. A particular consequence
of this is that the conductivity of graphene with this type of
disorder is equal to the value $\sigma_U^*$ of the longitudinal
conductivity $\sigma_{xx}$ at the critical point of the IQH transition
multiplied by four (because of spin and valleys). 
Note that the conclusion of IQH criticality formally
holds for arbitrary energy $\varepsilon$, although in reality it only
works near 
half-filling; for other $\varepsilon$ the critical point would only be
approached for unrealistic system sizes and temperatures.

If a uniform transverse magnetic field is applied, the topological angle
$\theta$ becomes energy-dependent. However, at the Dirac point 
($E = 0$), where $\sigma_{xy}
= 0$, its value remains unchanged, $\theta = \pi$. This implies the emergence
of the half-integer quantum Hall effect, with a plateau transition point at
$E = 0$.

\subsubsection{Preserved $C_0$ chirality:  Random abelian and
  non-abelian vector potentials}
\label{s6.7.3}

\paragraph{Generalities.}
Let us consider a type of disorder which preserves the $C_0$-chirality, $H =
-\sigma_3 H \sigma_3$; according to the Table~\ref{Tab:sym}, the
corresponding couplings are $\alpha_\perp$, $\beta_\perp$, and
$\gamma_\perp$.
The one-loop RG equations then read
\begin{align}
 \partial{\alpha_\perp}/\partial \log L
  &= 0, 
\label{e6.7.6a} \\
 \partial{\beta_\perp}/\partial \log L
  &= 4 \beta_\perp \gamma_\perp, 
\label{e6.7.6b} \\
 \partial{\gamma_\perp}/\partial \log L
  &= \beta_\perp^2\ . 
\label{e6.7.6c}
\end{align}
The specifics of the $C_0$-chirality is that it corresponds to the
more common meaning of the term ``chirality'', which refers to a
distinction between ``left'' and ``right'' particles. The  
model takes then the form of the euclidean version of $1+1$ theory of
massless Dirac fermions coupled to a gauge field. Indeed, according to
(\ref{ham}) the matrices $\sigma_1\tau_3$ and $\sigma_2\tau_3$
play the role of Dirac $\gamma$-matrices, and  the matrix
$\sigma_3$ determining the $C_0$-chirality is nothing but $\gamma_5$. 
Depending on further symmetries, three different $C_0$-chiral
models arise:

(i) The only coupling present is $\alpha_\perp$, which corresponds to
the random abelian vector potential. In this case the
nodes are decoupled, and the Hamiltonian decomposes in two copies of
a model of the class AIII. This model has already been mentioned in
Sec.\ref{s6.7.2}.

(ii) If the time-reversal symmetry $T_0$ is preserved, only the
couplings $\beta_\perp$ and $\gamma_\perp$ are allowed, and the
problem is in the symmetry class CI. The model
describes then fermions coupled to a SU(2) non-abelian gauge field,
and is a particular case of analogous SU(N) models.

(iii) All three couplings are present. This describes Dirac fermions
coupled to both abelian U(1) and non-abelian SU(2) gauge fields. This
model is in the AIII symmetry class.

Remarkably, all these critical $C_0$-chiral models are exactly solvable: one
can calculate exactly the spectra of multifractal exponents, the critical index
of the DOS, and the value of conductivity. They have been studied in a
large number of works, and we summarize the key ideas and results. 

\paragraph{Abelian vector potential.}
The model of 2D Dirac fermions moving in a random vector  potential,
\be
H = v_0\sigma_\mu (i\partial_\mu - A_\mu)\ ,
\label{e6.7.7}
\ee 
is exactly solvable at zero energy and characterized by a line of 
fixed points \cite{ludwig94}. Decomposing the vector potential in the
longitudinal (pure gauge) and transverse parts,
\be
A_\nu = \epsilon_{\nu\rho}\partial_\rho \phi({\bf r}) + \partial_\nu
\chi({\bf r})\
,
\label{e6.7.8}
\ee
and assuming that the total magnetic flux is zero\footnote{For a
nonzero total flux there will be additional zero modes, in view of
the Atiyah-Singer index theorem; see, e.g. \textcite{aharonov79}.}, one
can explicitly write the zero-energy wave functions
\cite{ludwig94,castillo97}:
\bea
&& \hspace*{-1cm} 
\Psi_{+}({\bf r}) = \left( \begin{array}{c} 0 \\ \psi_+({\bf r})
  \end{array} \right ) ; \qquad 
\Psi_{-}({\bf r}) = \left( \begin{array}{c} \psi_-({\bf r}) \\ 0
  \end{array} \right )\ ; 
\label{e6.7.9} \\
&& \hspace*{-1cm} 
\psi_{\pm}({\bf r}) = B_{\pm}^{-1/2} e^{-i\chi({\bf
    r})}e^{\pm\phi({\bf r})} \ , 
\label{e6.7.10}
\eea
where $B_\pm = \int d^2{\bf r} e^{\pm 2\phi({\bf r})}$ is the
normalization factor.
We consider the first of the functions (\ref{e6.7.9}) for
definiteness and analyze the statistical properties of 
$|\psi_+({\bf r})|^2$. They are governed by fluctuations of
$\phi({\bf r})$. Assuming a white-noise distribution of the random
vector potential with $\langle A_\mu ({\bf r}))A_\nu({\bf r'})\rangle
= 2\pi g_A \delta_{\mu\nu}\delta({\bf r}-{\bf r'})$, one gets the 
following statistics of $\phi({\bf r})$,
\be
\label{e6.7.11}
{\cal P}[\phi] \propto \exp\left\{ {-1 \over 4\pi g_A} \int d^2{\bf r}
  (\mathbb{\nabla}\phi)^2 \right\} \ .
\ee
Equation (\ref{e6.7.11}) implies that $\phi({\bf r})$ is a free
massless bosonic field characterized by the correlation function
\be
\label{e6.7.12}
\langle \phi({\bf r})  \phi({\bf r'}) \rangle = g_A \ln (L/|{\bf
  r}-{\bf r'}| )\ .
\ee

\paragraph{Multifractality.}
To find the multifractal spectrum, one considers the moments 
$\langle |\psi_+({\bf r})|\rangle^{2q}$, see Sec.~\ref{s2.3.1}. 
For not too strong disorder ($g_A \le 1$), it turns out to be
sufficient to average separately the exponential $e^{2q\phi({\bf r})}$
and each of the $q$ normalization factors $\int d^2{\bf r} e^{\pm
  2\phi({\bf r})}$. The resulting spectrum \cite{ludwig94}
\be
\label{e6.7.13}
\tau_q = 2(q-1)(1-g_A q)
\ee
has an exactly parabolic form. The corresponding $f(\alpha)$ spectrum
is given by Eq.~(\ref{e2.31}) with $d=2$, $\gamma=2g_A$:
\be
\label{e6.7.14}
f(\alpha) = 2 - (\alpha-2-2g_A)^2/8g_A\ .
\ee
In view of the exact parabolicity, the spectrum necessarily contains a
termination point at $\alpha=0$, see Sec.~\ref{s2.3.6}. The
corresponding value of $q$ is $q_c = (1+g_A)/2g_A$. Thus,
Eq.~(\ref{e6.7.13}) is only valid for $q\le q_c$; for larger $q$ the
exponent $\tau_q$ saturates at a constant value,  $\tau_{q\ge q_c}
= -f(0) = (1-g_A)^2/2g_A$. It is worth recalling that we
consider the spectrum describing the ensemble-averaged IPRs, $\langle
P_q\rangle$,  see Sec.~\ref{s2.3.5}. If one looks at the scaling of
the typical IPR, $P_q^{\rm typ}$, the information about rare events
encoded in the part of the spectrum with 
$f(\alpha)<0$ gets lost, and only the range 
$\alpha_+ < \alpha < \alpha_-$ is probed, where $\alpha_\pm =
2(g_A^{1/2}\mp 1)^2$. The corresponding $\tau_q^{\rm typ}$ spectrum
reproduces 
$\tau_q$ in the range $q_-<q<q_+$, where $q_\pm = g_A^{-1/2}$, and
becomes linear outside this range, see Eq.~(\ref{e2.36}). This
behavior was found for the random vector potential problem in
\textcite{chamon96,castillo97,carpentier00} where  the spectrum
$\tau_q^{\rm typ}$ was studied (in these works the notation $q_c$ is
used for our $q_+$). 

\paragraph{Freezing.}
As was found in \textcite{chamon96,castillo97,carpentier00}, with
increasing disorder the system undergoes a phase transition
(``freezing'') at
$g_A=1$. In the strong disorder phase, $g_A>1$,
the spectrum takes the form
\bea
\label{e6.7.15}
&& \hspace*{-1cm} \tau(q) = - 2 (1 - g_A^{1/2} q)^2\ , \qquad q<q_c =
g_A^{-1/2}\ ;\\ 
&& \hspace*{-1cm} f(\alpha) = \alpha (8  g_A^{1/2} - \alpha)/ 8g_A\ .
\label{e6.7.16}
\eea
In this phase $f(0)=0$, which implies that there is a probability of
order unity to find a point in the sample where the wave function
is of order unity. Correspondingly,  
the saturation value of $\tau_q$ for $q> q_c$ is
 $\tau_{q> q_c}=0$. 
At first sight, this may seem to imply that the wave functions are
localized. The situation is, however, 
not so simple: a non-trivial multifractal spectrum 
shows a complex structure of the wave functions. Further, it can be
shown that the probability to find a secondary spike in the wave
function of approximately 
the same magnitude as the main one and separated by a distance
comparable to the system size $L$ is of order unity
\cite{carpentier00,fukui02}. The nature of these ``quasilocalized''
wave functions is therefore similar to that of critical states in 1D
systems of chiral symmetry, see Sec.~\ref{s5.4}. 

\paragraph{Density of states.}
The scaling of the DOS is governed by the dynamical exponent $z$ via
\be
\label{e6.7.17}
\rho(E)\propto E^{(2-z)/z} \ .
\ee 
In the weak-disorder
phase ($g_A<1$) the value of $z$ is straightforwardly found to be
$z=1+2g_A$ \cite{ludwig94,nersesyan95,altland02}. However, as was already
discussed in Sec.~\ref{s6.6.2}, the freezing has also impact on the
dynamical exponent; the value of $z$ in the strong-disorder phase is
given by the second line of Eq.~(\ref{e6.6.14}).  

\paragraph{Non-abelian random gauge field.}
The problem remains exactly solvable in the case of a SU(N) non-abelian
gauge filed. However, in contrast to the abelian case, where one finds
a line of fixed points, the theory flows now into an isolated fixed
point, which is a WZW theory on the level $k=-2N$
\cite{nersesyan95,mudry96,caux98b,caux98c}. The spectrum of
multifractality is parabolic with $\gamma=(N-1)/N^2$ and is given by
Eqs.~(\ref{e6.7.13}), (\ref{e6.7.14}) with a replacement of 
$2 g_A$  by $(N-1)/N^2$. The DOS is given by Eq.~(\ref{e6.7.17}) with
a dynamical exponent $z= 2 - 1/N^2$. In the case $N=2$ that arises in
the two-node model one thus finds $z=7/4$, yielding the DOS scaling
$\rho(E)\propto E^{1/7}$. 

When both abelian and non-abelian random
gauge fields are present, they contribute additively to the exponents
(in the non-frozen regime), as the two sectors of the theory
decouple. This yields \cite{mudry96} the multifractal scaling
~(\ref{e6.7.13}), (\ref{e6.7.14}) with $\gamma =2g_A \to 2g_A +
(N-1)/N^2$ and the DOS scaling (\ref{e6.7.17}) with $z=1+2g_A +
(N^2-1)/N^2$. 

\paragraph{Conductivity.}
A further remarkable feature of the models with $C_0$-chiral randomness
is the independence of conductivity on the disorder strength. We
sketch the proof of this statement given in \textcite{ostrovsky06}.  The
conductivity is given by the Kubo formula 
\begin{equation}
 \sigma
  = -\frac 1{2\pi\hbar}\mathop{\mathrm{Tr}} \Bigl[
      j^x ( G^R - G^A )
      j^x ( G^R - G^A )
    \Bigr] \ ,
 \label{KuboFull}
\end{equation}
where `Tr' includes both the matrix trace and the spatial
integration. The chiral symmetry implies the identity
\begin{equation}
 \sigma_3 G^{R(A)}(E) \sigma_3
  = - G^{A(R)}(-E)\ ,
 \label{GRA}
\end{equation}
allowing one to trade all
advanced Green functions in Eq.\ (\ref{KuboFull}) for retarded ones and thus to
present the conductivity at zero energy in terms of retarded Green functions.
Further, in view of the Dirac character of the spectrum,
the components of the current operator are related via
\begin{equation}
 \sigma_3 j^x
  = -j^x \sigma_3
  = i j^y \ .
 \label{current-x-y}
\end{equation}
By making use of Eqs.\ (\ref{GRA}) and (\ref{current-x-y}) 
the Kubo formula can be cast in the following form:
\begin{equation}
 \sigma(E=0)
  = -\frac 1{\pi\hbar} \sum_{\alpha = x,y}
    \mathop{\mathrm{Tr}} \Bigl[
      j^\alpha G^R
      j^\alpha G^R
    \Bigr].
 \label{KuboRR}
\end{equation}
At first glance, this expression may be thought to be 
zero due to gauge invariance. Indeed, the
right-hand side of Eq.\ (\ref{KuboRR}) is proportional to the second derivative
of the partition function $Z[\mathbf{A}] = \mathop{\mathrm{Tr}} \log
G^R[\mathbf{A}]$ (or, equivalently, first derivative of the current
$\mathop{\mathrm{Tr}} j^\alpha G^R[\mathbf{A}]$) with respect to the constant
vector potential $\mathbf{A}$. The gauge invariance implies that a constant
vector potential does not affect gauge-invariant quantities like the partition
function or the current, so that the derivative is zero. 
This argument is,
however, not fully correct, in view of a quantum anomaly. 
The elimination of $\mathbf{A}$ amounts technically to a shift in the
momentum space $\mathbf{k} \to \mathbf{k} - e\mathbf{A}$, which naively does
not change the momentum integral. If we consider a formal expansion in the
disorder strength, this argument will indeed hold for all terms involving
disorder but not for the zero-order contribution. The momentum integral $\int
d^2 k \mathop{\mathrm{Tr}} j^\alpha G^R_0(\mathbf{k})$ is
ultraviolet-divergent and the shift of variable is illegitimate. This anomaly
was first identified in \textcite{schwinger62} for $1+1$-dimensional
massless Dirac fermions. In the Schwinger model, the polarization operator is
not affected  by an arbitrary external vector potential $\mathbf{A}(x,t)$ and
is given by the anomalous contribution, yielding a photon mass in the $1+1$
electrodynamics \cite{schwinger62,peskin}. In the present context, the role of
$\mathbf{A}(x,t)$ is played by the chiral disorder. The explicit calculation of
the zero-order (disorder-free) diagram 
yields (including a factor of two accounting for spin)
\begin{equation}
 \sigma
  = -\frac{8 e^2 v_0^2}{\pi\hbar} \int \frac{d^2k}{(2\pi)^2}\,
    \frac{\delta^2}{(v_0^2 k^2 + \delta^2)^2}
  = \frac{4 e^2}{\pi h}.
 \label{universal}
\end{equation}
Therefore, the dimensionless conductivity acquires the universal value
$4/\pi$, with no corrections at any order in the disorder strength. 
This result was earlier obtained in \textcite{ludwig94} for the
abelian case and in \textcite{tsvelik95} for a certain model of
non-abelian gauge field.

\subsubsection{Disorders preserving $C_z$ chirality:  Gade-Wegner 
criticality}
\label{s6.7.4}

Let us now turn to the disorder which preserves the $C_z$-chirality, $H =
-\sigma_3 \tau_3 H \sigma_3 \tau_3$; according to Table~\ref{Tab:sym},
the corresponding coupling
constants are  $\beta_0$, $\alpha_\perp$, $\gamma_\perp$ and
$\beta_z$. If no time-reversal symmetries are preserved, the system
belongs to the chiral unitary (AIII) class. 
The combination of $C_z$-chirality and the time reversal invariance
$T_0$ (only couplings $\gamma_\perp$ and $\beta_z$ are present) corresponds the
chiral orthogonal 
symmetry class BDI; this model has already been discussed in
Sec.~\ref{s6.6.2}.  Finally, the combination of $C_z$-chirality and
$T_z$-symmetry (couplings $\gamma_\perp$ and $\beta_0$) falls into the
chiral symplectic symmetry class CII. The RG flow and DOS in these
models have been analyzed in \textcite{guruswamy99} in notations 
$g_A=\gamma_\perp/2$, $g_m=\beta_z/2$, $g_\mu=\beta_0/2$,
$g_{A'}=\alpha_\perp/2$. .  In all the cases, the
resulting theory is of the Gade-Wegner type, see Sec.~\ref{s6.6}.

In contrast to random gauge field models ($C_0$ chirality), the proof
of the universality of the conductivity based on gauge-invariance
argument does not apply to models with $C_z$ chirality. 
Nevertheless, the zero-energy conductivity is found to have the same
universal value, $\sigma = 4e^2 /\pi h$, in the limit of weak disorder
\cite{ostrovsky06,ryu07a}. In this case, however, there are
corrections to this value 
perturbative in the disorder strength. In particular, the leading
correction is found in the second order \cite{ostrovsky06},
 $\delta\sigma^{(2)}= (e^2/\pi h)(\beta_0 - \beta_z)^2$.

The following comment (applicable both to
$C_0$ and $C_z$ chiral models) on the infrared 
regularization of the problem is in order here. This role is played in
Eq.~(\ref{universal}) by 
$\delta$, which is an infinitesimal imaginary part in the denominator of the
Green function. Physically, it has a meaning of the inverse
electron lifetime or,
alternatively, a dephasing rate, and can be thought of as modelling processes
of escape of electrons in some reservoir or some dephasing mechanism. 
As discussed in \textcite{ostrovsky06}, for a different type of the infrared
regularizations (i.e. by finite frequency) the critical conductivity will take
a different value, while still being of order $e^2/h$.

\subsubsection{Dirac Hamiltonians for dirty $d$-wave superconductors.} 
\label{s6.7.5}

We briefly discuss the application of Dirac fermion theory to
disordered $d$-wave spin-singlet superconductors with $d_{x^2-y^2}$
symmetry. In such systems the superconducting gap vanishes at four
points in the Brillouin zone, and the dispersion relation near these
points of the Dirac type. One is thus led to analyze the low-energy
physics of the problem in terms of a four-flavor Dirac Hamiltonian. 
The investigation of symmetries of this problem in the presence of
different types of impurities and the corresponding RG treatment were
pioneered in \textcite{nersesyan95}; for a recent detailed analysis
the reader is 
referred to \textcite{altland02}. The main physical quantity of interest
for this problem is the low-energy behavior of DOS. 

Following \textcite{nersesyan95,altland02},
we specialize on potential disorder, assume that the spin-rotation
invariance is preserved, but allow for a possibility of time-reversal
symmetry breaking (i.e. by magnetic field in vortices in type-II
superconductors). In full analogy with the two-flavor model,
Sec.~\ref{s6.7.1}, it is crucially important, whether the disorder
couples the nodes or not.  
Depending on the range of disorder and the interval
of energies considered, one can distinguish three different
situations:

(i) short-range disorder: all four nodes are coupled;

(ii)  long-range disorder: inter-node scattering is negligible, and
the nodes are decoupled. The problem
acquires then a single-node character;

(iii) each node is coupled to the opposite one but not to the other
two. The system then decomposes in two parts -- each of them
describing a pair of the oppositely located nodes. It was shown in
\textcite{nersesyan95} that, in view of strong anisotropy of the nodal Dirac
Hamiltonians, there is an intermediate energy range where this model
becomes physically justified. 

Combining these three types of disorder with possibilities of preserved
or broken time-reversal invariance, one gets six distinct models.
Their symmetries, emerging types of criticality, and the corresponding
behavior of DOS are summarized in Table~\ref{Tab:d-waves}. When all
four nodes are coupled, the system is in the conventional localized
regime of symmetry classes CI and C describing spin-singlet
superconductors \cite{senthil98,senthil99a}. 
When only the opposite nodes are coupled, the
T-invariant problem becomes a model of non-abelian random gauge field,
and the theory acquires the WZ term, leading to the
$E^{1/7}$ scaling of the DOS, see Sec.~\ref{s6.7.3}. For
broken $T$-invariance, the two-node problem is described by the
class-$C$  $\sigma$-model with the $\theta=\pi$ topological term,
i.e. it is in the SQH transition universality class, Sec.~\ref{s6.4}.
The DOS scales as $E^{1/7}$ in this case as well. 
Finally, when the nodes are completely decoupled, the Hamiltonian for
each of them takes the form analyzed in Sec.~\ref{s6.7.2}.
More specifically, when T-invariance is preserved, the problem reduces
to the class-AIII model of random (abelian) vector potential,
Sec.~\ref{s6.7.3}, 
with the DOS following the non-universal power law (\ref{e6.7.17}), 
$\rho(E)\sim|E|^{(1-2g_A)/(1+2g_A)}$. In the case
of broken $T$-invariance, the problem belongs to the class A and is
described by the Pruisken $\sigma$-model with topological term, 
i.e. it is in the universality class of the IQH critical point. In the
latter situation, the DOS is uncritical. 

We also mention two further types of disorder that have been studied 
in detail in the literature. In the case of magnetic impurities botn
$T$-invariance and spin rotation symmetry are broken and thus the
system belongs to class D \cite{bocquet00,senthil00}, see Sec. 
\ref{ss:tqhe}. For the case of perfect nesting, where the chemical
potential $\mu{=}0$, and very strong potential scatterers
one encounters the symmetry class AIII leading to the Gade phase
\cite{altland02a}.

\begin{table}
\begin{center}
\begin{tabular}{ccccc}
\hline\hline
nodes coupled\  & \ T \  &\ Class\ &\ Criticality\ & \ DOS
\\ \hline
1 & $+$ & AIII & WZW           & $|E|^{(1-2g_A)/(1+2g_A)}$ \\
2 & $+$ & CI   & WZW           & $|E|^{1/7}$ \\
4 & $+$ & CI   & ---           & $|E|$ \\
1 & $-$ & A    & $\theta$-term &  non-critical \\
2 & $-$ & C    & $\theta$-term &   $|E|^{1/7}$ \\
4 & $-$ & C    & ---           & $E^2$
\\ \hline\hline
\end{tabular}
\end{center}
\caption{Criticality and DOS in $d$-wave superconductors with
  preserved spin rotation invariance and different
types of randomness. First column: number of nodes coupled. Second
column: presence or absence of time-reversal invariance. Third column:
symmetry class. Fourth column: type of criticality. Fifth column:
low-energy scaling of DOS.}
\label{Tab:d-waves}
\end{table}

\section{Concluding remarks}
\label{s7}

This concludes our review of physics of Anderson transitions between
localized and metallic phases in disordered electronic systems and of
associated critical phenomena. In our opinion, during the recent years
a quite detailed understanding of these transtions  has emerged,
including such salient features as symmetry and
topology classification, mechanisms of criticality in quasi-1D and 2D
systems, and wavefunction multifractality. We have attempted to give a
systematic exposition of these issues and hope that this will 
help interested researchers to navigate in this  remarkably
rich field. 

For several reasons (including limits in space, time, and -- last but
not least -- our expertise), we were not able to discuss all aspects
of the problem. The selection of material in the review certainly
reflects the authors' personal 
perspective of the field. We apologize to all researchers whose work
is not sufficiently discussed in the review. 
%
%
Probably, the most important issue out of those that we largely left
apart in the review is the effects of electron-electron
interactions. (We only briefly duscussed it in Sec.~\ref{s6.3} in the
context of the IQH transition.) This by itself is a very rich and complicated
problem; we restrict ourselves to just a few remarks and references.

\subsection{A few words about interaction effects}
\label{s7.1}

Physically, the impact of interaction effects onto low-temperature
transport and localization in
disordered electronic systems can be subdivided into two distinct
effects: (i) renormalization and (ii) dephasing. 

\paragraph{Renormalization.}
The renormalization
effects, which are governed by virtual processes, become increasingly
more pronounced with lowering temperature. The importance of such
effects in diffusive low-dimensional systems was demonstrated in
by Altshuler and Aronov, see the review \cite{altshuler84}. To resum 
the arising singular contributions, Finkelstein developed the RG
approach based on the $\sigma$-model for an interacting system, see
\textcite{finkelstein90} for a review. This made possible an analysis of the
critical behavior at the localization transition in $2+\epsilon$
dimensions in the situations when spin-rotation invariance is broken (by
spin-orbit scattering, magnetic field, or magnetic impurities). 
However, in the case of preserved spin-rotation symmetry 
it was found that the strength of the interaction  
in spin-triplet channel scales to infinity at certain RG scale. 
This was interpreted as some kind of magnetic instability of the
system; for a detailed exposition of proposed scenarios  
see the review \cite{belitz94}. 

Recently, the problem has attracted a
great deal of attention in connection with experiments on
high-mobility low-density 2D electron structures (Si MOSFETs) giving an
evidence in favor of a metal-insulator transition \cite{abrahams01}. 
Whether these results are due to a true metallic phase existing  in
these systems or, else, are explained by interaction effects at
intermediate (``ballistic'') temperatures  remains a
debated issue. In a recent work \cite{punnoose05} the RG for $\sigma$-model
for interacting 2D electrons with a number of valleys $N>1$ was
analyzed on the two-loop level. It was shown that in
the limit of large number of valleys $N$ (in practice, $N=2$ as in Si
is already sufficient) the temperature of magnetic instability is
suppressed down to unrealistically low temperatures, and a
metal-insulator transition emerges. 

The interaction-induced renormalization effects become extremely
strong for correlated 1D systems (Luttinger liquids). While 1D systems
provide a paradigmatic example of strong Anderson localization,  a sufficiently
strong attractive interaction can lead to delocalization in such
systems. An RG treatment of the corresponding localization transition in
a disordered interacting 1D systems was developed in
\textcite{giamarchi88}, see also the book \cite{giamarchi04}. 

\paragraph{Dephasing.}
We turn now to effects of dephasing governed by inelastic processes of
electron-electron scattering at
finite temperature $T$.  The dephasing has been studied in great
detail for metallic systems where it provides a cutoff for weak 
localization effects \cite{altshuler84}. As to the Anderson
transitions, they are quantum (zero-$T$) phase transitions, and dephasing
contributes to their smearing at finite $T$. This has been 
discussed in Sec.~\ref{s6.3.10} in the context of dynamical scaling at
the IQH transition. There is, however, an interesting situation when
dephasing processes can create a localization transition. We mean 
the systems where all states are localized in the absence of
interaction, such as wires or 2D systems. At high temperatures, 
when the dephasing is strong, so that the dephasing rate
$\tau_\phi^{-1}(T)$ is larger than mean level spacing in the
localization volume, the system is a good metal and its conductivity
is given by the quasiclassical Drude conductivity with relatively
small weak localization correction \cite{altshuler84}. With lowering
temperature the dephasing gets progressively less efficient, the
localization effects proliferate, and eventually the system becomes an
Anderson insulator. What is the nature of this state? A natural
question is whether the interaction of an electron with other
electrons will be
sufficient to provide a kind of thermal bath that would assist the
variable-range hopping transport \cite{fleishman78}, 
as it happens in the presence of a phonon bath. The answer to this
question was given by \textcite{fleishman80}, and
it is negative. Fleishman and Anderson found that at low $T$ the
interaction of a ``short-range class'' (which includes a finite-tange
interaction in any dimensionality $d$ and Coulomb interaction in
$d<3$) is not sufficient to delocalize otherwise localized electrons, 
so that the conductivity remains strictly zero. In combination with
the Drude conductivity at high-$T$ this implies the existence of
transition at some temperature $T_c$. 

This conclusion was recently
corroborated by an analysis \cite{gornyi05,basko06} in the framework
of the idea of Anderson localization in Fock space \cite{altshuler97a}.
In these works the temperature dependence of conductivity $\sigma(T)$  
in systems with localized states and weak electron-electron
interaction was studied. It was found that with decreasing $T$ the
system first shows a crossover from the weak-localization regime 
into that of ``power-law hopping'' over localized states 
(where $\sigma$ is a power-law function of $T$), and then undergoes a
localization transition. The transition  is obtained both within a
self-consistent Born approximation \cite{basko06} and an approximate
mapping onto a model on the 
Bethe lattice \cite{gornyi05}. The latter yields also
a critical behavior of $\sigma(T)$ above $T_c$, which has a
characteristic for the Bethe lattice non-power-law form $\ln\sigma(T)
\sim (T-T_c)^{-1/2}$, see Sec.~\ref{s2.3.8}. Up to now, this
transition has not been observed in experiments\footnote{Of course, 
in a real system, phonons are always present and provide a bath
necessary to support the hopping conductivity at low $T$, so that
there is no true transition. However, when the coupling to phonons is
weak, this hopping conductivity will have a small prefactor, yielding 
a ``quasi-transition''.}, which indicate
instead a smooth crossover from the metallic to the insulating phase
with lowering $T$ \cite{hsu95,vankeuls97,khavin98,minkov07}. 
The reason for this discrepancy remains unclear. 
A recent attempt to detect the transition in numerical simulations 
also did not give a clear
confirmation of the theory \cite{oganesyan07}, possibly because of
strong restrictions on the size of an interacting system that can be
numerically diagonalized. 
 
\subsection{Experimental studies of localization transitions}
\label{s7.2}

Of all the localization transitions, the best studied experimentally is the IQH
transition. We have discussed the corresponding experimental findings in
Sec.~\ref{s6.3.10}. Its superconducting counterparts (SQH and TQH transitions)
have not been observed yet, although several physical realizations of
them were proposed, Sec.~\ref{s6.4.1} and \ref{sss:tqhe-1}. Much
effort has been invested in research on strongly interacting
disordered 2D systems in zero magnetic field but it remains
controversial whether what is observed there is a true metal-insulator
transition, see Sec.~\ref{s7.1}.  

Below we briefly review the
situation with the experimental observation of the Anderson transition in
3D electronic and optical
systems. For electronic systems (doped semiconductors,
Sec.~\ref{s7.2.1}) 
the localization transition has been observed unambiguously. However, a
theoretical analysis of the critical behavior is complicated by the
presence of Coulomb interaction, which modifies the critical
behavior. As a result, one can not expect that the experimentally
extracted critical exponents agree with numerical values obtained from
computer simulations on non-interacting systems. The 
localization of light (Sec.~\ref{s7.2.2}) has an advantage in this respect,
since the photon-photon interaction 
is negligibly small. 
However, it turns out that implementation of sufficiently strong 
disorder so as to reach the Anderson transition
and strong localization of light in 3D is a remarkably complicated 
endeavor. 
The second major difficulty is posed by the absorption.

\subsubsection{Anderson transition in doped semiconductors} 
\label{s7.2.1}

The 3D localization transition was extensively studied on doped semiconductor
systems, such as Si:P, Si:B, Si:As, Ge:Sb. In most of the works, samples with a
substantial degree of compensation [i.e. acceptors in addition to donors,
e.g. Si:(P,B)] were used, which allows one to vary the amount of disorder and
the electron concentration independently. On these samples, 
values of the conductivity exponent $s$ in the vicinity 
of $s\approx 1$ were reported 
\cite{thomas82,zabrodskii84,field85,hirsch88} with scattering of values and
the uncertainties of the order of $10\%$. A similar result was obtained
for an amorphous material Nb$_x$Si$_{1-x}$. 
(Recall that in 3D $s$ is expected to be equal in 3D to
the localization length exponent $\nu$ according to the scaling relation
$s=\nu(d-2)$) 

On the other hand, the early study of the transition in undoped Si:P 
\cite{paalanen82,thomas83,rosenbaum83} gave an essentially different result,
$s\approx 0.5$. A resolution of this discrepancy was proposed in
\textcite{stupp93,stupp94} where it was found that the actual
critical region in an uncompensated Si:P is rather narrow and that the scaling
analysis restricted to this range yields $s\approx 1.3$. A more recent study
\cite{waffenschmidt99}, 
which employed uniaxial stress to tune through the transition (as used in
\textcite{paalanen82,thomas83,rosenbaum83}), has essentially confirmed these
conclusions, yielding $s=1.0\pm 0.1$, in agreement with the values obtained for
samples with compensation. Further, in this work dynamical scaling near
the transition with varying temperature was demonstrated; the corresponding
dynamical exponent was found to be $z=2.94\pm 0.3$. Good
scaling was also observed in a similar experiment on uncompensated
Si:B \cite{sarachik99}, however with somewhat different critical exponents
($s\approx 1.6$, $z\approx 2$). A possible explanation for this
discrepancy is that the
temperatures reached in this work were not sufficiently low. Another
possibility is that Si:B belongs to a different universality class, in
view of stronger spin-orbit scattering.

The fact that the experimental value $s\approx 1$ found in the
majority of works  differs from what one
would expect based on numerical studies for non-interacting systems 
is not surprising, since the
Coulomb interaction affects the critical exponents. For a more detailed
discussion of the experimental data and their comparison with theoretical
expectations for the Anderson transitions in the presence of Coulomb
interaction  the reader is refered to \textcite{belitz94}.

\subsubsection{Anderson localization of light}
\label{s7.2.2}

Experimental efforts to observe the Anderson localization with light
turned out to be very challenging. The main difficulty is, that the 
characteristic signature of wave localization, the exponential
decrease  $e^{-\xi/L}$ of the transmission with the system size $L$,
is often hard to disentangle from another exponential $e^{-\ell_a/L}$ 
originating from photon absorption.  
 
\textcite{garcia91,genack91} studied the transmission of microwave radiation
through a random mixture of aluminum and Teflon spheres inside a
copper tube. In the vicinity of the localization
transition, they report power law scaling of the
effective diffusion constant $D(L)\sim L^{-1}$.
The presence of strong absorption in the microwave system, makes it
complicated to interpret the data that were reported as 
evidence for the localization of light.

To diminish photon losses, 
\textcite{wiersma97} used a powder from 
$\mu$m gallium arsenide crystals employing near infrared radiation. 
However, still the exponential signal of localization reported in this 
work is not undisputed, again due to the presence of residual 
absorption \cite{scheffold99}. A similar study in silicon powders did
not show any hint of localization \cite{gomezrivas99}.

In order to overcome the notorious problem of separating localization
from absorption, an interesting proposal has been made by
\textcite{chabanov00}. Localization is accompanied by 
very large mesoscopic fluctuations in the spectral function which
carry over also to the frequency dependent 
transmission function $T(\omega)$. Hence, the measurement of the statistical
properties of $T(\omega)$, e.g. its variance, could provide a
sensitive means to uncover quantum interference effects. Using this
method the authors were able to confirm localization in
quasi-one-dimensional waveguides; an application to a
three-dimensional system, a mixture of aluminum spheres,
did not show signatures of strong localization.

To summarize, up to now there does not seem to be a convincing
demonstration for the Anderson localization and transition
in 3D systems. However, in very recent work
\textcite{stoerzer06} report the observation of an anomalous
time dependency of the light diffusion in a TiO$_2$ powder
indicating the vicinity of the Anderson critical point. 
Another very promising research direction has
appeared  with the advent of photonic crystals \cite{busch07}, where in the presence
of disorder a localization transition should take place for states
near a photonic band edge \cite{john84,john87,busch99}.

\section{Acknowledgments}

We express our gratitude to J. Chalker, Y. Fyodorov, I. Gornyi, I. Gruzberg, 
A. Ludwig, R. Narayanan, P. Ostrovsky, D. Polyakov, 
A. Subramaniam, and, especially, A. Mildenberger  
for enjoyable collaborations over many years on the 
topics included in this review.
We are grateful to I. Gornyi, I. Gruzberg, Y. Fyodorov,
A. Mildenberger,  A. Subramaniam, and P. W\"olfle 
for reading the manuscript and many
valuable comments. In addition, this review has benefitted from discussions 
with I. Burmistrov, K. Busch, and A. Finkelstein.  Finally, we thank many of 
our colleagues, especially   
A. Altland, D. Belitz, M. Feigelman, M. Jan{\ss}en, V. Kagalovsky,
R. Klesse, V. Kravtsov, C. Mudry, L. Schweitzer, A. Tsvelik, P. W\"olfle, 
and M. Zirnbauer, for numerous illuminating discussions over the past years.  

This work has received support from the 
Center for Functional Nanostructures of the Deutsche
Forschungsgemeinschaft. 

\bibliographystyle{apsrmp}

\bibliography{./rmp.bib}

\end{document}